\documentclass[twocolumn,showpacs,preprintnumbers,superscriptaddress]{revtex4-2}            
\usepackage[T1]{fontenc}    
\usepackage{amsfonts,amsmath,amsbsy,amssymb,graphicx,enumerate,latexsym,bm,ulem,multirow,listings,here,qcircuit}                              
\usepackage{times}  
\usepackage{color}

\begin{document}                 
\title{ Quantum Circuits for Collective Amplitude Damping in Two-Qubit Systems }           
\author{Yusuke~Hama}  
\affiliation{Quemix Inc., 2-11-2 Nihombashi, Chuo-ku, Tokyo 103-0027, Japan}

\date{\today}  

\begin{abstract}{ 
Quantum computers have now appeared in our society and are utilized for the investigation of science and engineering. 
At present, they have been built as intermediate-size computers containing about fifty qubits and are weak against noise effects. Hence, they are called noisy-intermediate scale quantum devices.           
In order to accomplish efficient quantum computation with using these machines, a key issue is going to be the coherent control of individual and collective quantum noises.  
In this work, we focus on a latter type and investigate formulations of the collective quantum noises represented as quantum circuits. 
To simplify our discussions and make them concrete, we analyze collective amplitude damping processes in two-qubit systems.
As verifications of our formalisms and the quantum circuits,        
we demonstrate digital quantum simulations of the collective amplitude damping by examining six different initial conditions  
with varying the number of execution of an overall operation for our quantum simulations.   
We observe that our results show good numerical matching with the solution of quantum master equation for the two-qubit systems as we increase such a number.   
In addition, we explain the essence of the way to extend our formalisms to analyze the collective amplitude damping in larger qubit systems.  
These results pave the way for establishing systematic approaches to control the quantum noises and designing large-scale quantum computers.
}   
\end{abstract}

\maketitle

\section{Introdunction}\label{Intro}  
Our quantum technologies have become matured such that we are now becoming able to engineer the computational machines working under  laws of quantum mechanics, namely, quantum computers \cite{QCFeynman,QCDeutsch,QCLloyd,DiVincenzoQC,QCQINandC,QSRMP2014}.
 Toward the realization of quantum computers, historically many different types of implementations (hardware)  as well as working schemes have been proposed and realized ranging from 
 solid-state-system setups (nuclear spin systems manipulated by the nuclear magnetic resonance, silicon-based systems, quantum dots controlled by electron spin resonance and exchange interaction, superconducting circuits constructed with Josephson junctions)  \cite{QCNMR1,KanesiliconQCNature1998,ItohgroupsiliconQC,Qcsilicon1,quantumdotreview1,QCQD1,NakamuraTsaigroupSC,SCQRMP2001,circuitqedreview1,SCQRPP2017,SCQNISQ20191,SCQNISQ20192}
to atomic-molecular and optical setups (trapped ions using laser pulse, single photons with beamsplitters, phase shifters, and the Kerr medium, cavity quantum electrodynamic setups) \cite{CiracZollergate,trappedionsreview1,trappedionsreview2,trappedionNISQ2019,linearoptQC1,JKimbleCQED,cavityqedreview1}. 
In parallel, numerous quantum algorithms have been proposed and established for problems which are considered to be effective to solve by using the quantum computer,
e.g., prime factorization and database retrieval \cite{ShorQA,GroverQA,QAreview1,QAreview2}.  
Owing to these great efforts, we are now in the era where the quantum computers based on quantum circuits (circuit models) have been built with using several types of elements such as  superconducting circuits and trapped ions \cite{SCQNISQ20191,SCQNISQ20192,trappedionNISQ2019}. 
 By using the properties inherent in the quantum systems like quantum coherence and quantum entanglement,  
 we expect that the quantum computer becomes the machine which can solve quantum mechanical problems more efficiently than the existing (classical) computer, i.e., quantum supremacy \cite{PreskillNISQ2018,Qsup2}.    
  When the quantum computers become daily used and being applied to the investigation of science and engineering, we expect the advancement of quantum technologies which can support our daily lives and industries, for example,     
  material development  and drug discovery.   
 At present, however, our quantum computers have been built as intermediate-size computers including about fifty qubits and are not robust against noises yet.
  Hence, they are called noisy-intermediate scale quantum (NISQ) devices  \cite{PreskillNISQ2018,SCQNISQ20192,trappedionNISQ2019}.  
To improve the computational results, some tasks like an error mitigation  \cite{EMPRL2017andEMNature2019,EMPRX2017and2018,EMarxiv2018,EMarxiv2020}
are needed to be practiced, which are different schemes from quantum error correction \cite{QCQINandC,NemotogroupQEC,LidarWhaleygroupcollectivenoises}.
 On the other hand, from practical points of view, toward the improvement of the qualities of quantum computers (NISQ devices),
  in recent years a lot of great efforts have been made for such a goal by using NISQ devices to compute simpler quantum mechanical problems with analyzing the obtained results as well as understanding their present qualities so as to make them into more useful machines.   
 For instance,  the analysis and computation of many-body electronic systems such as  many-body wave functions and the energies of ground states for hydrogen molecules 
 as well as those of other types of molecules like LiH have been conducted \cite{SCQRPP2017,SCQNISQ20191,SCQNISQ20192,trappedionNISQ2019,QCMBEVQ2018}.
  Further, the computation of Green's functions has been investigated \cite{HMQCPRA2015,DFTDMRGarXiv2019,QCGreensfuncPRX2016,QCGreensfunction1,QCGreensfunction2}.   
Throughout these investigations, numerous quantum algorithms which suit in NISQ devices have been proposed or established, e.g., the Variational Quantum Eigensolver and Quantum Approximate Optimization Algorithm \cite{VQE0,trappedionQS2018,OpenQVQE1,SCQRPP2017,SCQNISQ20191,SCQNISQ20192,trappedionNISQ2019,QCMBEVQ2018,DFTDMRGarXiv2019,NISQchemistrymaterialsciencearXiv2020,QCGreensfunction1,OpenQVQE1,OpenQVQE2,QAOA2014}.
Another important theme of quantum computing is the quantum simulation of Ising model \cite{TIsing1,TIsing2,TIsing3}. 
This simple model can be applied to study many types of problems including the simulation of quantum magnets and optimization problems like the traveling salesman problem.  
For the simulation of transverse Ising model based on quantum annealing method, see for instance \cite{quantumannealing}. 

The quantum computers can be used to study not only for the analysis of physics and chemistry of many-body electronic systems represented as the unitary processes.  
They also enables us to study non-unitary processes of systems under consideration, i.e., the dynamics in open quantum systems (quantum dynamics under dissipation) \cite{OpenQVQE1,OpenQVQE2,carmichaeltxb,Agarwalltxb,opendynamicstext,IOPreviewopendynamicsentanglement}.  
In nature, many systems exhibit the open quantum dynamics originating in the interaction between an environment and systems under consideration.     
Examples include the transverse and longitudinal relaxation processes (thermalization and dephasing processes) of electron and nuclear spins in solid-state systems such as GaAs semiconductors \cite{quantumdotreview1,electronnuclear1,electronnuclear2,electronnuclear3,solidsuperradiance2016},
spontaneous emission in atomic-molecular and optical systems like cavity quantum electrodynamic systems \cite{carmichaeltxb,Agarwalltxb,opendynamicstext,superradiance1,GH82,cavityqedreview2},
and further, the open quantum dynamics emerge even in biological systems such as excitonic dynamics (transport phenomena) in photosynthetic systems \cite{QBiology1,QBiology2,QBiology3,QBiology4}.
The investigation of open quantum dynamics is also important from the quantum-engineering perspectives.     
The qubits constituting the quantum devices (hardware) are affected by many kinds of quantum noise effects, for instance, thermal noise, bit and phase flips, and depolarization \cite{EkertgroupQCdissipation}
.   
Thus, the qubits as components of the quantum computers are actually showing the non-unitary dynamics (amplitude damping, dephasing, and decoherence). 
To improve the qualities of present quantum computers (NISQ devices), we must understand deeply the ways qubits are subjected to the quantum noises and how to deal with them.  
The first possibility we can think of is the qubits couple with the environments individually and experience the noise effects (individual noises).
In contrast, another possibility is that the qubits couple collectively with a single (common) environment and experience collective noises \cite{Nathanmastereqsim}.  
Examples include collective decoherence or dephasing \cite{Nathanmastereqsim,DuancollectivedecohePRA1998,LidarWhaleygroupcollectivenoises,ViolagroupNJP2002,YamamotogroupcollectivenoiseQA}. 
Other important examples are superradiant and subradiant effects \cite{carmichaeltxb,Agarwalltxb,opendynamicstext,YamamotogroupcollectivenoiseQA,superradiance1,GH82,cavityqedreview2,WallraffgroupSCsupsub,solidsuperradiance2016,superradianceYavuz,NVsuperradiantdecay,NakamuragroupsubradianceSC,superradianceSCPRL2020}.
Futher,  the environments can become either uncorrelated or correlated \cite{TCADPRA2002R,TCADPRA2003,EMarxiv2018,EMarxiv2020,Cnoisepra2005,NoviasprlCnoise,KitaevgroupCnoiseprl2006}.
Since the quantum computation are performed in terms of the single- and two-qubit gate operations, we can think that the qubits are going to get influenced by both the individual and collective noises.
Thus, in order to conduct the error analysis of the quantum computation we must establish some schemes for the coherent control of these two types of noises. 
Besides the quantum error correction \cite{QCQINandC,NemotogroupQEC,LidarWhaleygroupcollectivenoises} and error mitigation \cite{EMPRL2017andEMNature2019,EMPRX2017and2018,EMarxiv2018,EMarxiv2020}, 
an alternative approach we can think of is to model the quantum noises by representing them as the quantum circuits \cite{QCQINandC}. 
In this way, we expect that it brings insight into the coherent control of the individual and collective noises as well as the conduction of the error analysis by the single- and two-qubit gate operations and measurements.
Further, it enable us to design architectures of large-scale quantum computers.

 In this paper, we focus on the collective quantum noise effects and investigate the formalisms as well as the quantum algorithms for their generations.
 Here, we analyze in detail the amplitude damping processes (relaxation or energy dissipation processes) at zero temperature.       
 To have a concreteness and make our argument simple, we analyze this problem throughout the collective amplitude damping processes in two-qubit systems.

 The structure of this is paper is given as follows. 
 It begins in Sec. \ref{OpenQKrausRep} with presenting the basics of the amplitude damping in the single-qubit system at zero temperature. 
 We start with this description so that before demonstrating the complex analysis of 
 the collective amplitude damping in the two-qubit systems, throughout this simpler case 
  we can understand and get used to the physical essence of the amplitude damping (or more broadly the open quantum dynamics) with using both
   quantum master equation and Kraus representation.  
  In this way, we can see clearly how the analysis of the amplitude damping in the single-qubit system is extended to that of the two-qubit system and enable us to construct the quantum circuit for the collective amplitude damping in the two-qubit systems.
 Then in Sec. \ref{BSCR}, which presents the main result of this paper, 
 we discuss the analysis of the collective amplitude damping in the two-qubit systems.
 First, we analyze the effective representation of the Hilbert space for the description of such processes. 
 We show that it is the direct-sum-spin-space representation obtained by the composition of spin angular momenta between the two qubits.
Based on it, we construct the formalisms as well as the quantum algorithms which generate the collective amplitude damping
and represent them as the quantum circuits by using the Kraus representation. To demonstrate numerical verifications of our quantum circuits,   
we present our quantum-simulation results obtained by the usage of qiskit \cite{IBMQqiskiturl}. 
We discuss in detail the interpretations of these results from both physical and computational perspectives.  
In Sec. \ref{largeQextension}, we explain the essence of extending our formalisms to analyze the collective amplitude damping in larger qubit systems.
Sec. \ref{conclusionanddiscussion} is devoted to the conclusion and discussion of this paper.

\section{ Single-Qubit Amplitude Damping} \label{OpenQKrausRep}
In this section, we discuss the amplitude damping of a single qubit  at zero temperature.    
We give descriptions with using both the quantum master equation and  Kraus representation and show how they are connected to each other.
Then, we construct the quantum circuit and present the digital quantum simulation result.          

\subsection{Quantum Master Equation} \label{QMeqsingleQ}
\begin{figure}[t!] 
\includegraphics[width=0.4 \textwidth]{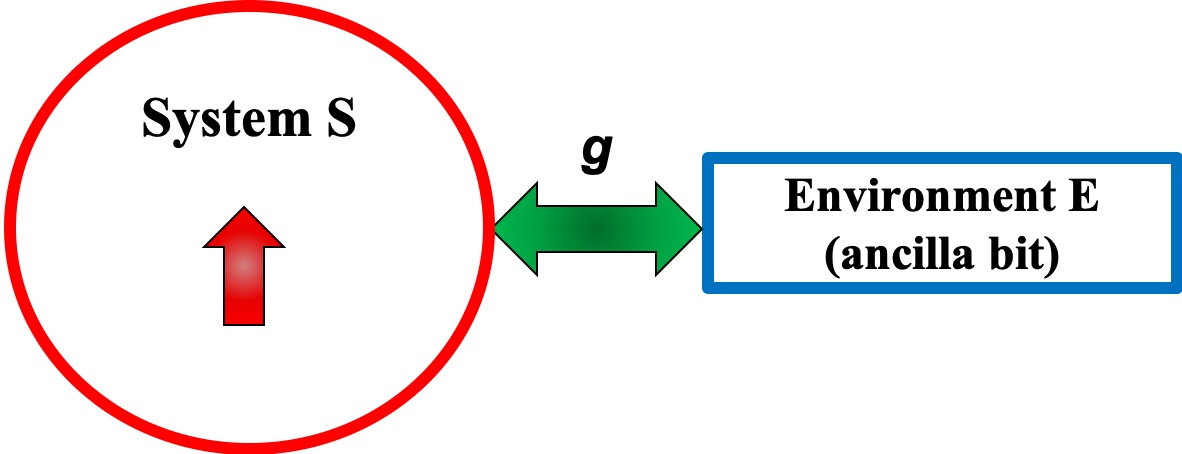}
\caption{ Schematic illustration of open quantum system for the single-qubit amplitude damping consisting of the system S and the environment E. 
They are interacting with a coupling constant $g$ and the system S exhibits the open quantum dynamics represented as the non-unitary dynamics. }  
\label{singleQandE} 
\end{figure}

The open quantum dynamics is the quantum dynamics such that the system under consideration time evolutes as a non-unitary process.      
Such a situation emerges when the system is interacting with an environment (a system which we do not know about its information).       
Namely, the composite system of the environment and the system under consideration is called the open quantum system.
 The quantum dynamics of system under consideration is showing the dissipation process due to the coupling with the environment.
  In other words, it is an information-exchange process between the system and the environment.  
  In Fig. \ref{singleQandE}, we show the schematic illustration of the open quantum system.  
The examples of the open quantum dynamics include the amplitude damping (the relaxation process), dephasing process, and decoherence.

Let us describe mathematically the open quantum dynamics. We denote the density matrix of total system (the system S plus the environment E) as $\rho_{\text{SE}}$.
The open quantum dynamics is described as the non-unitary time evolution of a reduced density matrix of the system defined by  $\rho_{\text{S}} \equiv \text{Tr}_{\text{E}} [\rho_{\text{SE}}] $.  
Here, $\text{Tr}_{\text{E}} $ denotes the trace operation with respect to the environmental degrees of freedom.   
By definition, the reduced density matrix $ \rho_{\text{S}} $ is solely described by the system degrees of freedom. 
The time evolution of  $ \rho_{\text{S}} $ is represented by quantum master equation \cite{QCQINandC,carmichaeltxb,Agarwalltxb,opendynamicstext}
\begin{align}
\frac { \partial \rho^{\text{I}}_{\text{S}} (t)}{\partial t } =   \sum_j \left[  L_j     \rho^{\text{I}}_{\text{S}} (t)   L^\dagger_j  - \frac{1}{2} \big{\{}   L^\dagger_j L_j,  \rho^{\text{I}}_{\text{S}} (t)   \big{\}} \right],
\label{masterequation1}
\end{align}
where $  \rho^{\text{I}}_{\text{S}} (t) $ is the reduced density matrix of the system in the interaction picture at time $t.$ 
By abbreviating the system Hamiltonian as $  H_{\text{S}},$ 
$  \rho^{\text{I}}_{\text{S}} (t) $ is defined by  $ \rho^{\text{I}}_{\text{S}} (t) = e^{ \frac{i  H_{\text{S}} t}{\hbar} }    \rho^{\text{S}}_{\text{S}}  (t)    e^{ \frac{ - i  H_{\text{S}} t}{\hbar} }$, 
where  $ \rho^{\text{S}}_{\text{S}}  $ is the reduced density matrix in the Schr$\ddot{\text{o}}$dinger picture. 
The right-hand-side of Eq.  \eqref{masterequation1} describes the non-unitary processes given by Lindblad operators $L_j$.
 The index $j$ denotes the number of channels of the non-unitary processes.   
The parenthesis in the second  term represents the anti-commutator of two operators defined by $ \big{\{}   A, B   \big{\}}  \equiv AB+BA. $
Let us now focus on the case of amplitude damping in the single-qubit (spin) system  at zero temperature.
The system Hamiltonian is given by $ H_{\text{S}}  =  \hbar \omega \frac{ Z  }{2} $, where $Z$ is the $Z$ gate, and the quantum master equation is given by \cite{carmichaeltxb,Agarwalltxb,opendynamicstext}
\begin{align}
\frac { \partial \rho^{\text{I}}_{\text{S}} (t)}{\partial t } =  \gamma \left[  \sigma^-    \rho^{\text{I}}_{\text{S}} (t)   \sigma^+
 - \frac{1}{2} \big{\{}   \sigma^+  \sigma^- ,  \rho^{\text{I}}_{\text{S}} (t)   \big{\}} \right],
\label{masterequation2}
\end{align} 
where $\sigma^\pm = \sigma^x \pm i \sigma^y$ with $ \sigma^x$ and $\sigma^y$ are  $x$ and $y$ components of Pauli matrices, respectively. 
The Pauli matrices $ \sigma^x$ and $\sigma^y$ are identical to the single-qubit gates $X$ and $Y$, respectively. 
We present the matrix representations of $X,Y,$ and $Z$ gates in the computational basis in Eq. \eqref{XYZ}  in Appendix. \ref{Qgates}.
$\gamma$ is the decay rate of the qubit. In this case, there is only one dissipation channel and the Lindblad operator is $\sqrt{\gamma}\sigma^-$.
Here the up-spin state and down-spin state are identical to $ | 0 \rangle $ and $ | 1 \rangle $ of the computational basis, respectively. 
They are represented in vector forms as $ | 0 \rangle = (1,0)^{\text{T}}$ and $ | 1 \rangle  = (0,1)^{\text{T}}$, where the superscript T denotes a transposition.
 \subsection{ Kraus Representation and Quantum Circuit}\label{KRQCsingleQ}
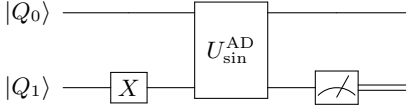
\begin{figure}[H] 
\centering
\mbox{ 
\Qcircuit @C=2em @R=2em { 
    \lstick{| Q_0 \rangle} & \qw      & \multigate{1}{U_{\mathrm{sin}}^{\mathrm{AD}}} & \qw & \qw \\
    \lstick{| Q_1 \rangle} & \gate{X} & \ghost{U_{\mathrm{sin}}^{\mathrm{AD}}}        & \meter & \cw
} 
} 
\caption{ Schematic quantum circuit for the amplitude damping of the single qubit.  
The qubit $Q_0$ is regarded as the system S while $Q_1$  (ancilla bit) as the environment E.
By applying the $X$ gate to the ancilla bit $Q_1$, the environment E ($Q_1$) is initialized such that it is initially in the ground state   
$ \rho^{\text{in}}_{\text{E}} =   | 1 \rangle_{\text{E}} \langle 1 |.$ After then, the overall unitary transformation $U^{\text{AD}}_{\text{sin}} $ is executed on both  $Q_0$ and $Q_1$. 
It describes the energy-exchange process between the system S and the environment E. At the end, the measurement is performed on $Q_1$. 
As a result, we obtain the reduced density matrix $ \rho^{\text{out}}_{\text{S}} $ describing the amplitude damping of $Q_0$.
}
\label{BellQCsingleQrelax} 
\end{figure}

\begin{figure}[H] 
\centering
\mbox{ 
\Qcircuit @C=1.5em @R=1.5em { 
    \lstick{| Q_0 \rangle} & \qw        & \qw      & \ctrl{1}     &    \gate{R_y \big{(} \vartheta (t)\big{)}}   &  \ctrl{1}   & \qw      & \qw      \\
    \lstick{| Q_1 \rangle} & \gate{X} & \qw      & \targ        &  \ctrl{-1}  & \targ       & \qw      & \meter & \cw \\
} 
} 
\caption{
Quantum circuit for the amplitude damping of the single qubit at the time $t$. It is constructed by 
Eqs.  \eqref{singleQrelaxationunitary2} - \eqref{singleQrelaxDMKrausrep2}
with Eqs.  \eqref{singleQrelaxQC}  and \eqref{singleQrelaxationunitary3}.}
\label{overallsingleQrelaxQC} 
\end{figure}
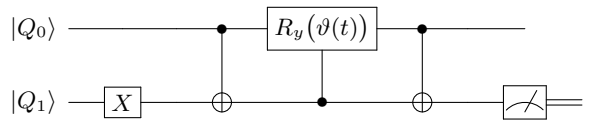

Next, let us construct the quantum circuit for the amplitude damping of the single qubit. This is done in the following way.  
First, we regard the single qubit  $Q_{0}$ as the system S while an ancilla bit $Q_{1}$ as the environment E. 
The quantum state of the total system is described by four tensor-product vectors
\begin{align}
 | 0 \rangle_{\text{S}} \otimes  | 0 \rangle_{\text{E}} & =
	\left ( \begin{array}{c} 
		 1  \\
		 0 \\
		  0  \\
		   0 
		\end{array} \right), \quad 
  | 0 \rangle_{\text{S}} \otimes  | 1 \rangle_{\text{E}} =
	\left ( \begin{array}{c} 
		 0  \\
		 1 \\
		  0  \\
		   0 
		\end{array} \right), \notag\\
		   | 1 \rangle_{\text{S}} \otimes  | 0 \rangle_{\text{E}} & =
	\left ( \begin{array}{c} 
		 0  \\
		 0 \\
		  1  \\
		   0 
		\end{array} \right), \quad 
   | 1 \rangle_{\text{S}} \otimes  | 1 \rangle_{\text{E}} = 
	\left ( \begin{array}{c} 
		 0  \\
		 0 \\
		  0  \\
		   1 
		\end{array} \right).
\label{singleQrelaxbasisvector1}
\end{align}
Hereinafter, let us rephrase the above four states as 
$   | 0 \rangle_{\text{S}} \otimes  | 0 \rangle_{\text{E}} = |0 0 \rangle_{\text{SE}}, | 0 \rangle_{\text{S}} \otimes  | 1 \rangle_{\text{E}} = |0 1 \rangle_{\text{SE}},
 | 1 \rangle_{\text{S}} \otimes  | 0 \rangle_{\text{E}} = |1 0 \rangle_{\text{SE}}, | 1 \rangle_{\text{S}} \otimes  | 1 \rangle_{\text{E}} = |11 \rangle_{\text{SE}} $.
To formulate the amplitude damping, first we initialize the ancilla bit $Q_{1}$. 
Physically, the amplitude damping of the qubit $Q_{0}$ (the system S) is induced by the environment in the ground state, to which we assign $ | 1 \rangle_{\text{E}}$. 
Therefore, in order to describe such a circumstance, 
we operate the $X$ gate on $Q_{1}$  and we have $ \rho^{\text{in}}_{\text{E}} =  | 1 \rangle_{\text{E}} \langle 1 |. $  
Here $ \rho^{\text{in}}_{\text{E}} $ is the density matrix which is expressed solely by the environmental degrees of freedom.
We have written the superscript ``in" to represent that it is the density matrix at an initial time.  
On the other hand, we take the initial state of $Q_{0}$ to be $ | 0 \rangle$ state (excited state).
Hence, the initial state of the total system is $ \rho^{\text{in}}_{\text{tot}} = \rho^{\text{in}}_{\text{S}} \otimes \rho^{\text{in}}_{\text{E}} =   | 0 \rangle_{\text{S}} \langle 0 | \otimes  | 1 \rangle_{\text{E}} \langle 1 |.$ 
Second, we formulate the unitary transformation acting on the total system of S and E so that it describes the amplitude damping of S.   
The amplitude damping at zero temperature is equivalent to the decay process from $ | 0 \rangle$ to $ | 1 \rangle$, where the system S is exhibiting the emission process.  
Therefore, we formulate the considering unitary transformation acting on the total system so that it represents the energy-exchange processes between the system S and the environment E.  
Note here that the dynamics of the total systems is represented as the unitary time evolution while the dynamics of the system S is non unitary.     
The two states $|0 0 \rangle_{\text{SE}}$ and $|11 \rangle_{\text{SE}}$ do not show the energy-exchange process because for both cases the system S 
and the environment E are in the same state: they are either in $ |0  \rangle$ state (excited state) or $ |1  \rangle$ state (ground state).   
In contrast, the energy-exchange process occurs between $|0 1 \rangle_{\text{SE}}$ and $|10 \rangle_{\text{SE}}$. 
By abbreviating the considering unitary transformation as $U^{\text{AD}}_{\text{sin}}$, its matrix representation based on the four vectors in Eq. \eqref{singleQrelaxbasisvector1} is 
\begin{align} 
	U^{\text{AD}}_{\text{sin}} &   =
	\left [
		\begin{array}{cccc} 
		 1 & 0 & 0 & 0 \\
		 0 & [ U^{\text{AD}}_{\text{sin}} ]_{2,2} & [ U^{\text{AD}}_{\text{sin}} ]_{2,3} & 0 \\
		  0 & [ U^{\text{AD}}_{\text{sin}} ]_{3,2} & [ U^{\text{AD}}_{\text{sin}} ]_{3,3} & 0 \\
		   0 & 0 & 0 & 1 
		\end{array}
	\right ]  .  \label{singleQrelaxationunitary1} 
\end{align} 
The matrix component $ [ U^{\text{AD}}_{\text{sin}} ]_{2,3} $ $( [ U^{\text{AD}}_{\text{sin}} ]_{3,2} )$ represents the transition amplitude from the state $ |0 1 \rangle_{\text{SE}} $  $( |10 \rangle_{\text{SE}} )$ to 
the state $ |10 \rangle_{\text{SE}} $  $( |01 \rangle_{\text{SE}} )$.
After the operation of $U^{\text{AD}}_{\text{sin}}$,  third, we perform the measurement on the ancilla bit $Q_{1}$.
Such a procedure is equivalent to the generation of the reduced density matrix  $\rho_{\text{S}} $. 
As a result, the dynamics of the system S is described effectively as the non-unitary dynamics.  
In Fig. \ref{BellQCsingleQrelax}, we summarize the above three procedures for the generation of the amplitude damping by representing them as the quantum circuit. 
Based on the above argument, 
we now construct the mathematical representation for $\rho_{\text{S}} $ describing the amplitude damping of the system S (Kraus representation)  \cite{QCQINandC,KrausRepresentationref}.  
As explained previously, first we choose the initial state $ \rho^{\text{in}}_{\text{tot}} = | 0 \rangle_{\text{S}} \langle 0 | \otimes  | 1 \rangle_{\text{E}} \langle 1 |.$
Next, we perform the unitary transformation $U^{\text{AD}}_{\text{sin}}$ on $Q_{0}$ and  $Q_{1}$  
and the quantum state of the total system becomes 
$ \rho^{\text{out}}_{\text{tot}} = U^{\text{AD}}_{\text{sin}}   \rho^{\text{in}}_{\text{tot}}   (U^{\text{AD}}_{\text{sin}})^\dagger.$
After then, we measure the ancilla bit $Q_{1}$ with respect to the computational basis  ($ |0 \rangle_{\text{E}} $ and $ |1 \rangle_{\text{E}} $) and we obtain the reduced density matrix $\rho^{\text{out}}_{\text{S}} $.
Such a measurement procedure is mathematically represented as 
   \begin{align}
\rho^{\text{out}}_{\text{S}}   = \text{Tr}_{\text{E}} \left[  \rho^{\text{out}}_{\text{tot}}     \right] 
  = \sum_{ n_{\text{E}}  =0,1} \mathcal{M}_{n_{\text{E}}}   \rho^{\text{in}}_{\text{S}}  \mathcal{M}^\dagger_{n_{\text{E}}} ,
\label{singleQrelaxKraus1}  
\end{align}
   where the operators $\mathcal{M}_{n_{\text{E}}} $ $(n_{\text{E}} = 0,1)$ are defined by 
    \begin{align}
 \mathcal{M}_{0}  & =  \sum_{ n_{\text{S}} , n^\prime_{\text{S}}  =0,1}  
   {}_{\text{S}}\langle n_{\text{S}}   | \otimes {}_{\text{E}}\langle 0   |   \left[  U^{\text{AD}}_{\text{sin}}         \right]  | n^\prime_{\text{S}} \rangle_{\text{S}} \otimes | 1  \rangle_{\text{E}}   
 \cdot   | n_{\text{S}} \rangle_{\text{S}} \langle n^\prime_{\text{S}} |, \notag\\
  \mathcal{M}_{1}  & =  \sum_{ n_{\text{S}} , n^\prime_{\text{S}}  =0,1}  
    {}_{\text{S}}\langle n_{\text{S}}   | \otimes {}_{\text{E}}\langle 1   |   \left[ U^{\text{AD}}_{\text{sin}}          \right]  | n^\prime_{\text{S}} \rangle_{\text{S}} \otimes | 1  \rangle_{\text{E}}   
 \cdot   | n_{\text{s}} \rangle_{\text{S}} \langle n^\prime_{\text{S}} |.
\label{singleQrelaxKraus2} 
\end{align}
  The operators $ \mathcal{M}_{ n_{\text{E}} }  $  in the above equation are the Kraus operators.
   They satisfy the condition $\sum_{ n_{\text{E}} =0,1} \mathcal{M}^\dagger_{ n_{\text{E}} }   \mathcal{M}_{ n_{\text{E}} } = \boldsymbol{1}_{2 \times 2}$ with $\boldsymbol{1}_{2 \times 2}$ denoting the two by two identity matrix. 
  In order to construct the unitary transformation $U^{\text{AD}}_{\text{sin}}$ (or the Kraus operators $ \mathcal{M}_{ n_{\text{E}} }  $) 
  so that the reduced density matrix $ \rho^{\text{out}}_{\text{s}} $ represents the time evolution,   
  we need to formulate the unitary transformation $U^{\text{AD}}_{\text{sin}}$ so that it is parametrized by the single real number which has one-to-one correspondence with a time $t$.
  Further,  the elements of $ \rho^{\text{out}}_{\text{s}} $ have to be represented by the exponential damping factor as a function of $t$ and with the decay rate $\gamma$ appearing in the quantum master equation \eqref{masterequation2}. 
  To do this, we use a controlled-rotational gate around $y$ axis with an angle naming as $\vartheta(t)$  \cite{QCQINandC}.  
  For the matrix representations of the single-qubit rotational gates, see  Eq. \eqref{xyzrotations}.  
  Here, we have written the argument $t$ for the angle $\vartheta$, because as we show later, we formulate the rotational gate so that the angle $\vartheta$ has the one-to-one correspondence with the time $t$. 
  Correspondingly, to describe explicitly the time dependence let us rewrite $U^{\text{AD}}_{\text{sin}} $, $ \mathcal{M}_{ n_{\text{E}} }  $, and $\rho^{\text{out}}_{\text{S}} $  
  as $U^{\text{AD}}_{\text{sin}}  \big{(}\vartheta(t)\big{)} $, $ \mathcal{M}_{n_{\text{E}} } \big{(}\vartheta(t)\big{)}$,  and $\rho^{\text{out}}_{\text{S}} \big{(}\vartheta(t)\big{)} $, respectively.
   Then,  $U^{\text{AD}}_{\text{sin}} \big{(}\vartheta(t)\big{)} $ and $ \mathcal{M}_{ n_{\text{E}} } \big{(}\vartheta(t)\big{)} $ are expressed as
        \begin{align} 
	U^{\text{AD}}_{\text{sin}} \big{(}\vartheta(t)\big{)} &   =
	\left [
		\begin{array}{cccc} 
		 1 & 0 & 0 & 0 \\
		 0 & \cos\big{(} \frac{\vartheta(t)}{2}\big{)} & -\sin\big{(} \frac{\vartheta(t)}{2}\big{)}  & 0 \\
		  0 & \sin\big{(} \frac{\vartheta(t)}{2}\big{)}  &\cos\big{(} \frac{\vartheta(t)}{2}\big{)}  & 0 \\
		   0 & 0 & 0 & 1 
		\end{array}
	\right ],    \label{singleQrelaxationunitary2} \\
    \mathcal{M}_{0} \big{(}\vartheta(t)\big{)}  & =    \left [
		\begin{array}{cc} 
		 0 & 0 \\
		  \sqrt{ \Gamma \big{(}\vartheta(t)\big{)}  } & 0
		\end{array}
	\right ], \notag\\
	 \mathcal{M}_{1} \big{(}\vartheta(t)\big{)} &  =  \left [
		\begin{array}{cc} 
		  \sqrt{1 -\Gamma \big{(}\vartheta(t)\big{)}  }   & 0 \\
		0  & 1
		\end{array}
	\right ], \label{singleQKraus3} 
\end{align} 
where we have set $ \sqrt{1 -\Gamma \big{(}\vartheta(t)\big{)}  }   = \cos\big{(} \frac{\vartheta(t)}{2}\big{)}  $  and  $ \sqrt{\Gamma \big{(}\vartheta(t)\big{)}  }   = \sin\big{(} \frac{\vartheta(t)}{2}\big{)}  $.
From the Kraus operator $\mathcal{M}_{0} \big{(}\vartheta(t)\big{)} $ in Eq. \eqref{singleQKraus3}, we see that
the quantity $ \Gamma \big{(}\vartheta(t)\big{)} $ represents the strength of decay from  $| 0  \rangle $ to  $| 1  \rangle $. 
In contrast, the diagonal component $ \sqrt{1 -\Gamma \big{(}\vartheta(t)\big{)}  } $ in $ \mathcal{M}_{1} \big{(}\vartheta(t)\big{)} $ describes the probability amplitude such that 
the system S remains to be in the $| 0 \rangle$ state at the time  $t$.
From Eqs. \eqref{singleQrelaxKraus1} and \eqref{singleQKraus3}  the reduced density matrix  $\rho^{\text{out}}_{\text{S}} \big{(}\vartheta(t)\big{)} $ is expressed as the function of the time $t$ as
       \begin{align}
  \rho^{\text{out}}_{\text{S}} \big{(}\vartheta(t)\big{)}   & =    \left [
		\begin{array}{cc} 
		\cos^2\big{(} \frac{\vartheta(t)}{2}\big{)}   & 0  \\
		0  &  \sin^2\big{(} \frac{\vartheta(t)}{2}\big{)}
		\end{array}
	\right ].
  \label{singleQrelaxDMKrausrep} 
\end{align} 
  Mathematically, the amplitude damping is represented as the time evolution of the expectation value of the $Z$ gate (the $z$ component of the spin operator of S). 
  Let us write it as $   \langle J^z _{ \text{S} } \rangle (t).$
    From the above equation, the  amplitude damping is represented as the function of $\vartheta(t) $ as
      \begin{align}
     \langle J^z _{ \text{S} } \rangle (t) = \text{Tr}_{ \text{S} }  \left(Z   \rho^{\text{out}}_{\text{S}} \big{(}\vartheta(t)\big{)}  \right) 
     =  \cos^2\left( \frac{\vartheta(t)}{2} \right)  - \frac{1}{2}.           \label{singleQrelaxDMKrausrep2} 
\end{align} 
 Let us take  $ \sqrt{1 -\Gamma \big{(}\vartheta(t)\big{)}  }  = e^{- \frac{\gamma t}{2}}.$  
 As a result, the reduced density matrix  $\rho^{\text{out}}_{\text{S}} \big{(}\vartheta(t)\big{)} $  in Eq.  \eqref{singleQrelaxDMKrausrep} becomes equivalent to the solution of the quantum master equation  \eqref{masterequation2}. 
 Subsequently, we have the correspondence between the rotation angle $\vartheta(t)$ and the time $t$:  $\gamma t = - 2 \log \left( \cos\big{(} \frac{\vartheta(t)}{2}\big{)}       \right)$.
 To summarize the above discussion, let us represent our operations for the generation of the single-qubit amplitude damping process at the  time $t$
 described by Eqs.  \eqref{singleQrelaxationunitary2} - \eqref{singleQrelaxDMKrausrep2} as the quantum circuit 
 and show it in Fig. \ref{overallsingleQrelaxQC}.
 This quantum circuit is given by the unitary transformation
\begin{align} 
U^{\text{AD,sin}}_{\text{QC}} \big{(}\vartheta(t)\big{)} & = U^{\text{AD}}_{\text{sin}} \big{(}\vartheta(t)\big{)} \cdot  \left[  \boldsymbol{1}_{Q_0} \otimes  X_{Q_1} \right], 
  \label{singleQrelaxQC}  \\
	U^{\text{AD}}_{\text{sin}} \big{(}\vartheta(t)\big{)}  &  =
       U_{\text{C}X}[Q_0;Q_1]   \cdot U_{\text{C} R_y \big{(} \vartheta(t) \big{)} }  [Q_1;Q_0] 
      \notag\\ &      \cdot  U_{\text{C}X}[Q_0;Q_1] .
  \label{singleQrelaxationunitary3} 
\end{align} 
The controlled-unitary operators $U_{\text{C}X}[Q_0;Q_1] $  is the CNOT (or C$X$) gate. 
Here the controlled bit is $Q_0$ while the target bit is $Q_1$.
   $ U_{\text{C} R_y ( \vartheta(t) ) }  [Q_1;Q_0]  $ is the controlled-unitary operator composed of the control bit  $Q_1$ and the target bit $Q_0$, 
    on which the rotational operation $R_y ( \vartheta(t) )$ acts.  
For the details on the controlled-unitary operators, see Eqs. \eqref{CNOT2}-\eqref{CHSTRy} in Appendix \ref{Qgates}.

\subsection{ Digital Quantum Simulation} \label{QSsingleQ} 
Let us discuss the digital quantum simulation of the single-qubit amplitude damping based on 
Eqs.  \eqref{singleQrelaxDMKrausrep}, \eqref{singleQrelaxDMKrausrep2},  \eqref{singleQrelaxQC}, and  \eqref{singleQrelaxationunitary3}
with setting $\gamma = 1$. The procedures are the following. 
First, we choose ten rotation angles $\vartheta_i = \frac{\pi}{10}i$  ($i=0,1,\ldots, 9$). 
Each of them corresponds to a time $t_i = - 2 \log \left( \cos\big{(} \frac{\vartheta_i}{2}\big{)}     \right).$
By performing the unitary transformation $ U^{\text{AD,sin}}_{\text{QC}} \big{(}\vartheta_i\big{)} $ given by Eq. \eqref{singleQrelaxQC} and the measurement on the ancilla bit $Q_1$, we obtain the reduced density matrix  
$ \rho^{\text{out}}_{\text{S}} \big{(}\vartheta_i \big{)} $ expressed by  Eq. \eqref{singleQrelaxDMKrausrep}. 
It contains the information of the amplitude damping at the time $t_i$ and they are the diagonal components of $ \rho^{\text{out}}_{\text{S}} \big{(}\vartheta_i \big{)} $, 
which are equivalent to the probability weights of states $ | 0 \rangle_{  \text{S}} $ and $ | 1 \rangle_{  \text{S}} $.
Let us denote the probability weights of $ | 0 \rangle _{  \text{S}  }$ and $ | 1 \rangle _{  \text{S}  } $ at the time $t_i$ as 
$ w_{   | 0 \rangle_{  \text{S}  }} (t_i) $ and $ w_{   | 1 \rangle_{  \text{S}  }} (t_i) $, respectively. 
From Eqs.  \eqref{singleQrelaxDMKrausrep} and  \eqref{singleQrelaxDMKrausrep2},  
we have $  \langle J^z _{ \text{S} } \rangle (t_i) =  
\frac{   w_{   | 0 \rangle_{ \text{S} }} (t_i) -   w_{   | 1 \rangle_{ \text{S} }} (t_i) }{2}$, 
where we have used  $ w_{   | 0 \rangle_{  \text{S}  }} (t_i) =  \cos^2\left( \frac{\vartheta_i }{2} \right)$ and 
$w_{   | 1 \rangle_{  \text{S}  }} (t_i) =  \sin^2\left( \frac{\vartheta_i }{2} \right) $. 
On the other side, what we can compute with qiskit  are the numerical values of 
$ w_{   | 0 \rangle_{  \text{S}  }} (t_i) $ and  $w_{   | 1 \rangle_{  \text{S}  }} (t_i) $. 
These are obtained by the overall operation given in Fig. \ref{overallsingleQrelaxQC} plus the measurement on $Q_0$.
\begin{figure}[t!] 
\includegraphics[width=0.4 \textwidth]{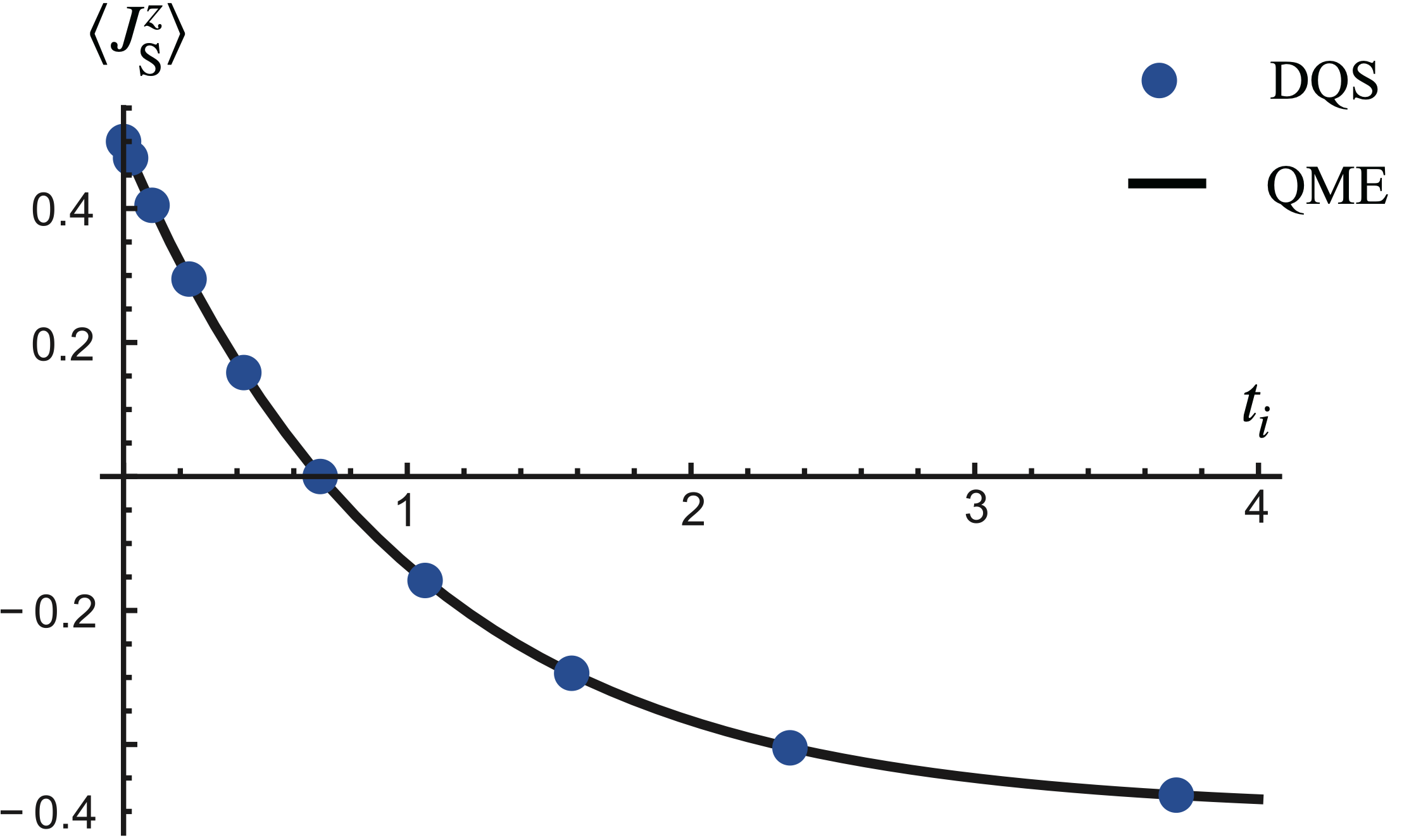}
\caption{The simulation results of the single-qubit amplitude damping  given with the ten blue points of
$ \big{(}t_i,   \overline{ \langle J^z _{ \text{S} } \rangle } (t_i) \big{)}$  ($i=0,1,\ldots, 9)$ for $ N_{ \text{shots}} = 2^{14}$ and $N_{ \text{ave}} = 25$.  
In addition, we present the solution of the quantum master equation \eqref{masterequation2} with the black curve.
The short-hand notation ``DQS" denotes the digital quantum simulation while ``QME" describes the quantum master equation.}  
\label{singleQzeroTrelax}
\end{figure} 
Let us say that we have performed this whole task $ N_{ \text{shots}} $ times.  
Then, we obtain the data of how many times the quantum state $ | 0 \rangle _{ \text{S} }  $ or $ | 1 \rangle _{ \text{S} }  $ has been generated as an output state. 
More accurately, we obtain the number of each four quantum states which have been generated as the output states at $t_i$: they are either $ | 00 \rangle_{ Q_0 Q_1  }, | 01 \rangle_{ Q_0 Q_1  }, | 10 \rangle_{ Q_0 Q_1  } $, or $ | 11 \rangle_{ Q_0 Q_1  } $ in Eq. \eqref{singleQrelaxbasisvector1}. 
Here, we have rewritten the subscripts S and E for the kets as $Q_0$ and  $Q_1$, respectively. 
Let us write these four numbers as $ N_{  | 00 \rangle } (t_i) , N_{  | 01 \rangle} (t_i),  N_{  | 10 \rangle } (t_i),$ and $N_{  | 11 \rangle } (t_i).$
The quantity $ N_{  | 00 \rangle } (t_i)$ describes the number of quantum state $ | 00 \rangle_{ Q_0 Q_1 } $ which has been obtained as the output state; 
similar definitions are given for the other three numbers. 
They satisfy $ N_{  | 00 \rangle } (t_i) + N_{  | 01 \rangle} (t_i) +  N_{  | 10 \rangle } (t_i) + N_{  | 11 \rangle } (t_i) = N_{ \text{shots}}.$
We have put the argument $ t_i $ on these four numbers to express that these are the quantities given at the time $t_i$.
The probability weights  $ w_{   | 0 \rangle_{ \text{S}}  } (t_i) $ and $ w_{   | 1 \rangle_{ \text{S}}  } (t_i) $  are described in terms of these four numbers as
  \begin{align}  
  w_{   | 0 \rangle_{ \text{S} }} (t_i) & = \frac {  N_{  | 00 \rangle } (t_i) + N_{  | 01 \rangle } (t_i)  }{ N_{ \text{shots} }} , \notag\\
  w_{   | 1 \rangle_{ \text{S} }} (t_i) & = \frac {  N_{  | 10 \rangle } (t_i) + N_{  | 11 \rangle } (t_i)  }{ N_{ \text{shots} }} .
   \label{probpolidentifysingleQ}
  \end{align}  
   From Eq. \eqref{probpolidentifysingleQ}, we have 
  \begin{align}  
   \langle J^z _{ \text{S} } \rangle (t_i) = \frac {  N_{  | 00 \rangle } (t_i) + N_{  | 01 \rangle } (t_i)  
 -  \big{(}  N_{  | 10 \rangle } (t_i) +  N_{  | 11 \rangle } (t_i)  \big{)}
   }{2 N_{ \text{shots} }}.
     \label{relaxprobweightssingleQ}
  \end{align} 
 By computing the numerical value of the right-hand side of Eq.  \eqref{relaxprobweightssingleQ} with qiskit, as a result, 
 we obtain the digital quantum simulation result of the single-qubit amplitude damping.  
 To make our result more trustable, we calculate the ``averaged expectation value" defined by
   \begin{align}  
  \overline{ \langle J^z _{ \text{S} } \rangle } (t_i) = \frac { \sum_{\alpha = 1}^{ N_{ \text{ave}} } \langle J^z _{ \text{S} } \rangle  (\alpha, t_i)  } { N_{ \text{ave}} }.
     \label{averagedADsingleQ}   
  \end{align} 
  As explained previously,
 to calculate the expectation value  $ \langle J^z _{ \text{S} } \rangle (t_i) $ given in Eq.  \eqref{relaxprobweightssingleQ}, 
  we perform $ N_{ \text{shots} } $ times the series of operations described by the quantum circuit in Fig. \ref{overallsingleQrelaxQC}  plus the measurement on the qubit $Q_0$. 
  Then, we repeat this overall task $N_{ \text{ave}} $ times in order to obtain the averaged expectation value  
 $ \overline{ \langle J^z _{ \text{S} } \rangle } (t_i) $  in Eq. \eqref{averagedADsingleQ}. In our simulation, 
 we have taken $N_{ \text{ave}} = 25$ and $ N_{ \text{shots}} =  2^{14}$. 
 The quantity $ \langle J^z _{ \text{S} } \rangle  (\alpha, t_i)  $ is the expectation value which is obtained in the $\alpha$-th round of the overall task for the calculation of  $ \langle J^z _{ \text{S} } \rangle (t_i) $. 
 In Fig. \ref{singleQzeroTrelax}, we present our simulation result of the single-qubit amplitude damping as two-dimensional plots of 
 $\big{(} t_i ,    \overline{ \langle J^z _{ \text{S} } \rangle } (t_i)   \big{)}$, which are given as ten blue points. 
To make a numerical comparison, we have presented the solution of the quantum master equation \eqref{masterequation2} given by the black curve. 
We do this by not only evaluating the averaged expectation value  $ \overline{ \langle J^z _{ \text{S} } \rangle } (t_i)   $
but also evaluating the variance defined by
 \begin{align}  
  \sigma^2 ( t_i) & = \overline{  \langle J^z _{ \text{S} }   \rangle^2 } ( t_i) -  \left( \overline{   \langle J^z _{ \text{S} }   \rangle }  \right)^2 ( t_i) ,
   \label{variance1singleQ}
    \end{align}  
where $ \overline{  \langle J^z _{ \text{S} }   \rangle^2 } ( t_i)  $ is given by
  \begin{align}  
\overline{  \langle J^z _{ \text{S} }   \rangle^2 } ( t_i) = 
 \frac { \sum_{\alpha = 1}^{ N_{ \text{ave}} } \langle J^z _{ \text{S} } \rangle^2  (\alpha,  t_i)  }
  { N_{ \text{ave}} }.
   \label{variance2singleQ}
    \end{align}  
  We have verified that the variance $\sigma^2 ( t_i) $ in Eq.  \eqref{variance1singleQ} is in the order of $10^{-5}-10^{-6}$. 
Therefore, our simulation result shows good numerical matching with the solution of the quantum master equation \eqref{masterequation2}.
At the end, let us analyze in detail the physics of our simulation result. 
What we have obtained are the two quantities $ N_{  | 01  \rangle } (t_i)$  and $ N_{  | 10  \rangle } (t_i)$.   
This fact is clearly representing the energy-exchange process between S and E: $  | 0 \rangle_{\text{S}} \otimes  | 1 \rangle_{\text{E}} \rightarrow    | 1 \rangle_{\text{S}} \otimes  | 0 \rangle_{\text{E}}$. 
Initially,  we have the values $ N_{  | 01  \rangle } (t_0) = N_{ \text{shots}} $  and $ N_{  | 10  \rangle } (t_0) = 0.$
As time goes by, the value of  $ N_{  | 01  \rangle } (t_i)$ (the probability weight  $w_{   | 0 \rangle_{ \text{S} }} (t_i)$) decreases 
while that of  $ N_{  | 10  \rangle } (t_i)$ (the probability weight  $w_{   | 1 \rangle_{ \text{S} }} (t_i)$)  increases.  
This is nothing but the amplitude damping of the system S as shown in Fig.  \ref{singleQzeroTrelax}. 

\section{Collective Amplitude Damping in Two-Qubit Systems}
\label{BSCR}
In the previous section, we have demonstrated the way to digital quantum simulate the single-qubit amplitude damping based on the Kraus representation. 
In this section, we extend this method to the case of two-qubit systems at zero temperature.    
 First, we analyze the Hilbert-space structure of the two-qubit systems. Then, we examine the effective Hilbert-space representation for the description of the collective 
 amplitude damping. Based on it, we formulate the unitary transformation for such processes with the single- and two-qubit gates
and construct the quantum circuits. 
Finally,  as the numerical verification of our formalisms and the quantum circuits we show our simulation results and discuss them in detail from both physics and computational perspectives.     

\subsection{Hilbert-Space Structure }\label{HSStruc}
Let us construct the Hilbert space of the two-qubit system. We name the two qubits as $Q_0$ and  $Q_1$. 
 We write the two states of the qubit $Q_0$ $(Q_1)$ in the computational basis as $ |0 \rangle_{ Q_0 (Q_1)} $ and  $ |1 \rangle_{ Q_0 (Q_1)} $.
With using them, we represent the quantum state of the two-qubit system.   
By denoting the total Hilbert space of the two-qubit system as $V_{ Q_0Q_1 } $, it is represented as 
\begin{align}
V_{ Q_0Q_1 }  = V_{ Q_0} \otimes V_{ Q_1}  = V_{j = 0} \oplus V_{j = 1}.  \label{twoQrelaxHspace1}
\end{align}
The total Hilbert space $V_{ Q_0Q_1 }$ in Eq.  \eqref{twoQrelaxHspace1} is represented in two different ways.
The first representation $ V_{ Q_0} \otimes V_{ Q_1} $  is spanned by the tensor products of  $ | n_{Q_0}  \rangle_{ Q_0 } $ and  $ | n_{Q_1}  \rangle_{ Q_1 } $  with $ n_{Q_0} , n_{Q_1} =0,1$ (tensor-product representation).
The basis vectors of this representation are given by the four vectors 
\begin{align}
  | \boldsymbol{e}_{ 1} \rangle & = | 0 \rangle_{ Q_0 } \otimes | 0  \rangle_{  Q_1}
\equiv  |0 0 \rangle_{ Q_0 Q_1}, \notag\\ 
  | \boldsymbol{e}_{ 2}  \rangle  & = | 0 \rangle_{ Q_0 } \otimes | 1  \rangle_{  Q_1}
\equiv  |0 1 \rangle_{ Q_0 Q_1}, \notag\\ 
 | \boldsymbol{e}_{ 3 }  \rangle & = | 1 \rangle_{ Q_0 } \otimes | 0  \rangle_{  Q_1}
\equiv  |1 0 \rangle_{ Q_0 Q_1}, \notag\\ 
| \boldsymbol{e}_{ 4 }  \rangle  & = | 1 \rangle_{ Q_0 } \otimes | 1  \rangle_{  Q_1}
\equiv  | 1 1 \rangle_{ Q_0 Q_1}.
\label{twoQrelaxbasisvector1} 
\end{align}
The second one, $V_{j = 0} \oplus V_{j = 1}$,  we call it as direct-sum-spin-space representation. 
It is obtained by the spin-angular-momentum composition between the two qubits $Q_0$  and $Q_1$ 
and is expressed by two subspaces; the meaning of two subscripts $j = 0$ and $j = 1$ are explained later.  
The subspace $V_{j = 0} $ is spanned by a spin state, namely,  $| 0, 0 \rangle$ state. 
It is the quantum state whose spin magnitude is zero and the degree of its $z$ component is zero.
Mathematically, this is described as follows. 
Let us write the quantum state of  the direct-sum spin space as $ | j, m \rangle $.
Then, we introduce total-spin operators defined by $J^x = \frac{ X_{ Q_0}  + X_{ Q_1} }{2},
 J^y =  \frac{ Y_{ Q_0}  + Y_{ Q_1} }{2}, J^z =  \frac{ Z_{ Q_0}  + Z_{ Q_1} }{2}.$
By using these three operators, we define the operator $ \boldsymbol{J}^2 = (J^x)^2 + (J^y)^2 + (J^z)^2$.
The quantum state $| j, m \rangle$ satisfies $ \boldsymbol{J}^2  | j, m \rangle = j(j+1)  | j, m \rangle $ and  $ J^z   | j, m \rangle = m  | j, m \rangle.$
 For the state $| 0, 0  \rangle$ $(j = m = 0)$, we have  $ \boldsymbol{J}^2  | 0, 0 \rangle = 0 $ and  $ J^z   | j, m \rangle =0$, as explained previously. 
For later convenience, let us rewrite this state as $ | \check{\boldsymbol{0}} \rangle $.
Similarly, we can discuss the characteristics of the subspace  $V_{j = 1}$ with the operators  $ \boldsymbol{J}^2 $ and  $ J^z$. 
It is spanned by the three spin states $ | 1, 1 \rangle,  | 1, 0 \rangle,  | 1, -1 \rangle$. All of them are the quantum states whose spin magnitude is equal to one $(j = 1)$.
On the other hand, the degrees of the $z$ component (the eigenvalues of $ J^z $) of the spin states $ | 1, 1 \rangle,  | 1, 0 \rangle$,  and $| 1, -1 \rangle$ are equal to $1,0$ and $-1$, respectively. 
These two characteristics are mathematically expressed as $ \boldsymbol{J}^2  | 1, m \rangle = 2  | 1, m \rangle $ and $ J^z   | 1, m \rangle = m  | 1, m \rangle $  with $ m=\pm 1,0.$
Let us rename $ | 1, 1 \rangle,  | 1, 0 \rangle$,  and $| 1, -1 \rangle$  as $ | \check{\boldsymbol{1}} \rangle, | \check{\boldsymbol{2}} \rangle$, and $| \check{\boldsymbol{3}} \rangle,$ respectively.
The relations between the basis vectors of the direct-sum spin space, $  | \check{\boldsymbol{0}} \rangle, | \check{\boldsymbol{1}} \rangle, | \check{\boldsymbol{2}} \rangle, | \check{\boldsymbol{3}} \rangle,$ 
and the basis vectors of the tensor-product space given in Eq. \eqref{twoQrelaxbasisvector1} are
\begin{align}
  | \check{\boldsymbol{0}} \rangle & =  \frac{ | \boldsymbol{e}_{ 2}  \rangle - | \boldsymbol{e}_{ 3}  \rangle }{ \sqrt{2} }, \quad 
  |\check{\boldsymbol{1}} \rangle  = | \boldsymbol{e}_{ 1}  \rangle,  \notag\\
 | \check{\boldsymbol{2}} \rangle & =  \frac{ | \boldsymbol{e}_{ 2}  \rangle + | \boldsymbol{e}_{ 3}  \rangle }{ \sqrt{2} } , \quad  
 | \check{\boldsymbol{3}} \rangle  = | \boldsymbol{e}_{ 4}  \rangle. \label{twoQrelaxbasisvector2}
\end{align}
Before ending this subsection, let us present the relations among  Bell states, the tensor-product states in Eq. \eqref{twoQrelaxbasisvector1}, and the direct-sum spin states in Eq. \eqref{twoQrelaxbasisvector2}.
By denoting the four Bell states as $| \Phi^\pm \rangle $ and $| \Psi^\pm \rangle $, they are expressed as 
\begin{align}
  | \Phi^+ \rangle  & =  \frac{  | \boldsymbol{e} _{ 1 } \rangle  +  | \boldsymbol{e} _{ 4 } \rangle }{\sqrt{2}} =  \frac{  | \check{\boldsymbol{1}} \rangle +  | \check{\boldsymbol{3}} \rangle  }{\sqrt{2}}, \notag\\ 
   | \Phi^- \rangle &  =  \frac{  | \boldsymbol{e} _{ 1 } \rangle  -  | \boldsymbol{e} _{ 4 } \rangle  }{\sqrt{2}} =  \frac{   | \check{\boldsymbol{1}} \rangle -  | \check{\boldsymbol{3}} \rangle  }{\sqrt{2}},  \notag\\
 | \Psi^+ \rangle  & =  \frac{  | \boldsymbol{e}  _{ 2 }  \rangle +  | \boldsymbol{e} _{ 3 }  \rangle }{\sqrt{2}} =  |  \check{\boldsymbol{2}}  \rangle , \notag\\ 
   | \Psi^- \rangle  & =  \frac{  | \boldsymbol{e}_{ 2 } \rangle  -  | \boldsymbol{e}_{ 3 }  \rangle }{\sqrt{2}} =   | \check{\boldsymbol{0}} \rangle.
\label{Bellstates1}
\end{align}
\subsection{ Quantum Master Equation and Effective Representation}\label{QMEERtwoQ}
\begin{figure}[t!]  
\includegraphics[width=0.4 \textwidth]{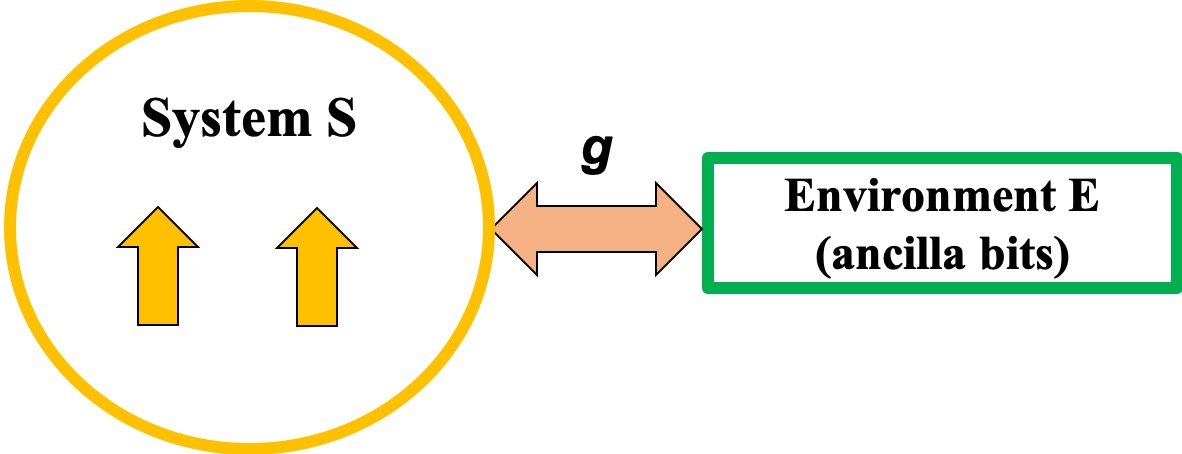}
\caption{ Schematic illustration of the two-qubit open quantum system.
The two qubits (system S) are collectively interacting with the single environment E with a coupling strength $g$.
As a result, the system S behaves as a single spin whose spin size is equal to one and exhibits the collective 
open quantum dynamics. }  
\label{twoQandE} 
\end{figure}

Since we have established the representations of Hilbert space for two qubits, 
we are going to formulate the unitary transformation in terms of single- and two-qubit gates, 
which generates the collective amplitude damping processes. Then, we construct the corresponding quantum circuits.
In order to do this, first let us analyze an effective Hilbert-space representation for such descriptions.   
The collective open quantum dynamics including the collective amplitude damping occur when the qubits 
interact equivalently with the single environment.  
In such circumstances, the qubits act as a single giant spin whose spin size is equal to $\frac{ N_{\text{Q}} }{2}$ with $N_{\text{Q}}$ denoting the total number of qubits.
Therefore, the collective open quantum dynamics are described by the total-spin operators. 
Let us illustrate such situations for the case of the two-qubit system in Fig. \ref{twoQandE}.   
Furthermore, the initial states are set to quantum states which are invariant under permutation between any two qubits, namely, permutational symmetric states \cite{LidarWhaleygroupcollectivenoises,carmichaeltxb,Agarwalltxb,opendynamicstext,superradiance1,GH82,Nathanmastereqsim,YamamotogroupcollectivenoiseQA}. The simple examples of the permutational symmetric states are the quantum states 
$ | \check{\boldsymbol{1}} \rangle, | \check{\boldsymbol{2}}  \rangle$, and $ | \check{\boldsymbol{3}}  \rangle$ 
in Eq. \eqref{twoQrelaxbasisvector2}. 
The Hilbert space spanned by the permutational symmetric states is equivalent to the subspace in the direct-sum spin space, 
which is comprised of spin states $| j, m  \rangle$ with $  j = \frac{ N_{\text{Q}} }{2}$ and $ m = j, j-1, \ldots, - (j-1), -j $,
and we name it as permutational symmetric subspace. The example is, as we have seen, the subspace $V_{j = 1}$ in Eq. \eqref{twoQrelaxHspace1}.  
The collective open quantum dynamics are, therefore, described by the permutational symmetric subspace and
 physical quantities which characterize the collective open quantum dynamics such as the relaxation time are expressed by $ N_{\text{Q}}$. 
The canonical and important example of the collective open quantum dynamics
is the superradiance (collective emission or radiation) \cite{carmichaeltxb,Agarwalltxb,opendynamicstext,superradiance1,GH82,cavityqedreview2,YamamotogroupcollectivenoiseQA,WallraffgroupSCsupsub,solidsuperradiance2016,NVsuperradiantdecay,superradianceYavuz,NakamuragroupsubradianceSC,superradianceSCPRL2020}.
The superradiance in the two-qubit system is essentially equivalent to the collective amplitude damping in the two-qubit system. 
In this work, all the collective amplitude damping processes which we analyze are the dynamics starting from the permutational symmetric states. 
Therefore, we use the direct-sum-spin-space representation ($V_{j = 1}$ in Eq. \eqref{twoQrelaxHspace1}) as the effective representation for the description of the collective amplitude damping.

Let us return to the discussion on the collective amplitude damping in the two-qubit system. 
In this case, the Hamiltonian of the two-qubit system is given by $H = \hbar \omega J^z $, where $\omega$ the frequency of $Q_0$ as well as  that of $Q_1$.
The four states   $ | \check{\boldsymbol{0}} \rangle,   | \check{\boldsymbol{1}} \rangle,   | \check{\boldsymbol{2}} \rangle, $ and  $  | \check{\boldsymbol{3}} \rangle $   
are the eigenstates of this Hamiltonian and their eigenvalues are $0,\hbar \omega,0,$ and $-\hbar \omega$, respectively. 
The quantum master equation which describes the collective amplitude damping is given by \cite{carmichaeltxb,opendynamicstext} 
 \begin{align}
& \frac { \partial \rho^{\text{I}} _{\text{S}} (t)}{\partial t }  =  \gamma \left[  J^-    \rho^{\text{I}}_{\text{S}} (t)   J^+ - \frac{1}{2} \big{\{}  J^+  J^- ,  \rho^{\text{I}}_{\text{S}} (t)   \big{\}} \right] ,\label{masterequation3}
\end{align} 
where $ \gamma$ is the decay rate.
  The total-spin operator $J^+ = J^x + i J^y$ in the above equation is the creation spin operator (ladder operator) and is  equivalent to 
  the Hermitian conjugate of the annihilation spin operator $J^- = J^x - i J^y$. 
  Note that the operator $J^z _{ \text{S} } $ in the interaction picture is mathematically equivalent to that in the Schr$\ddot{\text{o}}$dinger picture because the system Hamiltonian is 
given by  $H _{ \text{S} }  = \hbar \omega J^z _{ \text{S} } $.  
  The creation and annihilation spin operators act on the direct-sum spin state $| j, m \rangle$ as 
  $ J^+  | j, m \rangle  = \sqrt{ j(j+1) - m(m+1)}  | j, m + 1 \rangle,
  J^-  | j, m \rangle  = \sqrt{ j(j+1) - m(m-1)}  | j, m - 1 \rangle .$
  For $j=1,$ we have $ J^- | 1, m \rangle  = \sqrt{ 2 - m(m-1)}  | 1, m - 1 \rangle,$ with $m = 1,0,-1$. 
It is equivalent to the description of transition processes $ | \check{\boldsymbol{1}} \rangle \rightarrow | \check{\boldsymbol{2}} \rangle \rightarrow | \check{\boldsymbol{3}} \rangle$,
which are obtained by sequentially operating $ J^- $ from the state $ | \check{\boldsymbol{1}} \rangle $. 
This is essentially describing the decay processes occurring among the three states $ | \check{\boldsymbol{1}} \rangle,  | \check{\boldsymbol{2}} \rangle,  | \check{\boldsymbol{3}} \rangle$. 
Note that the quantum state  $ | \check{\boldsymbol{0}} \rangle $ does not show the decay process since $J^- | \check{\boldsymbol{0}} \rangle = J^-  |0,0 \rangle = 0.$
Let us represent the four states  $ | \check{\boldsymbol{0}} \rangle,  | \check{\boldsymbol{1}} \rangle,  | \check{\boldsymbol{2}} \rangle,  | \check{\boldsymbol{3}} \rangle$
in the vector form as
\begin{align}
  |\check{\boldsymbol{0}} \rangle  & =   
	\left ( \begin{array}{c} 
		 1  \\
		 0 \\
		  0  \\
		   0 
		\end{array} \right), \quad 
| \check{\boldsymbol{1}}   \rangle =  
	\left ( \begin{array}{c} 		
	 0  \\
		 1 \\
		  0  \\
		   0 
		\end{array} \right), \notag\\
| \check{\boldsymbol{2}}   \rangle  & = 
	\left ( \begin{array}{c} 
		 0  \\
		 0 \\
		  1  \\	   
		  0 
		\end{array} \right), \quad 
| \check{\boldsymbol{3}}  \rangle  =
	\left ( \begin{array}{c} 
		 0  \\
		 0 \\
		  0  \\
		   1 
		\end{array} \right).
\label{twoQrelaxbasisvector3} 
\end{align}
The matrix representations of the creation and annihilation spin operators in the direct-sum-spin-space representation are 
 \begin{align}
 J^+ & =  
	\left [
		\begin{array}{cccc} 
		 0 & 0 & 0 & 0 \\
		 0 & 0 & \sqrt{2} & 0 \\
		  0 & 0 & 0 & \sqrt{2} \\
		   0 & 0 & 0 & 0 
		\end{array}
	\right ], \notag\\
	J^-  & =  
	\left [
		\begin{array}{cccc} 
		 0 & 0 & 0 & 0 \\
		 0 & 0 & 0 & 0 \\
		  0 &   \sqrt{2}  & 0 & 0\\
		   0 & 0 &  \sqrt{2}  & 0 
		\end{array}
	\right ] ,
\label{creannispintwoQ}
\end{align}  
  By writing the elements of the reduced density matrix for the initial state as $ \left[  \rho^{\text{in}}_{\text{S}}  \right]_{i,j} $,   
  while we write the solution of quantum master equation \eqref{masterequation3} as $\left[  \rho^{\text{ME}}_{\text{S}}  \right]_{i,j} (t)$ ($i,j = \boldsymbol{\check{0}},\boldsymbol{\check{1}},\boldsymbol{\check{2}},\boldsymbol{\check{3}}$),
we have 
\begin{align}
 \left[  \rho^{\text{ME}}_{\text{S}}  \right]_{\boldsymbol{\check{0}},\boldsymbol{\check{1}}} (t)  &  =   \left[  \rho^{\text{in}}_{\text{S}}  \right]_{\boldsymbol{\check{0}},\boldsymbol{\check{1}}} e^{- \gamma t}, \ 
 \left[  \rho^{\text{ME}}_{\text{S}}  \right]_{\boldsymbol{\check{0}},\boldsymbol{\check{2}}} (t)    =   \left[  \rho^{\text{in}}_{\text{S}}  \right]_{\boldsymbol{\check{0}},\boldsymbol{\check{2}}} e^{- \gamma t}, \notag\\
 \left[  \rho^{\text{ME}}_{\text{S}}  \right]_{\boldsymbol{\check{1}},\boldsymbol{\check{1}}} (t)  &  =   \left[  \rho^{\text{in}}_{\text{S}}  \right]_{\boldsymbol{\check{1}},\boldsymbol{\check{1}}} e^{-2 \gamma t}
, \notag\\
\left[  \rho^{\text{ME}}_{\text{S}}  \right]_{\boldsymbol{\check{1}},\boldsymbol{\check{2}}} (t)  & =   \left[  \rho^{\text{in}}_{\text{S}}  \right]_{\boldsymbol{\check{1}},\boldsymbol{\check{2}}} e^{-2 \gamma t}, \notag\\
\left[  \rho^{\text{ME}}_{\text{S}}  \right]_{\boldsymbol{\check{1}},\boldsymbol{\check{3}}} (t)  &  =   \left[  \rho^{\text{in}}_{\text{S}}  \right]_{\boldsymbol{\check{1}},\boldsymbol{\check{3}}} e^{- \gamma t}, \notag\\
\left[  \rho^{\text{ME}}_{\text{S}}  \right]_{\boldsymbol{\check{2}},\boldsymbol{\check{2}}} (t) &  =   \left[  \rho^{\text{in}}_{\text{S}}  \right]_{\boldsymbol{\check{2}},\boldsymbol{\check{2}}} e^{-2 \gamma t}
+   \left[  \rho^{\text{in}}_{\text{S}}  \right]_{\boldsymbol{\check{1}},\boldsymbol{\check{1}}} \left( 2 \gamma t e^{-2 \gamma t} \right)
, \notag\\
\left[  \rho^{\text{ME}}_{\text{S}}  \right]_{\boldsymbol{\check{2}},\boldsymbol{\check{3}}} (t) &  =  
 \left[  \rho^{\text{in}}_{\text{S}}  \right]_{\boldsymbol{\check{2}},\boldsymbol{\check{3}}} e^{- \gamma t} + 
  \left[  \rho^{\text{in}}_{\text{S}}  \right]_{\boldsymbol{\check{1}},\boldsymbol{\check{2}}} \left( 2e^{- \gamma t} ( 1 - e^{- \gamma t} ) \right), \notag\\
   \left[  \rho^{\text{ME}}_{\text{S}}  \right]_{\boldsymbol{\check{3}},\boldsymbol{\check{3}}} (t) &  =
1 - \left[  \rho^{\text{ME}}_{\text{S}}  \right]_{\boldsymbol{\check{2}},\boldsymbol{\check{2}}} (t)    - \left[  \rho^{\text{ME}}_{\text{S}}  \right]_{\boldsymbol{\check{1}},\boldsymbol{\check{1}}} (t).
\label{masterequation4}
\end{align} 
To derive the formula of $ \left[  \rho^{\text{ME}}_{\text{S}}  \right]_{\boldsymbol{\check{3}},\boldsymbol{\check{3}}} (t)$ 
in the above equation, we have used the condition  $\text{Tr} _{\text{S}} [ \rho^{\text{ME}}_{\text{S}} ] = 1.$ 
Since the density matrix is a Hermitian operator, we have  $\left[ ( \rho^{\text{ME}}_{\text{S}} )^\dagger \right]_{i,j} = \left[  \rho^{\text{ME}}_{\text{S}}  \right]_{j,i} $. 
We note that the quantum master equation  \eqref{masterequation3} is also effectively describing the amplitude damping of a single qutrit \cite{qutirtKraus,qutrit1}.
This is because essentially the open quantum dynamics under consideration is described solely by the subspace $V_{j = 1}$.  
Energetically, the three quantum states $ | \check{\boldsymbol{1}} \rangle, | \check{\boldsymbol{2}} \rangle$ and $ | \check{\boldsymbol{3}} \rangle$ are equally separated with $ \frac{\hbar \omega}{2}.$ 
Such a circumstance is effectively describing a three-level system, and thus, 
we are describing the amplitude damping of the single qutrit.   
 \subsection{Kraus Representation and Quantum Circuit}\label{twoQKraus}
Our next task is to model the composite system of the two-qubit system (system under consideration) and the environment using the quantum states in Eq. \eqref{twoQrelaxbasisvector3}. Like we did in Sec. \ref{OpenQKrausRep}, we regard the two-qubit system of $Q_0$ 
and $Q_1$ as the system S, while we prepare two ancilla bits, $Q_2$ and $Q_3$ and constitute the single environment E with them:
  $Q_0 + Q_1 \equiv \text{S}$ and $Q_2 + Q_3 \equiv \text{E}$. 
 Then, we introduce the tensor-product states of the system S and the environment E represented by
\begin{align}
  | \boldsymbol{\tilde{\alpha}  } _{ i} \rangle & =  | \check{\boldsymbol{0}}  \rangle _{ \text{S} }   \otimes   | \boldsymbol{\check{i}}   \rangle _{ \text{E} } ,\notag\\ 
    | \boldsymbol{\tilde{\alpha} } _{ i + 4} \rangle &  =  | \check{\boldsymbol{1}}  \rangle _{ \text{S} }   \otimes   | \boldsymbol{\check{i}}  \rangle _{ \text{E} } ,\notag\\ 
      | \boldsymbol{\tilde{\alpha} }  _{ i + 8} \rangle & =  | \check{\boldsymbol{2}}   \rangle _{ \text{S} }   \otimes   |  \boldsymbol{\check{i}} \rangle _{ \text{E} } ,\notag\\
        | \boldsymbol{\tilde{\alpha}  }  _{ i + 12} \rangle &  =  | \check{\boldsymbol{3}}  \rangle _{ \text{S} }   \otimes   | \boldsymbol{\check{i}}  \rangle _{ \text{E} } ,
  \label{twoQrelaxbasisvector4}
\end{align}
where $ i = 1,2,3,4$ and $\boldsymbol{\check{i}} = \boldsymbol{\check{0}},\boldsymbol{\check{1}},\boldsymbol{\check{2}},\boldsymbol{\check{3}}.$ 
By using the basis vectors in Eq.  \eqref{twoQrelaxbasisvector4}, we formulate the collective amplitude damping in the following way. 
First, we initialize the quantum state of the environment E so that it is initially in the ground state:  
$\rho_{\text{E}} = | 11  \rangle _{ Q_2Q_3} \langle 11 | = | \check{\boldsymbol{3}}  \rangle _{ \text{E} } \langle \check{\boldsymbol{3}} |$.
On the other side, we perform a unitary transformation $U^{\text{in}}_{\text{S}} $ solely on the qubits $Q_0 $ and $ Q_1 $. 
This procedure is the initialization of the system S.   
As a result, the initial state of the total system is described as 
$ \rho^{\text{in}}_{\text{tot}} =  \rho^{\text{in}}_{\text{S}} \otimes   \rho^{\text{in}}_{\text{E}} = 
\left( U^{\text{in}}_{\text{S}}  | 00 \rangle_{\text{S}} \langle 00 | ( U^{\text{in}}_{\text{S}} )^\dagger     \right)   \otimes   | \check{\boldsymbol{3}} \rangle_{\text{E}} \langle \check{\boldsymbol{3}} |.$
Second, we construct the overall unitary transformation operating on both S and E  and write it as $U^{\text{AD}}_{\text{two}} $.  
We mathematically formulate $U^{\text{AD}}_{\text{two}} $ so that it describes the energy-exchange processes between the system S and the environment E.
Since the state $ | \check{\boldsymbol{0}}  \rangle$ satisfies $J^+  | \check{\boldsymbol{0}}  \rangle = J^-  | \check{\boldsymbol{0}}  \rangle =0$, 
 the energy-exchange processes occurs among the states  $  | \boldsymbol{\check{j}}  \rangle _{ \text{S} }  \otimes   | \boldsymbol{\check{k}}  \rangle _{ \text{E} } $,
  where $  | \boldsymbol{\check{j}}  \rangle _{ \text{S} },  | \boldsymbol{\check{k}}  \rangle _{ \text{E} } = 
  | \check{\boldsymbol{1}}  \rangle, | \check{\boldsymbol{2}}  \rangle, | \check{\boldsymbol{3}}  \rangle $. 
Further, we formulate $U^{\text{AD}}_{\text{two}} $ so that it is characterized by the time $t$. 
To describe this explicitly, let us rewrite  $U^{\text{AD}}_{\text{two}} $ as  $U^{\text{AD}}_{\text{two}} (t)$. 
Then, at the end of our procedures the measurement is performed on the two ancilla bits $Q_2$ and $Q_3$ (the environment E). 
As a result, we obtained the reduced density matrix $\rho^{\text{out}}_{\text{S}} (t) \equiv \text{Tr}_{\text{E}} [\rho^{\text{out}}_{\text{SE}}(t)] =
 \text{Tr}_{\text{E}} \left[ U^{\text{AD}}_{\text{two}} (t) \rho^{\text{in}}_{\text{tot}} (U^{\text{AD}}_{\text{two}})^\dagger  (t)  \right]$.   
It contains the information of the collective damping and can be extracted from the probability weights of 
$  | \check{\boldsymbol{1}}  \rangle, | \check{\boldsymbol{2}}  \rangle,$ and  $ | \check{\boldsymbol{3}}  \rangle$ at the time $t$. 
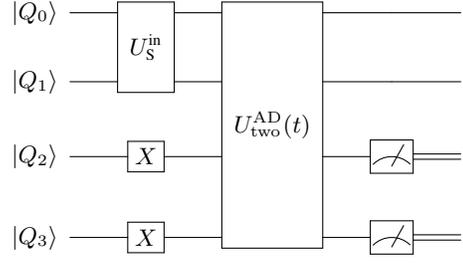
\begin{figure}[H] 
\centering
\mbox{ 
\Qcircuit @C=2.0em @R=2.0em { 
   \lstick{| Q_0 \rangle} &  \multigate{1}{  U^{\text{in}}_{\text{S}}  }      & \multigate{3}{U_{\mathrm{two}}^{\mathrm{AD}}  (t)}  & \qw & \qw \\
   \lstick{| Q_1 \rangle} & \ghost{  U^{\text{in}}_{\text{S}}  }      & \ghost{U_{\mathrm{two}}^{\mathrm{AD}}  (t)}        & \qw & \qw \\
    \lstick{| Q_2 \rangle} & \gate{X} & \ghost{U_{\mathrm{two}}^{\mathrm{AD}}  (t)}       & \meter & \cw \\
    \lstick{| Q_3 \rangle} & \gate{X} & \ghost{U_{\mathrm{two}}^{\mathrm{AD}}  (t)}       & \meter & \cw
} 
}  
\caption{Schematic quantum circuit for the simulation of collective amplitude damping.
The two qubits $Q_0$ and $Q_1$ are regarded as the system S while the two ancilla bits $Q_2$ and $Q_3$ as the environment E.
We apply the $X$ gate to the qubits $Q_2$ and $Q_3$ to formulate that the environment E  is initially in the ground state: 
$ \rho^{\text{in}}_{\text{E}} =   | 11 \rangle_{Q_2Q_3} \langle 11 | =  |  \check{\boldsymbol{3}} \rangle_{\text{E}} \langle  \check{\boldsymbol{3}} | $. 
 On the other hand, we initialize the system S by applying the unitary transformation $U^{\text{in}}_{\text{S}}$  to the qubits $Q_0$ and $Q_1$: 
 $ \rho^{\text{in}}_{\text{S}} = \left( U^{\text{in}}_{\text{S}}  | 00 \rangle_{\text{S}} \langle 00 | ( U^{\text{in}}_{\text{S}} )^\dagger     \right).$
 Hence, the initial state of the total system is described as 
 $ \rho^{\text{in}}_{\text{tot}} =  \rho^{\text{in}}_{\text{S}} \otimes   \rho^{\text{in}}_{\text{E}} .$
Next, we perform the overall unitary transformation $U^{\text{AD}}_{\text{sin}} (t)$ 
describing the energy-exchange process between the system S and the environment E at the time $t$ given in terms of the three states 
$ | \check{\boldsymbol{1}}  \rangle, | \check{\boldsymbol{2}}  \rangle, | \check{\boldsymbol{3}}  \rangle$. 
At the end, we perform the measurement on the two qubits $Q_2$ and $Q_3$.  
As a result, we obtain the reduced density matrix $ \rho^{\text{out}}_{\text{S}}  (t) $ describing the collective amplitude damping
at $t$. }
\label{BellQCtwoQCAD} 
\end{figure}  
Let us write the probability weights of the state $ | \boldsymbol{\check{i}}   \rangle _{ \text{S} } $ $( | \boldsymbol{\check{i}}   \rangle _{ \text{S} } =   | \check{\boldsymbol{1}}  \rangle_{ \text{S} }, | \check{\boldsymbol{2}}  \rangle_{ \text{S} },  | \check{\boldsymbol{3}}  \rangle_{ \text{S} } )$ at $t$ as $ w_{   |  \boldsymbol{\check{i}}  \rangle_{ \text{S} }} (t).$
Then, we can quantum simulate the collective amplitude damping process at the time $t$ 
represented by $ \langle J^z _{ \text{S} }   \rangle (t) =   \text{Tr}_{\text{S}} \left[  \rho^{\text{out}}_{\text{S}} (t) J^z _{\text{S}}  \right] 
=  w_{   |  \boldsymbol{\check{1}}  \rangle_{ \text{S} }} (t) - w_{   |  \boldsymbol{\check{3}}  \rangle_{ \text{S} }} (t)$. 
The operator  $J^z _{ \text{S} } =  \frac{ Z_{Q_0} + Z_{Q_1} }{2}$ is the  $z$-component of the total-spin operator of the system S.
The matrix representation of  $J^z _{ \text{S} } $ in the direct-sum spin space representation is $ J^z _{ \text{S} } =  \text{diag}(0,1,0,-1).$ 
The numerical values of the
probability weights $ w_{ |  \boldsymbol{\check{i}}  \rangle_{ \text{S} }} (t)$ are obtained by the qiskit simulation
and our digital quantum simulation of the collective amplitude damping completes. 
In Fig. \ref{BellQCtwoQCAD}, we summarize the above procedures and represent them as the schematic quantum circuit. 

\begin{widetext}
\begin{figure}[t!] 
\includegraphics[width=0.35 \textwidth]{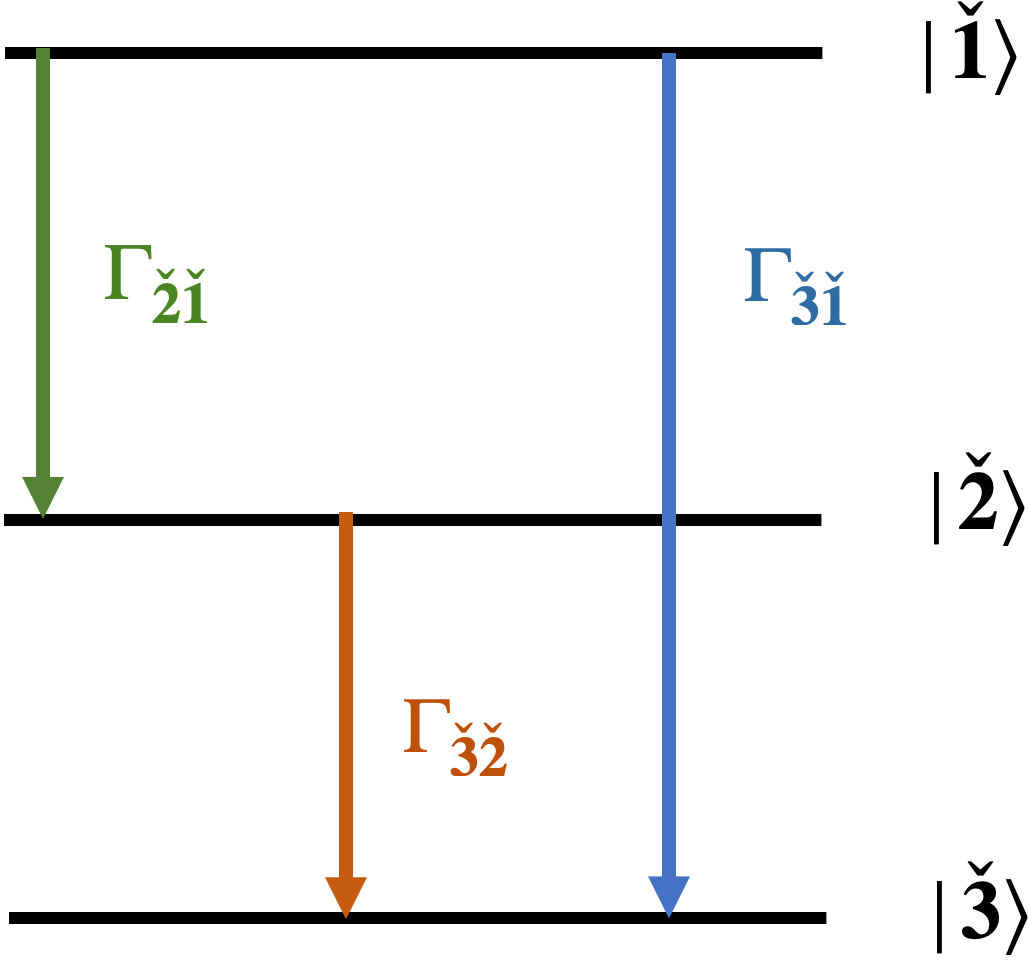}
\caption{ Three decay processes in the two-qubit system. The quantities 
$\Gamma_{  \check{\boldsymbol{2}} \check{\boldsymbol{1}} }, \Gamma_{  \check{\boldsymbol{3}} \check{\boldsymbol{2}} }$ and $\Gamma_{  \check{\boldsymbol{3}} \check{\boldsymbol{2}} } $
are the decay strengths of the processes  $ | \check{\boldsymbol{1}} \rangle \rightarrow  |  \check{\boldsymbol{2}} \rangle, | \check{\boldsymbol{2}} \rangle \rightarrow  | \check{\boldsymbol{3}} \rangle$, 
and $ | \check{\boldsymbol{1}}  \rangle \rightarrow |  \check{\boldsymbol{3}} \rangle $, respectively. } 
\label{Bellthreedecays} 
\end{figure}  \end{widetext}

Let us now return to the argument on the formulation of the overall unitary transformation $U^{\text{AD}}_{\text{two}} $ 
and the derivation of the reduced density matrix of the system using the Kraus representation.  
The collective amplitude damping under consideration shows three types of decay channels:
$\Gamma_{  \check{\boldsymbol{2}} \check{\boldsymbol{1}} }: | \check{\boldsymbol{1}}  \rangle  \rightarrow  | \check{\boldsymbol{2}}  \rangle$, 
$\Gamma_{  \check{\boldsymbol{3}} \check{\boldsymbol{2}} }: | \check{\boldsymbol{2}} \rangle  \rightarrow  |  \check{\boldsymbol{3}} \rangle$, and 
$\Gamma_{  \check{\boldsymbol{3}} \check{\boldsymbol{1}} }: | \check{\boldsymbol{1}}  \rangle \rightarrow  | \check{\boldsymbol{3}} \rangle$. 
The quantity $\Gamma_{  \check{\boldsymbol{2}} \check{\boldsymbol{1}} }$ 
represents the strength of the decay process  $ | \check{\boldsymbol{1}} \rangle \to | \check{\boldsymbol{2}} \rangle$.
The strengths of the other two decay channels, $\Gamma_{  \check{\boldsymbol{3}} \check{\boldsymbol{2}} }$ and $\Gamma_{  \check{\boldsymbol{3}} \check{\boldsymbol{1}} }$, 
are defined in the similar way. 
Let us present a schematic illustration for these three decay channels in Fig. \ref{Bellthreedecays}. 
For instance, when the total system is in the quantum state $   | \boldsymbol{\tilde{\alpha} } _{ 11 } \rangle  =  | \check{\boldsymbol{2}}  \rangle _{ \text{S} }   \otimes   | \boldsymbol{\check{3}}  \rangle _{ \text{E} } $ 
 in Eq.  \eqref{twoQrelaxbasisvector4},  due to the interaction between the system S and environment E the quantum-state transition  
$ | \check{\boldsymbol{2}}  \rangle _{ \text{S} }   \otimes   | \boldsymbol{\check{3}}  \rangle _{ \text{E} }  \rightarrow  | \check{\boldsymbol{3}}  \rangle _{ \text{S} }   \otimes   | \boldsymbol{\check{2}}  \rangle _{ \text{E} }  =  | \boldsymbol{\tilde{\alpha} } _{ 14 } \rangle$ occurs. 
This is equivalent to the decay process of the system S, $ | \check{\boldsymbol{2}}    \rangle _{ \text{S} }  \rightarrow  | \check{\boldsymbol{3}}   \rangle _{ \text{S} }$.
On the other hand, when the total system is in the quantum state $   | \boldsymbol{\tilde{\alpha} } _{ 7 } \rangle  =  | \check{\boldsymbol{1}}  \rangle _{ \text{S} }   \otimes   | \boldsymbol{\check{3}}  \rangle _{ \text{E} } $, 
the state-transition processes 
$ | \check{\boldsymbol{1}}  \rangle _{ \text{S} }   \otimes   | \boldsymbol{\check{3}}  \rangle _{ \text{E} }  \rightarrow  | \check{\boldsymbol{2}}  \rangle _{ \text{S} }   \otimes   | \boldsymbol{\check{2}}  \rangle _{ \text{E} }   \rightarrow
| \check{\boldsymbol{3}}  \rangle _{ \text{S} }   \otimes   | \boldsymbol{\check{1}}  \rangle _{ \text{E} } $
 $(| \boldsymbol{\tilde{\alpha} } _{ 7 } \rangle  \rightarrow | \boldsymbol{\tilde{\alpha} } _{ 10 } \rangle  \rightarrow
| \boldsymbol{\tilde{\alpha} } _{ 13 } \rangle) $ occur.   
In order to analyze and understand such circumstances, we need to consider all three decay channels.  
By taking account of the above consideration, we formulate the overall unitary transformation $U^{\text{AD}}_{\text{two}} $  as 
\begin{widetext}
\begin{align}
	U^{\text{AD}}_{\text{two}} &   =
	\left [
		\begin{array}{cccccccccccccccc} 
		 1 & 0 & 0 & 0 & 0 & 0 & 0 & 0 & 0 & 0 & 0 & 0 & 0 & 0 & 0 & 0 \\
		 0 & 1 & 0 & 0 & 0 & 0 & 0 & 0 & 0 & 0 & 0 & 0 & 0 & 0 & 0 & 0 \\
		 0 & 0 & 1 & 0 & 0 & 0 & 0 & 0 & 0 & 0 & 0 & 0 & 0 & 0 & 0 & 0 \\
		 0 & 0 & 0 & 1 & 0 & 0 & 0 & 0 & 0 & 0 & 0 & 0 & 0 & 0 & 0 & 0 \\
		 0 & 0 & 0 & 0 & 1 & 0 & 0 & 0 & 0 & 0 & 0 & 0 & 0 & 0 & 0 & 0 \\
		 0 & 0 & 0 & 0 & 0 & 1 & 0 & 0 & 0 & 0 & 0 & 0 & 0 & 0 & 0 & 0 \\
		 0 & 0 & 0 & 0 & 0 & 0 &  [ U^{\text{AD}}_{\text{two}} ]_{7,7} & 0 & 0 &  [ U^{\text{AD}}_{\text{two}} ]_{7,10} & 0 & 0 & 0 & 0 & 0 & 0 \\
		 0 & 0 & 0 & 0 & 0 & 0 & 0 &  [ U^{\text{AD}}_{\text{two}} ]_{8,8} & 0 & 0 &  [ U^{\text{AD}}_{\text{two}} ]_{8,11} & 0 & 0 &  [ U^{\text{AD}}_{\text{two}} ]_{8,14} & 0 & 0 \\
		 0 & 0 & 0 & 0 & 0 & 0 & 0 & 0 & 1 & 0 & 0 & 0 & 0 & 0 & 0 & 0 \\
		 0 & 0 & 0 & 0 & 0 & 0 &  [ U^{\text{AD}}_{\text{two}} ]_{7,10} & 0 & 0 &  [ U^{\text{AD}}_{\text{two}} ]_{10,10} & 0 & 0 & 0 & 0 & 0 & 0 \\
		 0 & 0 & 0 & 0 & 0 & 0 & 0 &  [ U^{\text{AD}}_{\text{two}} ]_{11,8} & 0 & 0 &  [ U^{\text{AD}}_{\text{two}} ]_{11,11}  & 0 & 0 &  [ U^{\text{AD}}_{\text{two}} ]_{11,14} & 0 & 0 \\
		 0 & 0 & 0 & 0 & 0 & 0 & 0 & 0 & 0 & 0 & 0 &  [ U^{\text{AD}}_{\text{two}} ]_{12,12} & 0 & 0 &  [ U^{\text{AD}}_{\text{two}} ]_{12,15} & 0 \\
		 0 & 0 & 0 & 0 & 0 & 0 & 0 & 0 & 0 & 0 & 0 & 0 & 1 & 0 & 0 & 0 \\
		 0 & 0 & 0 & 0 & 0 & 0 & 0 & [ U^{\text{AD}}_{\text{two}} ]_{14,8}  & 0 & 0 &  [ U^{\text{AD}}_{\text{two}} ]_{14,11} & 0 & 0 &  [ U^{\text{AD}}_{\text{two}} ]_{14,14} & 0 & 0 \\
		 0 & 0 & 0 & 0 & 0 & 0 & 0 & 0 & 0 & 0 & 0 &  [ U^{\text{AD}}_{\text{two}} ]_{15,12} & 0 & 0 &  [ U^{\text{AD}}_{\text{two}} ]_{15,15} & 0 \\
		 0 & 0 & 0 & 0 & 0 & 0 & 0 & 0 & 0 & 0 & 0 & 0 & 0 & 0 & 0 & 1 \\
		\end{array}
	\right ]  .  \label{twoQrelaxationunitary1} 
\end{align} \end{widetext}
In the above equation, we have omitted the argument $t$ for $ U^{\text{AD}}_{\text{two}} $ so as to write the equation shortly.  
Let us take a look at the physical meanings of  matrix elements of $ U^{\text{AD}}_{\text{two}} $.
Here, let us focus on particular components, for example, the elements $ [ U^{\text{AD}}_{\text{two}} ]_{14,8}$ and $[ U^{\text{AD}}_{\text{two}} ]_{8,8}$.  
The matrix element $ [ U^{\text{AD}}_{\text{two}} ]_{14,8}$ describes the probability amplitude of transition process between   
$  | \boldsymbol{\tilde{\alpha} } _{ 7 } \rangle  = | \boldsymbol{\check{1}}  \rangle _{ \text{S} }   \otimes   | \boldsymbol{\check{3}}  \rangle _{ \text{E} } $ and 
$  | \boldsymbol{\tilde{\alpha} } _{ 13 } \rangle  = | \boldsymbol{\check{3}}  \rangle _{ \text{S} }   \otimes   | \boldsymbol{\check{1}}  \rangle _{ \text{E} } $. 
In contrast, the element $  [ U^{\text{AD}}_{\text{two}} ]_{8,8} $ represents the amplitude of the quantum state $   | \boldsymbol{\tilde{\alpha} } _{ 8} \rangle =  | \boldsymbol{\check{1}}  \rangle _{ \text{S} }   \otimes   | \boldsymbol{\check{3}}  \rangle _{ \text{E} } $
to be remained the same with the probability less than one due to the decay effect.  The physical meanings of the other components can be understood in the same way.  
Next,  we derive the reduced density matrix 
$  \rho^{\text{out}}_{\text{S}} = \text{Tr}_{\text{E}} \left[  \rho^{\text{out}}_{\text{tot}}     \right] = \text{Tr}_{\text{E}} \left[ U^{\text{AD}}_{\text{two}} \cdot    \rho^{\text{in}}_{\text{tot}}  \cdot (U^{\text{AD}}_{\text{two}})^\dagger   \right] $.
It is described by the Kraus operators defined by 
 $ \mathcal{M}_{n_{\text{E}}} =  \sum_{ n_{\text{S}}  , n^\prime_{\text{S}}  }  {}_{\text{S}}\langle n_{\text{S}}   | \otimes {}_{\text{E}}\langle n_{\text{E}}   |   \left[  U^{\text{AD}}_{\text{two}}         \right]  | n^\prime_{\text{S}} \rangle_{\text{S}} \otimes | \check{\boldsymbol{3}}  \rangle_{\text{E}}   
 \cdot   | n_{\text{S}} \rangle_{\text{S}} \langle n^\prime_{\text{S}} | $.
  Here  $ n_{\text{E}}  , n_{\text{S}}  , n^\prime_{\text{S}} = \check{\boldsymbol{0}}, \check{\boldsymbol{1}},  \check{\boldsymbol{2}}, \check{\boldsymbol{3}}.$
The matrix representation of the Kraus operators $ \mathcal{M}_{n_{\text{E}}} $ are given by 
  \begin{align}
 \mathcal{M}_{\check{\boldsymbol{0}}}  &  =
	\left [
		\begin{array}{cccc} 
		 0 & 0 & 0 & 0 \\
		 0 & 0 & 0 & 0 \\
		  0 & 0 & 0 & 0 \\
		   0 & 0 & 0 & 0 
		\end{array}
	\right ] , \notag\\
  \mathcal{M}_{\check{\boldsymbol{1}}}  
&  =
	\left [
		\begin{array}{cccc} 
		 0 & 0 & 0 & 0 \\
		 0 & 0 & 0 & 0 \\
		  0 & 0 & 0 & 0 \\
		   0 & [ U^{\text{AD}}_{\text{two}} ]_{14,8}  & 0 & 0 
		\end{array}
	\right ] , \notag\\
 \mathcal{M}_{\check{\boldsymbol{2}}}  & =	\left [
		\begin{array}{cccc} 
		 0 & 0 & 0 & 0 \\
		 0 & 0 & 0 & 0 \\
		  0 &  [ U^{\text{AD}}_{\text{two}} ]_{11,8} & 0 & 0 \\
		   0 & 0 & [ U^{\text{AD}}_{\text{two}} ]_{15,12}  & 0 
		\end{array}
	\right ] ,
 \notag\\
  \mathcal{M}_{\check{\boldsymbol{3}}} & =	\left [
		\begin{array}{cccc} 
		 1 & 0 & 0 & 0 \\
		 0 &  [ U^{\text{AD}}_{\text{two}} ]_{8,8} & 0 & 0 \\
		  0 & 0 &  [ U^{\text{AD}}_{\text{two}} ]_{12,12} & 0 \\
		   0 &0  & 0 & 1 
		\end{array}
	\right ] .
\label{twoQrelaxKraus2} 
\end{align}
The Kraus operators in Eq. \eqref{twoQrelaxKraus2} satisfy the condition $ \sum_{ n_{\text{E}}  } \mathcal{M}^\dagger_{n_{\text{E}}}   \mathcal{M}_{n_{\text{E}}} = \boldsymbol{1}_{4 \times 4}$.
The matrix $\boldsymbol{1}_{4 \times 4}$ is the four by four identity matrix.  
Let us see the physical meanings of the matrix elements of the Kraus operators $  \mathcal{M}_{n_{\text{E}}}$ \cite{qutirtKraus}.
As we did for $ U^{\text{AD}}_{\text{two}} $ in Eq.  \eqref{twoQrelaxationunitary1}, we take a look at several components.
 The  $(3,2)$-th element of $  \mathcal{M}_{\check{\boldsymbol{2}}} $ represents the probability amplitude of the decay process  
 $ | \boldsymbol{\check{1}}  \rangle _{ \text{S} }   \rightarrow   | \boldsymbol{\check{2}}  \rangle _{ \text{S} } . $
 On the other hand,  $(3,3)$-th element of $  \mathcal{M}_{\check{\boldsymbol{3}}} $ describes the probability amplitude of the quantum state $  | \boldsymbol{\check{2}}  \rangle _{ \text{S} } $ to be unchanged under the decay effects.  
We see that the squares of  absolute $| [ U^{\text{AD}}_{\text{two}} ]_{11,8} |^2$, 
$| [ U^{\text{AD}}_{\text{two}} ]_{15,12} |^2$,  and $|  [ U^{\text{AD}}_{\text{two}} ]_{14,8} |^2$ can be identified with the decay strengths  
$\Gamma_{  \check{\boldsymbol{2}} \check{\boldsymbol{1}} }, \Gamma_{  \check{\boldsymbol{3}} \check{\boldsymbol{2}} }, $ and $\Gamma_{  \check{\boldsymbol{3}} \check{\boldsymbol{1}} }$, respectively. 
The reduced density matrix $  \rho^{\text{out}}_{\text{S}} $ are expressed by the Kraus operators in Eq. \eqref{twoQrelaxKraus2} as
$
\rho^{\text{out}}_{\text{S}}    = \sum_{ n_{\text{E}}   } {}_{\text{E}}\langle n_{\text{E}}   | \rho^{\text{out}}_{\text{tot}}  | n_{\text{E}}  \rangle_{\text{E}}
 = \sum_{ n_{\text{E}}  } \mathcal{M}_{n_{\text{E}}}  \rho^{\text{in}}_{\text{S}}  \mathcal{M}^\dagger_{n_{\text{E}}} $, or
 \begin{align}
 \left[ \rho^{\text{out}}_{\text{S}}  \right]_{  \check{\boldsymbol{0}} ,  \check{\boldsymbol{1}} }  & =    [ U^{\text{AD}}_{\text{two}} ]_{8,8}    \left[  \rho^{\text{in}}_{\text{S}}  \right]_{  \check{\boldsymbol{0}} ,  \check{\boldsymbol{1}} } , \notag\\
 \left[ \rho^{\text{out}}_{\text{S}}  \right]_{  \check{\boldsymbol{0}} ,  \check{\boldsymbol{2}} }  & =    [ U^{\text{AD}}_{\text{two}} ]_{12,12}    \left[  \rho^{\text{in}}_{\text{S}}  \right]_{  \check{\boldsymbol{0}} ,  \check{\boldsymbol{2}} } , \notag \\
\left[ \rho^{\text{out}}_{\text{S}}  \right]_{  \check{\boldsymbol{1}} ,  \check{\boldsymbol{1}} }  & =   | [ U^{\text{AD}}_{\text{two}} ]_{8,8} |^2   \left[  \rho^{\text{in}}_{\text{S}}  \right]_{  \check{\boldsymbol{1}} ,  \check{\boldsymbol{1}} } , \notag\\
\left[ \rho^{\text{out}}_{\text{S}}  \right]_{  \check{\boldsymbol{1}} ,  \check{\boldsymbol{2}} }  & = \left[ \left( U^{\text{AD}}_{\text{two}} \right)^\dagger \right]_{12,12}   \cdot   [ U^{\text{AD}}_{\text{two}} ]_{8,8}   \left[  \rho^{\text{in}}_{\text{S}}  \right]_{  \check{\boldsymbol{1}} ,  \check{\boldsymbol{2}} } , \notag\\
\left[ \rho^{\text{out}}_{\text{S}}  \right]_{  \check{\boldsymbol{1}} ,  \check{\boldsymbol{3}} } &  =     [ U^{\text{AD}}_{\text{two}} ]_{8,8}   \left[  \rho^{\text{in}}_{\text{S}}  \right]_{  \check{\boldsymbol{1}} ,  \check{\boldsymbol{3}} } , \notag\\ 
\left[ \rho^{\text{out}}_{\text{S}}  \right]_{  \check{\boldsymbol{2}} ,  \check{\boldsymbol{2}} }  & =   | [ U^{\text{AD}}_{\text{two}} ]_{12,12} |^2   \left[  \rho^{\text{in}}_{\text{S}}  \right]_{  \check{\boldsymbol{2}} ,  \check{\boldsymbol{2}} }  +
 | [ U^{\text{AD}}_{\text{two}} ]_{11,8} |^2   \left[  \rho^{\text{in}}_{\text{S}}  \right]_{  \check{\boldsymbol{1}} ,  \check{\boldsymbol{1}} } ,  \notag\\
 \left[ \rho^{\text{out}}_{\text{S}}  \right]_{  \check{\boldsymbol{2}} ,  \check{\boldsymbol{3}} }  & =  [ U^{\text{AD}}_{\text{two}} ]_{12,12}   \left[  \rho^{\text{in}}_{\text{S}}  \right]_{  \check{\boldsymbol{2}} ,  \check{\boldsymbol{3}} }  \notag\\
& + \left[ \left( U^{\text{AD}}_{\text{two}} \right)^\dagger \right]_{15,12}   \cdot     [ U^{\text{AD}}_{\text{two}} ]_{11,8}   \left[  \rho^{\text{in}}_{\text{S}}  \right]_{  \check{\boldsymbol{1}} ,  \check{\boldsymbol{2}} } , \notag\\
\left[ \rho^{\text{out}}_{\text{S}}  \right]_{  \check{\boldsymbol{3}} ,  \check{\boldsymbol{3}} }  & =
1 - \left[ \rho^{\text{out}}_{\text{S}}  \right]_{  \check{\boldsymbol{1}} ,  \check{\boldsymbol{1}} } 
 - \left[ \rho^{\text{out}}_{\text{S}}  \right]_{  \check{\boldsymbol{2}} \check{\boldsymbol{2}} } ,
\label{twoQrelaxKraus1}  
\end{align} 
 where  $  \left[  \rho^{\text{out}}_{\text{S}}  \right]_{ n_{\text{S}}, n^\prime_{\text{S}}}  =  
 {}_{\text{S}}\langle n_{\text{S}}   | \rho^{\text{out}}_{\text{S}}  | n^\prime_{\text{S}}  \rangle_{\text{S}}. $
 We note that the solution presented in Eq. \eqref{masterequation4} and 
the similar Kraus representations with the ones given in Eq. \eqref{twoQrelaxKraus2} have also been  derived in \cite{commonEDavidovichgroupPRA2009}. 

Let us now formulate the unitary transformation $U^{\text{AD}}_{\text{two}}$ with the single- and two-qubit gates,  
which generates the reduced density matrix $  \rho^{\text{out}}_{\text{S}} $ in Eq. \eqref{twoQrelaxKraus1}. 
This analysis is going to be the main issue of this paper.
We will do this by extending the formalism of the single-qubit amplitude damping. 
To describe this, we have prepared the single controlled-rotational gate which is given by the rotational operation  $R_y \big{(} \vartheta(t) \big{)}$ 
in Fig. \ref{overallsingleQrelaxQC}. We adopt the same approach and prepare some controlled-rotational operations. 
In this case, there are three types of decay channels.  
By making a correspondence with this, we prepare three different controlled-rotational operations. 
Let us name the three rotation angles for these controlled-rotational operations as  
$\vartheta_{  \check{\boldsymbol{2}} \check{\boldsymbol{1}} }, \vartheta_{  \check{\boldsymbol{3}} \check{\boldsymbol{2}} }, $ and $\vartheta_{  \check{\boldsymbol{3}} \check{\boldsymbol{1}} }$, which
 correspond to the three decay strengths 
$\Gamma_{  \check{\boldsymbol{2}} \check{\boldsymbol{1}} }, \Gamma_{  \check{\boldsymbol{3}} \check{\boldsymbol{2}} }, $ and $\Gamma_{  \check{\boldsymbol{3}} \check{\boldsymbol{1}} }$, respectively.
By writing these three controlled-rotational operations as $R \big{[}\vartheta_{ \check{\boldsymbol{2}} \check{\boldsymbol{1}} }  \big{]},   R \big{[}\vartheta_{ \check{\boldsymbol{3}} \check{\boldsymbol{2}} }  \big{]}$, and 
$ R \big{[}\vartheta_{ \check{\boldsymbol{3}} \check{\boldsymbol{1}} }  \big{]} $,  we formulate the unitary operation  $U^{\text{AD}}_{\text{two}}$ with them as 
\begin{align} 
 U^{\text{AD}}_{\text{two}} \left[    \vartheta_{ \check{\boldsymbol{2}} \check{\boldsymbol{1}} }, \vartheta_{ \check{\boldsymbol{3}} \check{\boldsymbol{1}} }, \vartheta_{ \check{\boldsymbol{3}} \check{\boldsymbol{2}} }    \right]
= R \big{[}\vartheta_{ \check{\boldsymbol{2}} \check{\boldsymbol{1}} }  \big{]} \cdot R \big{[}\vartheta_{  \check{\boldsymbol{3}} \check{\boldsymbol{1}} }  \big{]}  \cdot R \big{[}\vartheta_{ \check{\boldsymbol{3}} \check{\boldsymbol{2}}  }  \big{]}, 
 \label{twoQrelaxationrotations0} 
\end{align} 
where 
\begin{widetext}
\begin{align}
&  R \big{[}\vartheta_{ \check{\boldsymbol{2}} \check{\boldsymbol{1}}  }  \big{]}   =
 \tau_{11,16} \cdot \tau_{10,12}  \cdot \tau_{12,15}  \cdot U_{\text{C}_2R_y \left[  \vartheta_{  \check{\boldsymbol{2}} \check{\boldsymbol{1}} }  \right] }
  \left[ Q_1Q_2;Q_0 \right]  \cdot
 \tau_{12,15}    \cdot  \tau_{10,12} \cdot \tau_{11,16} , \notag\\
&  R \big{[}\vartheta_{ \check{\boldsymbol{3}} \check{\boldsymbol{1}}  }  \big{]} =
\tau_{14,16}   \cdot U_{\text{C}_3R_y \left[  \vartheta_{  \check{\boldsymbol{3}} \check{\boldsymbol{1}} }  \right] } \left[ Q_1Q_2Q_3;Q_0  \right]   \cdot   \tau_{14,16}, \notag\\
& R \big{[}\vartheta_{ \check{\boldsymbol{3}} \check{\boldsymbol{2}} }  \big{]}   =
 \tau_{14,16} \cdot \tau_{15,16}   \cdot U_{\text{C}_2R_y \left[  \vartheta_{  \check{\boldsymbol{3}} \check{\boldsymbol{2}} }  \right] } \left[ Q_0Q_2;Q_1 \right]    \cdot  \tau_{15,16} \cdot \tau_{14,16}.
 \label{twoQrelaxationrotations1} 
\end{align}  \end{widetext} 
Let us explain the characteristics of the controlled-unitary operators presented in the above equations.  
At first, the unitary operator $U_{\text{C}_2R_y \left[  \vartheta_{  \check{\boldsymbol{2}} \check{\boldsymbol{1}} }  \right] } \left[ Q_1Q_2;Q_0  \right] $ is the controlled-unitary operator such that  the two qubits $Q_1$ and $Q_2$ are the controlled bits while $Q_0$ is the target bit. 
The unitary operator to become performed is the rotational operation $R_y ( \vartheta_{  \check{\boldsymbol{2}} \check{\boldsymbol{1}} })$.
We have introduced the notation $\text{C}_2R_y \left[  \vartheta_{  \check{\boldsymbol{2}} \check{\boldsymbol{1}} }  \right] $ 
to express that it is the controlled-unitary operator composed of two controlled bits with the given unitary operator $R_y \left[  \vartheta_{  \check{\boldsymbol{2}} \check{\boldsymbol{1}} }  \right] $. 
In the argument of $U_{\text{C}_2R_y\left[  \vartheta_{  \check{\boldsymbol{2}} \check{\boldsymbol{1}} }  \right]} $, we have separated the controlled bits $Q_1$ and $Q_2$  and the target bit $Q_0$  with a semicolon.   
In the same way, $U_{\text{C}_3R_y\left[  \vartheta_{  \check{\boldsymbol{3}} \check{\boldsymbol{1}} }  \right]} \left[ Q_1Q_2Q_3;Q_0 \right] $
 is defined as the controlled-unitary operator comprised of  
the three controlled bits $Q_1,Q_2$ and $Q_3$ and the target bit $Q_0$. 
The unitary operation to be executed is $R_y ( \vartheta_{  \check{\boldsymbol{3}} \check{\boldsymbol{1}} })$; 
similar meaning is given for $U_{\text{C}_2R_y\left[  \vartheta_{  \check{\boldsymbol{3}} \check{\boldsymbol{2}} }  \right]} \left[ Q_0Q_2;Q_1 \right]  $.
On the other hand, $\tau_{15,16}$ is  the controlled-unitary operator 
such that when it is multiplied on the left-hand side of matrix $A$, its 15th and 16th rows  become exchanged
while we multiply $\tau_{15,16}$ on the right-hand side of $A$, then its 15th and 16th columns get exchanged; similar definitions are 
given for $\tau_{14,16}, \tau_{12,16},\tau_{10,12},\tau_{12,15},$ and $\tau_{11,16}$.  
We show the derivation of interexchange operators $\tau_{10,12},\tau_{12,15},$ and $\tau_{11,16}$ in Eq. \eqref{exchangeoperatorsderivation}.
The operator $\tau_{15,16}$ is equivalent to the controlled-$X$ operator comprised of three controlled bits $Q_0,Q_1,$ and $Q_2$ and the target bit $Q_3$:
$ \tau_{15,16} \equiv U_{\text{C}_3X} \left[ Q_0Q_1Q_2;Q_3  \right].$ 
 To construct the controlled-unitary operators given in the right-hand side of Eq.  \eqref{twoQrelaxationrotations1} 
 with the single- and two-qubit gates, we have followed the analysis shown in \cite{elemgates}.   
In Eq. \eqref{controlledunitaryoperatorsBellrelax} in Appendix \ref{Qgatesfourqubits},   
we present the mathematical representations for the controlled-unitary operators appearing in the right-hand side of the above equation
 in terms of single- and two-qubit gates and the Toffoli gates.
Based on the above formulation of $U^{\text{AD}}_{\text{two}} \left[    \vartheta_{ \check{\boldsymbol{2}} \check{\boldsymbol{1}} }, \vartheta_{ \check{\boldsymbol{3}} \check{\boldsymbol{2}} }, \vartheta_{ \check{\boldsymbol{3}} \check{\boldsymbol{1}} }    \right]$,
 from Eqs. \eqref{masterequation4},  \eqref{twoQrelaxationunitary1}, \eqref{twoQrelaxKraus2}, \eqref{twoQrelaxKraus1}, and \eqref{twoQrelaxationrotations1} 
 we can make an identification  
\ \begin{align}
  | [ U^{\text{AD}}_{\text{two}} ]_{11,8} |^2  & = \sin^2 \Big{(} \frac{\vartheta_{ \check{\boldsymbol{2}} \check{\boldsymbol{1}}}}{2} \Big{)}  
\cos^2 \Big{(} \frac{\vartheta_{ \check{\boldsymbol{3}} \check{\boldsymbol{1}}}}{2} \Big{)}  
\equiv \Gamma_{  \check{\boldsymbol{2}} \check{\boldsymbol{1}} }, \notag\\ 
  | [ U^{\text{AD}}_{\text{two}} ]_{15,12} |^2   & = \sin^2 \Big{(} \frac{\vartheta_{ \check{\boldsymbol{3}} \check{\boldsymbol{2}}}}{2} \Big{)}  \equiv \Gamma_{  \check{\boldsymbol{3}} \check{\boldsymbol{2}} },  \notag\\ 
  | [ U^{\text{AD}}_{\text{two}} ]_{14,8} |^2 &  = \sin^2 \Big{(} \frac{\vartheta_{ \check{\boldsymbol{3}} \check{\boldsymbol{1}}}}{2} \Big{)}  \equiv \Gamma_{  \check{\boldsymbol{3}} \check{\boldsymbol{1}} }, \notag\\ 
   | [ U^{\text{AD}}_{\text{two}} ]_{8,8} |^2 & = 1  -   | [ U^{\text{AD}}_{\text{two}} ]_{11,8} |^2 -  | [ U^{\text{AD}}_{\text{two}} ]_{14,8} |^2, \notag\\ 
& =  \cos^2 \Big{(} \frac{\vartheta_{ \check{\boldsymbol{2}} \check{\boldsymbol{1}}}}{2} \Big{)}  
\cos^2 \Big{(} \frac{\vartheta_{ \check{\boldsymbol{3}} \check{\boldsymbol{1}}}}{2} \Big{)}  
\equiv 1 - \Gamma_{  \check{\boldsymbol{2}} \check{\boldsymbol{1}} } -  \Gamma_{  \check{\boldsymbol{3}} \check{\boldsymbol{1}} }, \notag\\ 
  | [ U^{\text{AD}}_{\text{two}} ]_{12,12} |^2  & = 1 -   | [ U^{\text{AD}}_{\text{two}} ]_{15,12} |^2 ,  \notag\\ 
  & =  \cos^2 \Big{(} \frac{\vartheta_{ \check{\boldsymbol{3}} \check{\boldsymbol{2}}}}{2} \Big{)}   \equiv 1 - \Gamma_{  \check{\boldsymbol{3}} \check{\boldsymbol{2}} } ,
\label{Ugammaidentify}
\end{align}
where we have used the unitary condition $ \big{[} ( U^{\text{AD}}_{\text{two}} )^\dagger  \cdot U^{\text{AD}}_{\text{two}} \big{]}_{a,b}  =
  \big{[} U^{\text{AD}}_{\text{two}}  \cdot ( U^{\text{AD}}_{\text{two}} )^\dagger  \big{]}_{a,b}  =
 \delta_{a,b}$ $(a,b=1,2,\ldots,16 )$.
Next, we formulate the decay strengths $\Gamma_{  \check{\boldsymbol{2}} \check{\boldsymbol{1}} }, \Gamma_{  \check{\boldsymbol{3}} \check{\boldsymbol{2}} }, $ and $\Gamma_{  \check{\boldsymbol{3}} \check{\boldsymbol{1}} }$
as functions of the time $t$ so that  they match with the solution of quantum master equation \eqref{masterequation4} up to $ \mathcal{O} \left(   (\gamma t)^2 \right)$ \cite{qutirtKraus}.
From Eqs.  \eqref{masterequation4}, \eqref{twoQrelaxKraus1}, and \eqref{Ugammaidentify}, we have
\begin{align}
\Gamma_{  \check{\boldsymbol{2}} \check{\boldsymbol{1}} }  (t) & = 2 \gamma t - 4(\gamma t)^2 + \mathcal{O} \left(   (\gamma t)^3     \right), \notag\\
\Gamma_{  \check{\boldsymbol{3}} \check{\boldsymbol{2}} }  (t) &  = 2 \gamma t - 2(\gamma t)^2 + \mathcal{O} \left(   (\gamma t)^3     \right), \notag\\ 
\Gamma_{  \check{\boldsymbol{3}} \check{\boldsymbol{1}} }  (t)  & = 2(\gamma t)^2 + \mathcal{O} \left(   (\gamma t)^3     \right) ,
\label{twoQdecaychannel1}
\end{align}
or equivalently,
\begin{align}
 \vartheta_{  \check{\boldsymbol{2}} \check{\boldsymbol{1}} }   (t)  &  = 2 \arcsin \left[   \left( 2 \gamma t - 4(\gamma t )^2 \right)^{\frac{1}{2}} \right]   , \notag\\
 \vartheta_{  \check{\boldsymbol{3}} \check{\boldsymbol{2}} }  (t) & = 2 \arcsin \left[   \left(  2 \gamma t - 2(\gamma t )^2  \right)^{\frac{1}{2}}   \right]   , \notag\\
 \vartheta_{  \check{\boldsymbol{2}} \check{\boldsymbol{1}} }   (t) & =  2 \arcsin \left[   \left(  2(\gamma t)^2 \right)^{\frac{1}{2}} \right] , 
\label{twoQdecaychannel2}
\end{align}
where we have used  
$\sin^2 \Big{(} \frac{\vartheta_{ \check{\boldsymbol{2}} \check{\boldsymbol{1}}}}{2} \Big{)}  
\cos^2 \Big{(} \frac{\vartheta_{ \check{\boldsymbol{3}} \check{\boldsymbol{1}}}}{2} \Big{)}  \approx
\sin^2 \Big{(} \frac{\vartheta_{ \check{\boldsymbol{2}} \check{\boldsymbol{1}}}}{2} \Big{)} $.
 Indeed, such an approximation is valid as long as we take $\Gamma_{  \check{\boldsymbol{2}} \check{\boldsymbol{1}} }  (t)$ up to $\mathcal{O} \left(   (\gamma t)^2     \right)$.
Thus, the values of rotation angles  
$ \vartheta_{  \check{\boldsymbol{2}} \check{\boldsymbol{1}} }, \vartheta_{  \check{\boldsymbol{3}} \check{\boldsymbol{2}} },\vartheta_{  \check{\boldsymbol{3}} \check{\boldsymbol{1}} }$
are determined by Eq. \eqref{twoQdecaychannel2} and 
generate the collective amplitude damping at the time $t$. 
To summarize, the quantum circuit for our simulation of the collective  amplitude damping is given by the  unitary transformation
 \begin{align}
U^{\text{AD,two}}_{\text{QC}} \left[    \vartheta_{ \check{\boldsymbol{2}} \check{\boldsymbol{1}} }, \vartheta_{ \check{\boldsymbol{3}} \check{\boldsymbol{1}} }, \vartheta_{ \check{\boldsymbol{3}} \check{\boldsymbol{2}} }    \right] (t)
& =
U^{\text{AD}}_{\text{two}} \left[    \vartheta_{ \check{\boldsymbol{2}} \check{\boldsymbol{1}} }, \vartheta_{ \check{\boldsymbol{3}} \check{\boldsymbol{1}} }, \vartheta_{ \check{\boldsymbol{3}} \check{\boldsymbol{2}} }    \right] (t) \notag\\
&
\cdot  \left[  U^{\text{in}}_{\text{S}}  \otimes  X_{Q_2} \otimes X_{Q_3} \right],
  \label{twoQrelaxQC} 
\end{align} 
and with the measurement procedures on the ancilla bits $Q_2$ and  $Q_3$ (the environment E). 
Let us illustrate this quantum circuit in Fig. \ref{BellQCtwoQCAD2}.
\begin{figure}[H] 
\centering
\mbox{ 
\Qcircuit @C=1.3em @R=1.3em { 
    \lstick{| Q_0 \rangle} &  \multigate{1}{  U^{\text{in}}_{\text{S}}  }      & \multigate{3}{ 
   U^{\text{AD}}_{\text{two}} \left[    \vartheta_{ \check{\boldsymbol{2}} \check{\boldsymbol{1}} }, \vartheta_{ \check{\boldsymbol{3}} \check{\boldsymbol{1}} }, \vartheta_{ \check{\boldsymbol{3}} \check{\boldsymbol{2}} }    \right] (t)
    } & \qw & \qw \\
    \lstick{| Q_1 \rangle} &\ghost{  U^{\text{in}}_{\text{S}} }        & \ghost{
   U^{\text{AD}}_{\text{two}} \left[    \vartheta_{ \check{\boldsymbol{2}} \check{\boldsymbol{1}} }, \vartheta_{ \check{\boldsymbol{3}} \check{\boldsymbol{1}} }, \vartheta_{ \check{\boldsymbol{3}} \check{\boldsymbol{2}} }    \right] (t)
    }        & \qw & \qw \\
    \lstick{| Q_2 \rangle} & \gate{X} & \ghost{
    U^{\text{AD}}_{\text{two}} \left[    \vartheta_{ \check{\boldsymbol{2}} \check{\boldsymbol{1}} }, \vartheta_{ \check{\boldsymbol{3}} \check{\boldsymbol{1}} }, \vartheta_{ \check{\boldsymbol{3}} \check{\boldsymbol{2}} }    \right] (t)
    }        & \meter & \cw \\
    \lstick{| Q_3 \rangle} & \gate{X} & \ghost{
    U^{\text{AD}}_{\text{two}} \left[    \vartheta_{ \check{\boldsymbol{2}} \check{\boldsymbol{1}} }, \vartheta_{ \check{\boldsymbol{3}} \check{\boldsymbol{1}} }, \vartheta_{ \check{\boldsymbol{3}} \check{\boldsymbol{2}} }    \right] (t)
    }        & \meter & \cw
} 
} 
\caption{ Quantum circuit for the collective amplitude damping constructed by the unitary operator 
$ U^{\text{AD}}_{\text{two}} \left[    \vartheta_{ \check{\boldsymbol{2}} \check{\boldsymbol{1}} }, \vartheta_{ \check{\boldsymbol{3}} \check{\boldsymbol{1}} }, \vartheta_{ \check{\boldsymbol{3}} \check{\boldsymbol{2}} }    \right] (t) $
in Eq.  \eqref{twoQrelaxQC} and the measurement procedures on the ancilla bits $Q_2$ and $Q_3$.
}
\label{BellQCtwoQCAD2} 
\end{figure}
Consequently, we obtain the reduced density matrix $\rho^{\text{out}}_{\text{S}}$ given in Eq. \eqref{twoQrelaxKraus1} 
owing to the quantum circuit in Fig. \ref{BellQCtwoQCAD2}.
From  $\rho^{\text{out}}_{\text{S}}$, we can extract the probability weights of states 
$ | \check{\boldsymbol{1}}  \rangle_{\text{S}} , | \check{\boldsymbol{2}}  \rangle_{\text{S}} $, 
and $| \check{\boldsymbol{3}}  \rangle_{\text{S}}$ at the time $t$. 
This is done by performing the measurement on the qubits $Q_0$ and $Q_1$ (system S).
As a result,  we obtain the collective amplitude damping at the time $t$: $ \langle J^z _{ \text{S} }   \rangle (t) =  w_{   |  \boldsymbol{\check{1}}  \rangle_{ \text{S} }} (t) - w_{   |  \boldsymbol{\check{3}}  \rangle_{ \text{S} }} (t). $

\subsection{Digital Quantum Simulation }\label{twoQDQM} 
\subsubsection{Procedures}\label{twoQcomresultprocedures} 
 We demonstrate in detail the digital quantum simulation of the collective  amplitude damping based  
 on the quantum circuit presented in Fig. \ref{BellQCtwoQCAD2}. We take  $\gamma = 1$.
 First, we set the time interval  given by $t_i = 0.005 \times i$  with $i=0,1,\ldots 9$. 
 By inserting them into Eq. \eqref{twoQdecaychannel2}, we obtain three values of angles 
 $ \vartheta_{  \check{\boldsymbol{2}} \check{\boldsymbol{1}} }   (t_i),  \vartheta_{  \check{\boldsymbol{3}} \check{\boldsymbol{2}} }   (t_i),  \vartheta_{  \check{\boldsymbol{3}} \check{\boldsymbol{1}} }   (t_i) $.
  Then, owing to   Eq.  \eqref{twoQrelaxQC}, 
  the overall unitary transformation 
  $ U^{\text{AD,two}}_{\text{QC}} \left[    \vartheta_{ \check{\boldsymbol{2}} \check{\boldsymbol{1}} }, \vartheta_{ \check{\boldsymbol{3}} \check{\boldsymbol{2}} }, \vartheta_{ \check{\boldsymbol{3}} \check{\boldsymbol{1}} }    \right] $
    is determined as the function of the time $t_i$. 
    To explicitly represent the $t_i$-dependence of 
    $ U^{\text{AD,two}}_{\text{QC}} \left[    \vartheta_{ \check{\boldsymbol{2}} \check{\boldsymbol{1}} }, \vartheta_{ \check{\boldsymbol{3}} \check{\boldsymbol{2}} }, \vartheta_{ \check{\boldsymbol{3}} \check{\boldsymbol{1}} }    \right] $, 
    let us rewrite it as  
    $ U^{\text{AD,two}}_{\text{QC}} \left[    \vartheta_{ \check{\boldsymbol{2}} \check{\boldsymbol{1}} }, \vartheta_{ \check{\boldsymbol{3}} \check{\boldsymbol{2}} }, \vartheta_{ \check{\boldsymbol{3}} \check{\boldsymbol{1}} }    \right] (t_i).$ 
    After this unitary operation is executed, we perform the measurement on the ancilla bits $Q_2$ and $Q_3$, and as a result, 
    we obtain the reduced density matrix $\rho^{\text{out}}_{\text{S}}$ at the time $t_i$, namely, $ \rho^{\text{out}}_{\text{S}} (t_i)  $. 
The matrix elements of $ \rho^{\text{out}}_{\text{S}} (t_i)  $ are expressed by the trigonometric functions 
  $\left(   \cos \Big{(} \frac{\vartheta_{ \check{\boldsymbol{2}} \check{\boldsymbol{1}}  (t_i)}}{2} \Big{)},  \sin \left( \frac{\vartheta_{ \check{\boldsymbol{2}} \check{\boldsymbol{1}}  (t_i)}}{2} \right)
 \right),
 \left(   \cos \Big{(} \frac{\vartheta_{ \check{\boldsymbol{3}} \check{\boldsymbol{2}} (t_i)}}{2} \Big{)},  \sin \left( \frac{\vartheta_{ \check{\boldsymbol{3}} \check{\boldsymbol{2}}  (t_i)}}{2} \right)
 \right)$, 
  and  $\left(   \cos \Big{(} \frac{\vartheta_{ \check{\boldsymbol{3}} \check{\boldsymbol{1}} (t_i)}}{2} \Big{)},  \sin \left( \frac{\vartheta_{ \check{\boldsymbol{3}} \check{\boldsymbol{1}} (t_i)}}{2} \right)
 \right)$.  
 Next, with the usage of qiskit,  
 we execute $N_{ \text{shots}}$ times the whole operation 
 (the operation described by the quantum circuit in Fig. \ref{BellQCtwoQCAD2} plus the measurement on $Q_0$ and $Q_1$ (system S)).
 We obtain the data of how many times the quantum states given in Eq.  \eqref{twoQrelaxbasisvector4} have been generated as the output states,
 which are labeled by "0" and "1".
  In other words, the quantum state $  | \boldsymbol{\tilde{\alpha}  } _{a} \rangle$ $(a = 1,2,\ldots,16)$ in Eq.  \eqref{twoQrelaxbasisvector4}
  is redescribed as $ | n_{Q_0} n_{Q_1} n_{Q_2} n_{Q_3}  \rangle_{ Q_0 Q_1Q_2Q_3 } $ $( n_{Q_0}, n_{Q_1}, n_{Q_2}, n_{Q_3} = 0,1)$
   via the identification  $  | \check{\boldsymbol{0}} \rangle \equiv | 00 \rangle,  | \check{\boldsymbol{1}}  \rangle \equiv | 01 \rangle, 
    |  \check{\boldsymbol{2}} \rangle \equiv | 10 \rangle,  | \check{\boldsymbol{3}} \rangle \equiv | 11 \rangle$. 
    Here the number $n_{Q_0}$  describes that the quantum state of the qubit $Q_0$ is  $ | n_{Q_0}  \rangle_{ Q_0  } $ and  the rest of three numbers 
      $n_{Q_1}, n_{Q_2}$, and  $n_{Q_3}$ are defined in the same way. 
      For instance, the quantum state  $ | 0111  \rangle_{ Q_0 Q_1Q_2Q_3 } $ is identical to the quantum state
      $| \check{\boldsymbol{1}}  \rangle _{ \text{S} }   \otimes   | \boldsymbol{\check{3}}  \rangle _{ \text{E} } =  | \boldsymbol{\tilde{\alpha}  } _{7} \rangle$.
   Let us denote the number $N_{  |  n_{Q_0} n_{Q_1} n_{Q_2} n_{Q_3} \rangle } (t_i)$ 
      as the number of  $ | n_{Q_0} n_{Q_1} n_{Q_2} n_{Q_3}  \rangle_{ Q_0 Q_1Q_2Q_3 } $, which have been generated as the output state at $t_i$. 
      These numbers satisfy the condition 
      $ \sum_{ n_{Q_0}, n_{Q_1}, n_{Q_2} , n_{Q_3} = 0,1 } N_{  |  n_{Q_0} n_{Q_1} n_{Q_2} n_{Q_3} \rangle } (t_i) =  N_{ \text{shots}} .$ 
  The numerical values  of  the probability weights $ w_{   |  \boldsymbol{\check{0}}  \rangle_{ \text{S} }} (t_i), w_{   |  \boldsymbol{\check{1}}  \rangle_{ \text{S} }} (t_i), w_{   |  \boldsymbol{\check{2}}  \rangle_{ \text{S} }} (t_i)$
  and   $w_{   |  \boldsymbol{\check{3}}  \rangle_{ \text{S} }} (t_i)$ are given by $N_{  |  n_{Q_0} n_{Q_1} n_{Q_2} n_{Q_3} \rangle } (t_i)$ as
  \begin{align}  
  w_{   | \check{\boldsymbol{0}}  \rangle_{ \text{S} }} (t_i)& =  \sum_{n_{Q_2}, n_{Q_3}=0,1}    \frac { N_{  | 00 n_{Q_2} n_{Q_3}  \rangle }  (t_i)  }{ N_{ \text{shots}} }  , \notag\\
  w_{   | \check{\boldsymbol{1}}  \rangle_{ \text{S} }} (t_i) & =  \sum_{n_{Q_2}, n_{Q_3}=0,1}  \frac { N_{  | 01 n_{Q_2} n_{Q_3}  \rangle } (t_i) }{ N_{ \text{shots}} } , \notag\\
  w_{   | \check{\boldsymbol{2}}  \rangle_{ \text{S} }} (t_i) & =  \sum_{n_{Q_2}, n_{Q_3}=0,1}  \frac { N_{  | 10 n_{Q_2} n_{Q_3}  \rangle} (t_i) }{ N_{ \text{shots}} } , \notag\\
  w_{   | \check{\boldsymbol{3}}  \rangle_{ \text{S} }} (t_i) & =  \sum_{n_{Q_2}, n_{Q_3}=0,1}  \frac { N_{  | 11 n_{Q_2} n_{Q_3}  \rangle } (t_i) }{ N_{ \text{shots}} } .
   \label{probpolidentifytwoQ}
  \end{align}    
  By using these values and the relation $ \langle J^z _{ \text{S} }   \rangle (t_i) =  w_{   |  \boldsymbol{\check{1}}  \rangle_{ \text{S} }} (t_i) 
  - w_{   |  \boldsymbol{\check{3}}  \rangle_{ \text{S} }} (t_i) $,
  we complete the quantum simulation of the collective amplitude damping at $t_i$ described as 
   \begin{align}  
  \langle J^z _{ \text{S} }   \rangle (t_i) =    \sum_{n_{Q_2}, n_{Q_3}=0,1}  \frac { N_{  | 01 n_{Q_2} n_{Q_3}  \rangle } (t_i) - N_{  | 11 n_{Q_2} n_{Q_3}  \rangle } (t_i)
  }  { N_{ \text{shots}} }.  
   \label{qiskitBellrelaxationformula}
    \end{align}  
 For our simulation, we  examine six different initial conditions given by
 $ \rho^{\text{in},1}_{\text{S}} =  |  \check{\boldsymbol{0}} \rangle_{\text{S}} \langle  \check{\boldsymbol{0}} | =  | \Psi^- \rangle_{\text{S}} \langle  \Psi^-   |$,  
 $ \rho^{\text{in},2}_{\text{S}} =  |  \check{\boldsymbol{3}} \rangle_{\text{S}} \langle  \check{\boldsymbol{3}} | $,
  $ \rho^{\text{in},3}_{\text{S}} =  |  \check{\boldsymbol{1}} \rangle_{\text{S}} \langle  \check{\boldsymbol{1}} | $,  
  $ \rho^{\text{in},4}_{\text{S}} =  |  \check{\boldsymbol{2}} \rangle_{\text{S}} \langle  \check{\boldsymbol{2}} | =  | \Psi^+ \rangle_{\text{S}} \langle  \Psi^+   |$,
 $ \rho^{\text{in},5}_{\text{S}} =  | \Phi^+ \rangle_{\text{S}} \langle  \Phi^+   | $,  
 and $ \rho^{\text{in},6}_{\text{S}} =  |   \Phi^-   \rangle_{\text{S}} \langle   \Phi^-  | $.
To generate these six initial states,  we perform six types of unitary transformations 
$ U^{\text{in},1}_{\text{S}} =  \boldsymbol{1}_{Q_0} \otimes  \boldsymbol{1}_{Q_1}  $,  
$ U^{\text{in},2}_{\text{S}} =  X_{Q_0} \otimes  X_{Q_1}  $, 
$ U^{\text{in},3}_{\text{S}} =  \boldsymbol{1}_{Q_0} \otimes X_{Q_1} $, 
$ U^{\text{in},4}_{\text{S}} =  X_{Q_0} \otimes  \boldsymbol{1}_{Q_1}  $, 
$ U^{\text{in},5}_{\text{S}} = U^\Phi _{\text{S}} \cdot  ( \boldsymbol{1}_{Q_0} \otimes X_{Q_1} )$,  
and $ U^{\text{in},6}_{\text{S}} = U^\Phi _{\text{S}} \cdot  ( X_{Q_0} \otimes X_{Q_1} )$.
The initial state $ \rho^{\text{in},l}_{\text{S}} $ is generated by the unitary transformation $ U^{\text{in},l}_{\text{S}} $ $(l = 1,2,\dots,6)$.
The unitary operator $U^\Phi _{\text{S}} $ appearing in $ U^{\text{in},5}_{\text{S}}  $ and $ U^{\text{in},6}_{\text{S}}  $
is defined by  $U^\Phi _{\text{S}} = U_{\text{SWAP}} \cdot  U_{\text{C}H}   [Q_0;Q_1] \cdot U_{\text{SWAP}},$
where $U_{\text{C}H}   [Q_0;Q_1]$ is the controlled-Hadamard gate such that the control bit is $Q_0$ whereas $Q_1$ is the target bit.
$U_{\text{SWAP}}$ is the SWAP gate. 
The matrix representations of  $ U_{\text{C}H}   [Q_0;Q_1] $ and $ U_{\text{SWAP}} $ are given in Eqs. \eqref{CHSTRy} and \eqref{SWAP}, respectively.
 
 For the two initial conditions  $ \rho^{\text{in},1}_{\text{S}} $ and $ \rho^{\text{in},2}_{\text{S}} $,  
 we have verified that both of them do not exhibit the amplitude damping processes as expected:
 we obtain $ N_{  | 0011  \rangle } (t_i) = N_{\text{shots}} $ for the initial condition $ \rho^{\text{in},1}_{\text{S}} $ 
 and  $ N_{  | 1111  \rangle } (t_i) = N_{\text{shots}}$ for the initial condition $ \rho^{\text{in},2}_{\text{S}} $.  
  In this paper, we do not present the simulation results for these two cases. 
 In the following, we focus on the collective amplitude damping processes for  
 the four initial conditions, $ \rho^{\text{in},3}_{\text{S}} $, $ \rho^{\text{in},4}_{\text{S}} $, $ \rho^{\text{in},5}_{\text{S}} $, and $ \rho^{\text{in},6}_{\text{S}} $, 
 and give some detailed discussions on them.    
 \begin{figure*}[t]
\includegraphics[width=1.0 \textwidth]{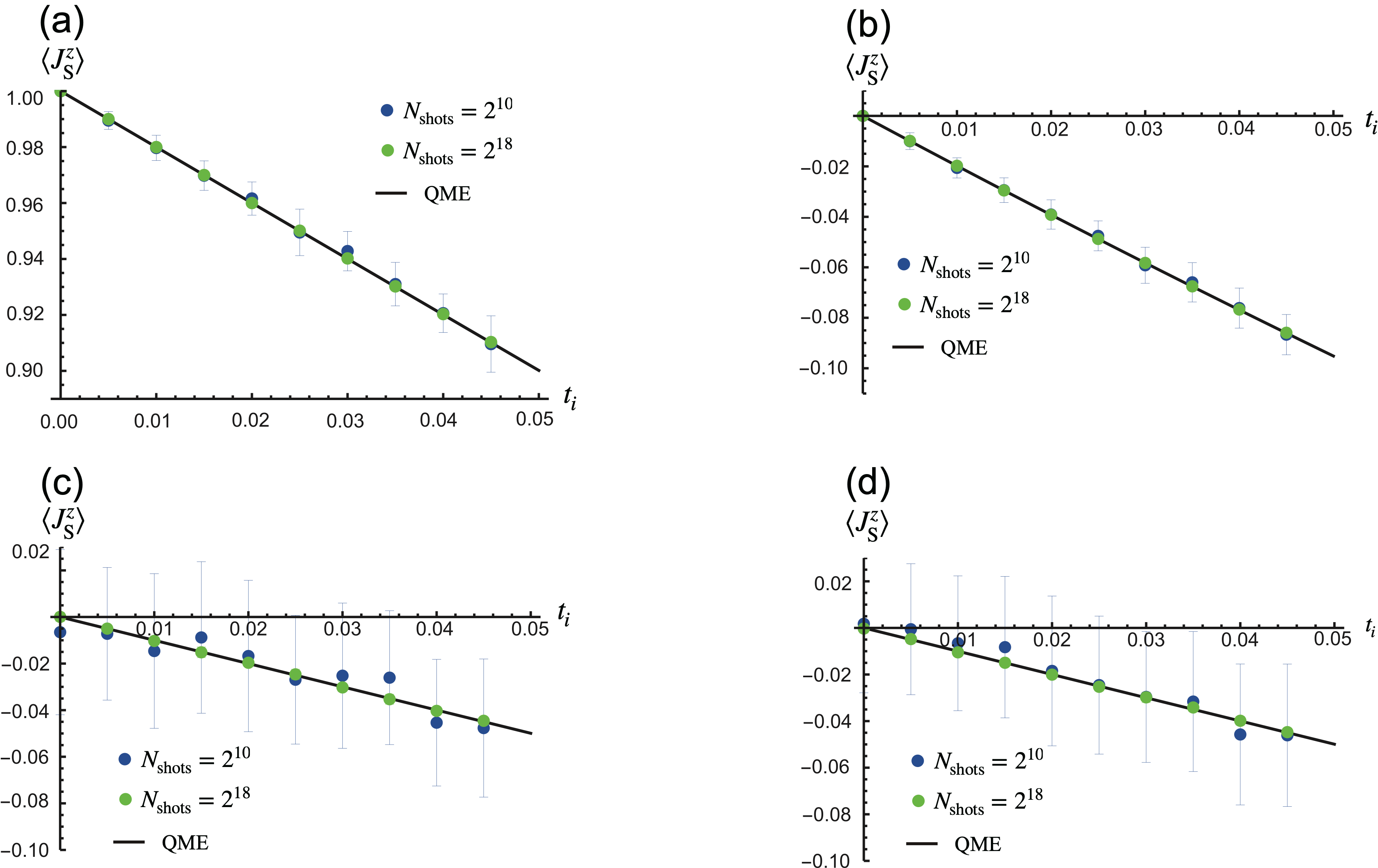}
\caption{ The quantum simulation results of the collective amplitude damping.  
The horizontal axis represents the time while the vertical axis is for the averaged expectation value $  \overline{ \langle J^z _{ \text{S} }   \rangle _{\text{in},l} } (N_{ \text{shots}},t_i) $. The data (a), (b), (c) and (d) are the results for the initial conditions $ \rho^{\text{in},3}_{\text{S}} , \rho^{\text{in},4}_{\text{S}} , \rho^{\text{in},5}_{\text{S}} , $ and $\rho^{\text{in},6}_{\text{S}}$, respectively, within the time interval $ t_i = 0.005 \times i$  $ (i=0,1,\ldots 9)$. 
The blue and green ten points are the data for  $ N_{ \text{shots}} = 2^{10}$ and $2^{18},$ respectively.
At each point, we present the error bar along the vertical direction with its size equal to 
the standard deviation $   \sigma_{\text{in},l} (N_{ \text{shots}}, t_i) = \sqrt{  \sigma^2_{\text{in},l} (N_{ \text{shots}}, t_i)  } $.
The black curves in each four figures represent the solution of  the quantum master equation given in  \eqref{masterequation4}.  
The short-hand notation ``QME" denotes the quantum master equation.}
\label{BellAD}  
\end{figure*}
\subsubsection{Results and Interpretations }\label{twoQcomresultinterpretations}  
We  display the results of the collective amplitude damping processes for the initial conditions $ \rho^{\text{in},3}_{\text{S}} , \rho^{\text{in},4}_{\text{S}} , \rho^{\text{in},5}_{\text{S}} , $ and $\rho^{\text{in},6}_{\text{S}}$ at the time interval $t_i = i  \times 0.005 $  in Figs. \ref{BellAD} (a), (b), (c), and (d), respectively. 
 The results for $ N_{ \text{shots}} = 2^{10}  $ and $2^{18}$  are plotted with blue and green points, respectively.  
 The black curves are the solution of the quantum master equation given in \eqref{masterequation4}.
  Like the simulation of the single-qubit amplitude damping, we plot the ``averaged expectation value" defined by
       \begin{align}  
  \overline{ \langle J^z _{ \text{S} }   \rangle _{\text{in},l} } (N_{ \text{shots}},t_i) 
  = \frac { \sum_{\alpha = 1}^{ N_{ \text{ave}} } \langle J^z _{ \text{S} } \rangle_{\text{in},l}  (\alpha, N_{ \text{shots}}, t_i)  } 
  { N_{ \text{ave}} },
     \label{averagedADtwoQ}   
  \end{align} 
    where $N_{ \text{ave}} $ is the repetition number of our quantum simulation for the collective amplitude damping based on the quantum circuit in Fig. \ref{BellQCtwoQCAD2}
 with given  $N_{ \text{shots}}$. We have taken $N_{ \text{ave}} = 50$. The quantity $ \langle J^z _{ \text{S} } \rangle_{\text{in},l}  (\alpha, N_{ \text{shots}}, t_i) $ is the expectation value obtained in the $\alpha$-th round of our simulation
 for the initial condition $ \rho^{\text{in},l}_{\text{S}} $  ($l = 3,4,5,6$). 
 In addition, we compute the variance defined by
 \begin{align}  
  \sigma^2_{\text{in},l} (N_{ \text{shots}}, t_i) & =  \overline{  \langle J^z _{ \text{S} }   \rangle_{\text{in},l}  ^2 } (N_{ \text{shots}}, t_i)  
  \notag\\
 & -  \left( \overline{   \langle J^z _{ \text{S} }   \rangle _{\text{in},l} } \right)^2 (N_{ \text{shots}}, t_i), 
   \label{variance1twoQ}
    \end{align}  
where the quantity $  \overline{  \langle J^z _{ \text{S} }   \rangle_{\text{in},l}  ^2 } (N_{ \text{shots}}, t_i)  $ is defined by
  \begin{align}  
 \overline{  \langle J^z _{ \text{S} }   \rangle_{\text{in},l}  ^2 } (N_{ \text{shots}}, t_i) = 
 \frac { \sum_{\alpha = 1}^{ N_{ \text{ave}} } \langle J^z _{ \text{S} } \rangle^2_{\text{in},l}  (\alpha, N_{ \text{shots}}, t_i)  } 
  { N_{ \text{ave}} }.
   \label{variance2twoQ}
    \end{align}  
  At each point, $ \big{(}t_i,   \overline{ \langle J^z _{ \text{S} }   \rangle _{\text{in},l} } (N_{ \text{shots}},t_i)  \big{)}$, 
   the bar parallel to the vertical axis is displayed with its size equal to 
  the standard deviation $   \sigma_{\text{in},l} (N_{ \text{shots}}, t_i) =
   \sqrt{  \sigma^2_{\text{in},l} (N_{ \text{shots}}, t_i)  } $. 

First, let us discuss our results from the physics point of view.   
For the case of the initial condition $ \rho^{\text{in},3}_{\text{S}} $ (Figs. \ref{BellAD} (a)),  
what we have obtained are the three quantities,  $ N_{  | 0111 \rangle } (t_i),N_{  | 1010 \rangle } (t_i)$, and $ N_{  | 1101 \rangle } (t_i)$. 
 Initially,  we have $ N_{  | 0111  \rangle } (t_0) = N_{\text{shots}} $ with all the other numbers equal to zero. 
 As time goes by, the number $ N_{  | 0111  \rangle } (t_i) $ decreases whereas both $ N_{  | 1010  \rangle } (t_i) $  and $ N_{  | 1101  \rangle } (t_i) $ increase.
 The decreasing of $ N_{  | 0111  \rangle } (t_i) $ while the increasing of $ N_{  | 1010  \rangle } (t_i) $ corresponds to the decay process 
 $  |  \check{\boldsymbol{1}} \rangle_{\text{S}} \otimes  |  \check{\boldsymbol{3}} \rangle_{\text{E}} \rightarrow   |  \check{\boldsymbol{2}} \rangle_{\text{S}} \otimes  |  \check{\boldsymbol{2}} \rangle_{\text{E}} $.
 On the other hand, the decreasing of $ N_{  | 0111  \rangle } (t_i) $ while the increasing of $ N_{  | 1101  \rangle } (t_i) $ 
 corresponds to the mixture of the two decay processes 
 $  |  \check{\boldsymbol{2}} \rangle_{\text{S}} \otimes  |  \check{\boldsymbol{2}} \rangle_{\text{E}} \rightarrow   |  \check{\boldsymbol{3}} \rangle_{\text{S}} \otimes  |  \check{\boldsymbol{2}} \rangle_{\text{E}} $
 and 
 $  |  \check{\boldsymbol{1}} \rangle_{\text{S}} \otimes  |  \check{\boldsymbol{3}} \rangle_{\text{E}} \rightarrow   |  \check{\boldsymbol{3}} \rangle_{\text{S}} \otimes  |  \check{\boldsymbol{1}} \rangle_{\text{E}} $.
 The increase of  $ N_{  | 1010  \rangle } (t_i) $ is larger than that of  $ N_{  | 1101  \rangle } (t_i) $.
 This is because owing to Eq. \eqref{twoQdecaychannel1}, in the short time regime the decay strength $ \Gamma_{  \check{\boldsymbol{2}} \check{\boldsymbol{1}} }  (t)  $ is stronger than 
 $ \Gamma_{  \check{\boldsymbol{3}} \check{\boldsymbol{1}} }  (t)  $.
 These facts clearly indicate that we have succeeded in the simulation of the decay processes
$  |  \check{\boldsymbol{1}} \rangle_{\text{S}} \otimes  |  \check{\boldsymbol{3}} \rangle_{\text{E}} \rightarrow  
 |  \check{\boldsymbol{2}} \rangle_{\text{S}} \otimes  |  \check{\boldsymbol{2}} \rangle_{\text{E}} \rightarrow  
 |  \check{\boldsymbol{3}} \rangle_{\text{S}} \otimes  |  \check{\boldsymbol{1}} \rangle_{\text{E}}$.  
For the initial condition $ \rho^{\text{in},4}_{\text{S}} $ (Figs. \ref{BellAD} (b)),  
we have obtained the values of  $ N_{  | 1011 \rangle } (t_i)$ and $ N_{  | 1110 \rangle } (t_i)$ with all the other numbers being zero. 
At first we have $ N_{  | 1011  \rangle } (t_0) = N_{\text{shots}} $.
Then as time goes by, the value of $ N_{  | 1011 \rangle } (t_i)$ decreases while the value of $ N_{  | 1110 \rangle } (t_i)$ increases. Thus, this clearly represents the simulation of the decay process   
$  |  \check{\boldsymbol{2}} \rangle_{\text{S}} \otimes  |  \check{\boldsymbol{3}} \rangle_{\text{E}} \rightarrow  |  \check{\boldsymbol{3}} \rangle_{\text{S}} \otimes  |  \check{\boldsymbol{2}} \rangle_{\text{E}}$.    
  Let us examine the collective amplitude damping processes for the initial conditions $ \rho^{\text{in},5}_{\text{S}} $ and $ \rho^{\text{in},6}_{\text{S}} $
  (Figs. \ref{BellAD} (c) for $ \rho^{\text{in},5}_{\text{S}} $ and (d) for $ \rho^{\text{in},6}_{\text{S}} $).
 As we see here, the averaged expectation value $  \overline{ \langle J^z _{ \text{S} }   \rangle _{\text{in},l} } (N_{ \text{shots}},t_i) $ for these two conditions  show 
 approximately the same behaviors. This is because $  \langle J^z_{ \text{S} } \rangle (t) $ is determined by the diagonal components of the reduced density matrix  $\rho^{\text{out}}_{\text{S}} (t)$. 
   For $ \rho^{\text{in},5}_{\text{S}} $ and $ \rho^{\text{in},6}_{\text{S}} $,  the diagonal components take the same values, and therefore, 
  from Eq. \eqref{masterequation4} or Eq.  \eqref{twoQrelaxKraus1} the two expectation values of $   J^z_{ \text{S} }  $  become theoretically the same. 
 In these cases, the data we have obtained are the numerical values of  $ N_{  | 0111  \rangle } (t_i) $, $ N_{  | 1010  \rangle } (t_i) $, $ N_{  | 1101  \rangle } (t_i) $, and $ N_{  | 1111  \rangle } (t_i) $.
 By staring from $ N_{  | 0111  \rangle } (t_0) + N_{  | 1111  \rangle } (t_0) = N_{ \text{shots}}$ with $ N_{  | 0111  \rangle } (t_0) , N_{  | 1111  \rangle } (t_0) > 0$ 
$($$N_{  | 0111  \rangle } (t_0)$ and  $ N_{  | 1111  \rangle } (t_0)$ are nearly equal while all the other numbers are zero$)$, 
 we have observed the decreasing of $ N_{  | 0111  \rangle } (t_i) $ while the increasing of $ N_{  | 1010  \rangle } (t_i) $ and $ N_{  | 1110  \rangle } (t_i) $.  
   Such a fact represents that we have correctly simulated the three decay processes  
   $ | \check{\boldsymbol{1}}  \rangle  \rightarrow  | \check{\boldsymbol{2}}  \rangle$,  $ | \check{\boldsymbol{2}}  \rangle  \rightarrow  | \check{\boldsymbol{3}}  \rangle$, and 
 $ | \check{\boldsymbol{3}}  \rangle  \rightarrow  | \check{\boldsymbol{1}}  \rangle$. 
 These processes occur owing to the initial condition $ \left[  \rho^{\text{in}}_{\text{S}}  \right]_{  \check{\boldsymbol{1}} ,  \check{\boldsymbol{1}} } = \frac{1}{2}. $
  On the other hand, the number  $ N_{  | 1111  \rangle } (t_i)  $ appears due to the initial condition $ \left[  \rho^{\text{in}}_{\text{S}}  \right]_{  \check{\boldsymbol{3}} ,  \check{\boldsymbol{3}} } = \frac{1}{2}. $
  These numerical behaviors of $ N_{  | 0111  \rangle } (t_i) N_{  | 1010  \rangle } (t_i),  N_{  | 1110  \rangle } (t_i) $,
  and  $ N_{  | 1111  \rangle } (t_i)  $ are consistent with Eq. \eqref{masterequation4}, and therefore,
  we have accomplished the quantum simulation of the collective damping for the initial conditions $ \rho^{\text{in},5}_{\text{S}} $ and $ \rho^{\text{in},6}_{\text{S}} $.
 
 Next, we discuss our results from computational perspective. 
 Let us focus on how the behaviors of our simulation plots become when $ N_{ \text{shots}} $  increases. 
 For all the four conditions and at any $t_i$, we observe that as we increase the value of $ N_{ \text{shots}} $ from $ 2^{10}$ to $2^{18},$ 
 the simulation plots become closer to the black curves (the solution of the quantum master equation), as expected. 
 Such behaviors are quantified by the variances defined in Eq. \eqref{variance1twoQ}. 
For any $ \rho^{\text{in},l}_{\text{S}} $ and $t_i$, we have examined the ordering
 $ \sigma^2_{\text{in},l} (2^{10}, t_i)  > \sigma^2_{\text{in},l} (2^{18}, t_i)$. 
 For the initial conditions $ \rho^{\text{in},3}_{\text{S}}$,
  the variances $ \sigma^2_{\text{in},3} (2^{10}, t_i) $ and   $ \sigma^2_{\text{in},3} (2^{18}, t_i)$  are in 
  the order of $10^{-4}-10^{-5}$ and $10^{-7}-10^{-8}$, respectively.
  For  $ \rho^{\text{in},4}_{\text{S}} $, $ \sigma^2_{\text{in},4} (2^{10}, t_i) $ and   $ \sigma^2_{\text{in},4} (2^{18}, t_i)$  are in 
  the order of $10^{-5}$ and $10^{-7}-10^{-8}$, respectively,   
  On the other hand, for $ \rho^{\text{in},5}_{\text{S}} $ and $ \rho^{\text{in},6}_{\text{S}} $, the variances
  $ \sigma^2_{\text{in},l} (2^{10}, t_i) $ and   $ \sigma^2_{\text{in},l} (2^{18}, t_i)$ $(l = 5,6)$ are in 
  the order of $10^{-3}-10^{-4}$ and $10^{-6}$, respectively; Note that in Fig. \ref{BellAD} the error bars for $ N_{ \text{shots}} = 2^{18}$ are also presented although they are hard to see since their sizes are too small.  
As a result, we have numerically examined that the variance $ \sigma_{\text{in},l} (N_{ \text{shots}}, t_i) $ decreases as 
$N_{ \text{shots}}$ increases, and therefore, this implies that our simulation results get closer to the solution of the quantum master equation as we increase the repetition number 
$ N_{\text{shots}} $. Such a behavior is quantum mechanically and statistically reasonable.
 
Consequently, we have completed the digital quantum simulation of the collective amplitude damping processes in the two-qubit systems. 
These results are our numerical verifications of our formalism given by 
Eqs.  \eqref{twoQrelaxationrotations0},  \eqref{twoQrelaxationrotations1}, and  \eqref{twoQrelaxQC} 
as well as the quantum circuit in Fig. \ref{BellQCtwoQCAD2}. 

 \section{Extension to Larger Qubit Systems }\label{largeQextension}  
 Let us explain the essence of extending our formalisms for the generation of the collective amplitude damping processes to larger qubit systems. 
  We prepare $N$ $(\geq 3)$ qubits and regard them as the system S while we prepare $N$ ancilla bits and constitute the single environment E.  
We perform a unitary transformation $  U^{\text{in}}_{\text{S}}$  on the system S  so as to prepare the initial state of the system we desire.    
We take this  initial state as the permutational symmetric state.
On the other hand, we apply the $X$ gate to all the $N$ ancilla bits so that the initial state of the environment E becomes the ground state. 
After such an initialization of the total system having been completed,  
we operate the overall unitary transformation which describes the energy-exchange processes between the system S and the environment E. 
We express both the quantum states of the system and the environment with the basis vectors of the direct-sum-spin space (permutational symmetric subspace)
whose dimension is equal to $N+1$. 
 They are  $ | j, j \rangle,  | j, j - 1 \rangle, \ldots,   | j, - j + 1 \rangle,  | j, - j  \rangle $, where $j = \frac{N}{2}. $
Let us rename them as $ | j, j \rangle \equiv  | \boldsymbol{1} \rangle, | j, j - 1 \rangle \equiv | \boldsymbol{2}\rangle, 
\ldots,  | j, -j + 1 \rangle \equiv | \boldsymbol{N}  \rangle, 
| j, -j  \rangle \equiv | \boldsymbol{N+1} \rangle.$     
By using them, we mathematically represent the quantum state of the total system as 
$   |  \boldsymbol{i}_{\text{S}}  \rangle_{\text{S}} \otimes  |  \tilde{\boldsymbol{i}}_{\text{E}}  \rangle_{\text{E}} $  
   with $ \boldsymbol{i}_{\text{S}} ,   \tilde{\boldsymbol{i}}_{\text{E}}  = \boldsymbol{1}, \boldsymbol{2}, \ldots, \boldsymbol{N+1}.$ 
 The considering overall unitary transformation, which we abbreviate as $ U^{\text{AD}}_{ \boldsymbol{N} } $,
 is expressed by the tensor-product  states $   |  \boldsymbol{i}_{\text{S}}  \rangle_{\text{S}} \otimes  |  \tilde{\boldsymbol{i}}_{\text{E}}  \rangle_{\text{E}} $. 
The formulation of  $ U^{\text{AD}}_{ \boldsymbol{N} } $ is equivalent to that of the decay processes of the system S.  
In this case, there are $
\left (
		\begin{array}{c} 
		 N +1  \\
		 2
		\end{array}
	\right )
$ collective decay channels, and correspondingly, 
we prepare the controlled-rotational operators with 
$
\left (
		\begin{array}{c} 
		 N + 1  \\
		 2
		\end{array}
	\right )
$ 
different angles naming as $ \vartheta_{ \boldsymbol{j} , \boldsymbol{i} } $, 
where $\boldsymbol{i},  \boldsymbol{j}  = \boldsymbol{1}, \boldsymbol{2}, \ldots, \boldsymbol{N+1}$  with  $\boldsymbol{i} \neq  \boldsymbol{j} $.
We parametrize them with a single real number corresponding to time $t$.
Correspondingly, to explicitly represent that we are formulating the decay process at the time $t$,
 we rephrase $ \vartheta_{ \boldsymbol{j} , \boldsymbol{i} } $ as $ \vartheta_{ \boldsymbol{j} , \boldsymbol{i} } (t) $ and  $ U^{\text{AD}}_{ \boldsymbol{N} } $ as  $ U^{\text{AD}}_{ \boldsymbol{N} } (t)$. 
     Like Eq. \eqref{twoQrelaxationrotations1},  the overall unitary transformation  $ U^{\text{AD}}_{ \boldsymbol{N} } (t)$ is expressed as the product of controlled-rotational operators introduced above.
    Let us write the controlled-rotational operator constructed by the rotational operation $ R_y \big{(}  \vartheta_{ \boldsymbol{j} , \boldsymbol{i} } (t)  \big{)}$
    as $ U_{ \text{C}_a R_y  } \big{[} \vartheta_{ \boldsymbol{j},  \boldsymbol{i} } \big{]} (t).$ 
    The subscript $a$ describes the number of controlled bit(s).
    This generates the decay process  $   |  \boldsymbol{i}_{\text{S}}  \rangle_{\text{S}} \rightarrow  |  \boldsymbol{j}_{\text{S}}  \rangle_{\text{S}} $ 
    with $ i < j. $ As we have shown in Sec. \ref{BSCR}, in order to obtain $ U^{\text{AD}}_{ \boldsymbol{N} } (t)$   
we need to introduce the controlled-unitary operators which exchange the rows and columns of the matrices under consideration like $\tau_{15,16}$ in Eq. \eqref{twoQrelaxationrotations1}.
By applying these interexchange operators to $ U_{ \text{C}_a R_y  } \big{[} \vartheta_{ \boldsymbol{j} , \boldsymbol{i} } \big{]} (t),$
 the orderings of its matrix elements get arranged. 
 We do such procedures for the other controlled-rotational operators and construct $ U^{\text{AD}}_{ \boldsymbol{N} } (t) $.
 After completing the formulation of $ U^{\text{AD}}_{ \boldsymbol{N} } (t)$, we perform the measurement on the $N$ ancilla bits 
 and we obtain $N+1$ Kraus operators $\mathcal{M}_{ \boldsymbol{k} } (t)$  $ ( \boldsymbol{k}  = \boldsymbol{1}, \boldsymbol{2}, \ldots, \boldsymbol{N+1}) $
 as well as the associated reduced density matrix of the system S. 
 Note that actually there are $2^{2N}$ Kraus operators. 
 The rest of all $2^{2N} - (N+1)$ Kraus operators are, however, zero matrices like $  \mathcal{M}_{\check{\boldsymbol{0}}} $  in Eq. \eqref{twoQrelaxKraus2} 
and do not contribute to the generation of the reduced density matrix.  
This is because the vector components of the total Hlibert space besides $ | \boldsymbol{1} \rangle,  | \boldsymbol{2}\rangle, \ldots,   | \boldsymbol{N+1} \rangle$ 
are not responsible for the collective decay processes.      
 The matrix elements of the Kraus operators $\mathcal{M}_{ \boldsymbol{k} } (t)$ are represented by the matrix elements of  $ U^{\text{AD}}_{ \boldsymbol{N} } (t)$. 
  We can express them with the decay strengths $\Gamma_{  \boldsymbol{j} ,  \boldsymbol{i}  } (t)$ ($i < j$). 
  For instance, the square absolute of $(N+1,1)$-th element of the Kraus operator $\mathcal{M}_{ \boldsymbol{1} } (t)$ can be identified with the decay strength $\Gamma_{  \boldsymbol{N+1}  , \boldsymbol{1}} (t)$. 
   In contrast, the $(1,1)$-th element of the Kraus operator $\mathcal{M}_{ \boldsymbol{N+1} } (t)$ is the probability amplitude of 
   the quantum state $  | \boldsymbol{1}\rangle_{\text{S}}   $ to be unchanged under the decay effects.    
   The elements of the reduced density matrix is described with those of the Kraus operators $\mathcal{M}_{ \boldsymbol{k} } (t)$, and therefore, 
   we can represent the elements of the reduced density matrix with the decay strengths $\Gamma_{  \boldsymbol{j} ,  \boldsymbol{i} } (t)$.
   Then,  the functional forms of $\Gamma_{  \boldsymbol{j}  , \boldsymbol{i} } (t)$ 
  given by the time $t$ are determined so that they match with the solution of the quantum master equation 
   at least up to $\mathcal{O}\big{(}  ( \gamma t )^N  \big{)}$. 
   Note that for sufficiently small $t$ the dominant term of the decay strength $\Gamma_{  \boldsymbol{j}  , \boldsymbol{i}  } (t)$ is in the order of $\mathcal{O}\big{(}  ( \gamma t )^{j - i}  \big{)}$  \cite{qutirtKraus}.
   This is why we have to include the terms up to $\mathcal{O}\big{(}  ( \gamma t )^N  \big{)}$  for all $\Gamma_{  \boldsymbol{j} ,  \boldsymbol{i}} (t)$ 
   because of  the decay strength $\Gamma_{  \boldsymbol{N+1}  , \boldsymbol{1} } (t)$.  
Subsequently,  we obtain the representations of  $\Gamma_{  \boldsymbol{j} ,  \boldsymbol{i}  } (t)$ in terms of  $ \sin \vartheta_{ \boldsymbol{j} , \boldsymbol{i} } (t) $ and $ \cos \vartheta_{ \boldsymbol{j},  \boldsymbol{i} } (t) $.
As a result, the functional forms of $ \vartheta_{ \boldsymbol{j} , \boldsymbol{i} } (t) $ get determined and
the execution of controlled-rotational operators with the angles $ \vartheta_{ \boldsymbol{j} , \boldsymbol{i} } (t) $ generate the collective
amplitude damping processes at the time $t$. 
We can then extract the probability weights of $ |  \boldsymbol{i}_{\text{S}}  \rangle_{\text{S}}  $ at the time $t$ and their numerical  values
can be obtained from the quantum simulation by collecting the data for the number of the output states. 
With repeating the same procedures for different values of time, we obtain the data of  expectation values $  \langle J^z_{ \text{S} } \rangle (t) $
and we complete the digital quantum simulation of the collective amplitude damping.  
 At the end, we comment that for the qubit systems with large $N$, the reduced density matrix obtained by the Kraus representation
 becomes closer to the solution of the quantum master equation as the function of the time $t$. 
 This is because as we explained above, we have to retain the terms of $\gamma t$ in the decay strengths $\Gamma_{  \boldsymbol{j}  , \boldsymbol{i} } (t)$ 
 at least up to $\mathcal{O}\big{(}  ( \gamma t )^N  \big{)}$. 
 This implies that for large $N$ we become able to quantum simulate the (collective) amplitude damping at any time regime.    

  \section{Conclusion and Discussion}\label{conclusionanddiscussion}  
In this paper, we have investigated the the formulation of the collective amplitude damping in the two-qubit systems
in terms of the single- and two-qubit gates and constructed the corresponding quantum circuits.   
First, we have analyzed the Hilbert-space structure and demonstrated the effectiveness of the direct-sum-spin-space representation for the description of the collective amplitude damping. 
Using this representation and the Kraus representation approach, 
we have analyzed the way to formulate the overall unitary transformation describing 
the energy-exchange processes occurring between the two-qubit system and the environment (two ancilla bits), which are equivalent to
 the collective decay processes of the two-qubit systems. This has been the main task of this work. 
In order to do this, we have naturally extended the method used in the analysis of the single-qubit amplitude damping in the following way. 
The two-qubit systems show three different collective decay channels.
By making a correspondence, we have introduced the three controlled-rotational operators.  
We have formulated them so that they become the functions of the time under consideration, 
and further, to become consistent with the solution of the quantum master equation up to the second order in the time.  
In addition to the controlled-rotational operators, 
we have introduced interexchange operators (for example, the operator $\tau_{15,16}$ in Eq. \eqref{twoQrelaxationrotations1}). 
The overall unitary transformation is constructed by the three controlled-rotational operators and the interexchange operators. 
After we have operated this overall unitary transformation,
we perform the measurements on the ancilla bits and we obtain the reduced density matrix of the two-qubit systems describing the collective amplitude damping. 
The corresponding quantum circuit is illustrated in Fig. \ref{BellQCtwoQCAD2}.  
The collective amplitude damping are represented quantitatively by the expectation values of $  \langle J^z_{ \text{S} } \rangle $ (the $z$-component of the total-spin operator).
In order to compute these values and conduct the digital quantum simulation, 
we have collected the data for the numbers of the quantum states in Eq. \eqref{twoQrelaxbasisvector4}, which have been generated as the output states.
For doing this, we have examined six different initial conditions 
and two different values of the number $ N_{ \text{shots}} $  $(= 2^{10}, 2^{18})$ for ten temporal points. 
These results are presented in Fig. \ref{BellAD}.
To make a comparison, we have analyzed the numerical closeness between our quantum simulation results  and the solution of the quantum master equation given in \eqref{masterequation4}.  
By evaluating the variances, we have examined that as we increase the value of repetition number $ N_{\text{shots}} $, 
our simulation results show better numerical agreement with the solution of the quantum master equation presented in Eq. \eqref{masterequation4}.
These results are quantum mechanically and statistically reasonable and are the numerical verifications of our formalism of the collective amplitude damping.
  
 Our formalism can be extended to the analysis of the collective amplitude damping in larger qubit systems. 
 This is done by preparing the controlled-rotational gates  with the number of the rotational angles equal to that of existing decay processes.
 We consider that this method is the natural extension of the formulation of the single-qubit amplitude damping.

In this work, we have focused on the collective amplitude damping in the two-qubit systems within the short time regime.  
We comment that our work can be applied to analyze the open quantum dynamics of real systems.
This is because the quantum behaviors are realized within the short time scale; as a guide, the quantum behaviors of the qubits remain within a coherence time or dephasing time.      
 In addition, our work is worthwhile from an engineering perspective because, for instance, the Bell states are one of the basic and important resources for performing the efficient quantum computation via the quantum entanglement.
Moreover, the Bell states are also applied in many other types of quantum technologies including the quantum communication processing.

We hope that our work presented in this paper becomes the basics for the quantum computation and exploration of open quantum dynamics
in many types of systems ranging from solids to atomic-molecular and optical systems, and even for the biological systems.
We expect that our quantum circuits for the collective amplitude damping bring insight into the improvement of the qualities of the NISQ devices and the performance of the quantum computation generating more trustable results.
Further, we expect that they pave the way for designing the architectures of large-scale quantum computers.
We anticipate that they also bring insight into the quantum error correction against both the individual and collective noises 
for the large qubit systems as well as the quantum error correction for the qutirt systems \cite{qutirtKraus,qutrit2}.

\acknowledgements
 We thank all the other members of Quemix Inc. for giving us the fruitful comments and reading this manuscript carefully.  
 Especially, we thank Taichi Kosugi for giving me some fruitful comments on the way to obtain the results presented in Figs. \ref{singleQzeroTrelax} and \ref{BellAD} 
 from statistics point of view and advising me the way to illustrate the quantum circuits in Figs. \ref{BellQCsingleQrelax}, \ref{overallsingleQrelaxQC}, \ref{BellQCtwoQCAD}, and \ref{BellQCtwoQCAD2} 
 and the ones in Figs. \ref{UC2Ry21q1q2q0}-\ref{UC2X12q2q3q1}.
 We thank Yu-ichiro Matsushita for giving me some fruitful comments on the description of the motivation of this work.
 \begin{widetext}  
  
 \appendix
   
\section{ Quantum Gates }\label{Qgates}
In this section, we list the quantum gates which are used for the quantum simulation in this work.         
\subsection{Single-Qubit Gates  }\label{singleQgates}
The single qubit has two internal degrees of freedom, namely, spin up and spin down. 
We describe them by two quantum states given in terms of 0 and 1: We denote a "0-state" as $|0 \rangle$ and it corresponds to the up-spin state $| \uparrow \rangle$.
The other one is the "1-state" and we write it as $|1 \rangle$, which corresponds to the down-spin state $| \downarrow \rangle$. 
They are orthogonal to each other. We take their vector representations as
\begin{align} 
	|0 \rangle  =
	\left (
		\begin{array}{c} 
		 1  \\
		 0
		\end{array}
	\right ), \qquad
	|1 \rangle  =
	\left (
		\begin{array}{c} 
		 0  \\
		 1
		\end{array}
	\right ). \label{singlequbitbasisvectorrepresentation}
\end{align} 
The vector representations given in Eq. \eqref{singlequbitbasisvectorrepresentation} are called the computational basis \cite{QCQINandC}.
In the following, we present the single-qubit gates with respect to the computational basis. 

Let us introduce the unitary (Hermitian) operators called $X,Y,Z$ gates. 
The matrix representations are given by   
\begin{align} 
	X = 
	\left [
		\begin{array}{cc} 
		 0 & 1 \\
		 1 & 0
		\end{array}
	\right ], \qquad 
	Y = 
	\left [
		\begin{array}{cc} 
		 0 & -i \\
		 i & 0
		\end{array}
	\right ],  \qquad
	Z = 
	\left [
		\begin{array}{cc} 
		 1 & 0 \\
		 0 & -1
		\end{array}
	\right ]
	. \label{XYZ}
\end{align} 
The matrix representations of  $X,Y,Z$ gates are identical to Pauli matrices $\sigma^x,\sigma^y,\sigma^z$, respectively.
They satisfy $X^2=Y^2=Z^2=\boldsymbol{1}_{2\times2}$, where $ \boldsymbol{1}_{2\times2}$ is the unit $2\times2$ matrix, 
and the commutation relation $ [ X, Y] = 2iZ,  [ Y,Z] = 2iX, [ Z,X] = 2iY$, with $ [A,B] \equiv AB-BA$.
 The computational basis $ |0 \rangle$  and  $|1 \rangle $ in Eq. \eqref{singlequbitbasisvectorrepresentation}  are the eigenvectors of $Z$ and their eigenvalues are $+1$ and $-1$, respectively.
On the other side,  the $X$ operator has a property such that it interchanges the states  $ |0 \rangle$  and  $|1 \rangle $: $ X |0 \rangle =  | 1 \rangle,  X |1 \rangle =  | 0 \rangle. $

Besides the $X,Y,Z$ gates, we use the single-qubit gate called a Hadamard gate $H$.  It is  represented by
\begin{align} 
	H = 
	\frac{1}{\sqrt{2}}  \left [
		\begin{array}{cc} 
		 1 & 1 \\
		 1 & -1
		\end{array}
	\right ].  \label{Hadamardgate}
\end{align} 
The Hadamard gate satisfies the relations  $H^\dagger =H$,  $H^\dagger H = HH^\dagger   = H^2 =  \boldsymbol{1}_{2\times2}$, $H^\dagger X H=X, H^\dagger Y H= -Y$,
 and $H^\dagger Z H=Z$. Furthermore, the Hadamard gate $H$ generates the superposition states $ | \pm \rangle = \frac{ |0 \rangle \pm  |1 \rangle }{\sqrt{2}}$.
They are obtained by applying $H$ to the $ |0 \rangle$  and  $|1 \rangle $ states as $ H |0 \rangle  = | + \rangle,  H |1 \rangle  = | - \rangle.$ 
The superposition states  $ | + \rangle $ and $ | - \rangle $ are the eigenstates of the $X$ operator and their eigenvalues are $0$ and $1$, respectively.   

In the rest, we introduce the other two types of single-qubit gates called a phase gate and a rotational gate. 
First, the phase gate is given by
\begin{align} 
	P_\phi = 
	\left [
		\begin{array}{cc} 
		 1 & 0 \\
		 0 & e^{ i \phi }
		\end{array}
	\right ].  \label{phasegate}
\end{align} 
The phase gate $P_\phi $ shifts the phase of the state $ | 1 \rangle$ with $\phi.$ 
In particular, the phase gates for $\phi = \pi/2$ and $\pi/4$ are called  $S$ gate and  $T$ gate ($\pi/8$ gate), respectively. 

There are three types of rotational gates: $ R_x(\theta_x),  R_y(\theta_y)$, and $R_z(\theta_z)$.  
$ R_a(\theta_a)$  $(a=x,y,z)$ is the operation which rotates the qubit around the $a$ axis with the angle $\theta_a/2$. 
The matrix representations are given by
\begin{align} 
	R_x (\theta_x) & =  \exp \left( -   \frac{i \theta_x }{2} X    \right) 
	= \cos  \left(    \frac{ \theta_x }{2}     \right) \boldsymbol{1}_{2\times2} -  i \sin  \left(    \frac{ \theta_x }{2}     \right) X =
	\left [
		\begin{array}{cc} 
		  \cos  \left(    \frac{ \theta_x }{2}     \right) & -i  \sin  \left(    \frac{ \theta_x }{2}    \right) \\
		- i  \sin  \left(     \frac{ \theta_x }{2}     \right) &  \cos  \left(     \frac{ \theta_x }{2}      \right)
		\end{array}
	\right ], \notag\\ 
	R_y (\theta_y) & = 
	 \exp \left( -   \frac{i \theta_y }{2} Y    \right) 
	= \cos  \left(    \frac{ \theta_y }{2}     \right) \boldsymbol{1}_{2\times2} -  i \sin  \left(    \frac{ \theta_y }{2}     \right) X =
	\left [
		\begin{array}{cc} 
		  \cos  \left(     \frac{ \theta_y }{2}      \right) & - \sin  \left(     \frac{ \theta_y }{2}      \right) \\
		   \sin  \left(      \frac{ \theta_y }{2}      \right) &  \cos  \left(      \frac{ \theta_y }{2}     \right)
		\end{array}
	\right ],  \notag\\ 
	R_z (\theta_z) & = 
	 \exp \left( -   \frac{i \theta_z }{2} Z    \right) 
	= \cos  \left(    \frac{ \theta_z }{2}     \right) \boldsymbol{1}_{2\times2} -  i \sin  \left(    \frac{ \theta_z }{2}     \right) Z =
	\left [
		\begin{array}{cc} 
		  \exp \left( - i  \frac{ \theta_z }{2}     \right) & 0  \\
		0 &  \exp \left(  i  \frac{ \theta_z }{2}     \right) 
		\end{array}
	\right ]. \label{xyzrotations}
\end{align} 

\subsection{Two-Qubit Gates  }\label{twoQgates}
Here we discuss the two-qubit gates. At first, let us introduce the computational basis of two-qubit states. There are four states given by Eq.  \eqref{twoQrelaxbasisvector1}  or
\begin{align} 
  &	|00 \rangle_{Q_0Q_1}    \equiv  |0 \rangle_{Q_0} \otimes | 0 \rangle_{Q_1} =
	\left (
		\begin{array}{c} 
		 1  \\
		 0 \\
		 0 \\
		 0 
		\end{array}
	\right ), \qquad 
	|01 \rangle_{Q_0Q_1}   \equiv  |0 \rangle_{Q_0} \otimes | 1 \rangle_{Q_1} =
	\left (
		\begin{array}{c} 
		 0  \\
		 1 \\
		 0 \\
		 0 
		\end{array}
	\right ), \notag\\
	&	|10 \rangle_{Q_0Q_1}   \equiv  |1 \rangle_{Q_0} \otimes | 0 \rangle_{Q_1} =
	\left (
		\begin{array}{c} 
		 0  \\
		 0 \\
		 1 \\
		 0 
		\end{array}
	\right ), \qquad
	|11 \rangle_{Q_0Q_1}   \equiv  |1 \rangle_{Q_0} \otimes | 1 \rangle_{Q_1} =
	\left (
		\begin{array}{c} 
		 0  \\
		 0 \\
		 0 \\
		 1 
		\end{array}
	\right ).
	 \label{twoqubitbasisvectorrepresentation}
\end{align} 
$Q_0$ denotes the ``zero-th" qubit while $Q_1$ represents the ``first" qubit.  
Next, let us list some two-qubit gates which are used in this paper. 
The first one is  the CNOT gate or the C$X$ gate and let us denote it as $U_{\text{CNOT}}.$ 
This is the two-qubit gate which operates on the two qubits $Q_0$ and $Q_1$ as
\begin{align} 
U_{\text{CNOT}}  |00 \rangle_{Q_0Q_1} & = |00 \rangle_{Q_0Q_1}, \qquad  U_{\text{CNOT}}  |01 \rangle_{Q_0Q_1}  = |01 \rangle_{Q_0Q_1}, \notag\\
U_{\text{CNOT}}  |10 \rangle_{Q_0Q_1} & = |11 \rangle_{Q_0Q_1}, \qquad  U_{\text{CNOT}}  |10 \rangle_{Q_0Q_1}  = |11 \rangle_{Q_0Q_1},
\label{CNOT2} 
		\end{align} 
or in a more compact form  
\begin{align} 
U_{\text{CNOT}}  | n_{Q_0} n_{Q_1}  \rangle_{Q_0Q_1}  = | n_{Q_0}  \rangle_{Q_0}  \otimes |  n_{Q_0}  \oplus n_{Q_1}  \rangle_{Q_1}   ,
\label{CNOT3} 
		\end{align} 
where the symbol $\oplus$ denotes the XOR gate of a logical circuit in the  classical  computer. 
When $n_{Q_0}  = 0$ then the value of $n_{Q_1}  $ remains unchanged under the  operation of $U_{\text{CNOT}} $.
In contrast, when $n_{Q_0}  = 1$  we obtain $U_{\text{CNOT}}  |1 n_{Q_1}  \rangle_{Q_0Q_1} \to  |1 \bar{n}_{Q_1}  \rangle_{Q_0Q_1}$
where $  \bar{n}_{Q_1}  = 1(0)$ when  $  n_{Q_1}  = 0(1)$.  Here the qubit $Q_0$  is called the controlled bit while $Q_1$ is the target bit.
For the later convenience, hereinafter let us rewrite this two-qubit operation as $U_{\text{C}X}[Q_0;Q_1].$
We have introduced this notation in the sense that the unitary operation is the C$X$ operation comprised of the controlled bit $Q_0$ and the target bit $Q_1$. 
The matrix representation is given by
\begin{align} 
	U_{\text{C}X}[Q_0;Q_1] &   =
	\left [
		\begin{array}{cccc} 
		 1 & 0 & 0 & 0 \\
		 0 & 1 & 0 & 0 \\
		  0 & 0 & 0 & 1 \\
		   0 & 0 & 1 & 0 
		\end{array}
	\right ] = 
	\left [
		\begin{array}{cccc} 
		 \boldsymbol{1}_{2\times2}  & \boldsymbol{0}_{2\times2}  \\
		\boldsymbol{0}_{2\times2}  & X
		\end{array} \right ], \label{CNOT} 
		\end{align} 
where $\boldsymbol{0}_{2\times2} $ is the two by two zero matrix.
In contrast to the above case, we can perform the CNOT gate such that the qubit $Q_0$ is the target bit while $Q_1$ is the controlled bit.  
Let us abbreviate it as $U_{\text{C}X}[Q_1;Q_0]$ and the matrix representation is 
\begin{align} 
	U_{\text{C}X}[Q_1;Q_0] &   =
	\left [
		\begin{array}{cccc} 
		 1 & 0 & 0 & 0 \\
		 0 & 0 & 0 & 1 \\
		  0 & 0 & 1 & 0 \\
		   0 & 0 & 0 & 1 
		\end{array}
	\right ] . \label{invCNOT} 
		\end{align} 
The $X$ gate contained in the matrix representation of $U_{\text{C}X}[Q_0;Q_1] $ in Eq. \eqref{CNOT} can be replaced with other unitary gates $U$ and it is called the controlled-unitary gate.
In this paper, we use $U=H,S,T,R_y (\theta_y)$. 
When the qubit $Q_0$ is the control bit while $Q_1$ is the target bit, the matrix representations of these controlled-unitary operators are 
		\begin{align} 
		U_{\text{C}H} [Q_0;Q_1]  & = 
	\left [
		\begin{array}{cc} 
		 \boldsymbol{1}_{2\times2}  & \boldsymbol{0}_{2\times2}  \\
		\boldsymbol{0}_{2\times2}  & H
		\end{array} \right ], \qquad 
		U_{\text{C}S} [Q_0;Q_1]  = 
	\left [
		\begin{array}{cc} 
		 \boldsymbol{1}_{2\times2}  & \boldsymbol{0}_{2\times2}  \\
		\boldsymbol{0}_{2\times2}  & S
		\end{array} \right ], \notag\\
		U_{\text{C}T}    [Q_0;Q_1] & = 
	\left [
		\begin{array}{cc} 
		 \boldsymbol{1}_{2\times2}  & \boldsymbol{0}_{2\times2}  \\
		\boldsymbol{0}_{2\times2}  & T
		\end{array} \right ], \quad
                U_{\text{C}R_y(\theta_y)}   [Q_0;Q_1]  = 
	       \left [
		\begin{array}{cc} 
		 \boldsymbol{1}_{2\times2}  & \boldsymbol{0}_{2\times2}  \\
		\boldsymbol{0}_{2\times2}  & R_y (\theta_y)
		\end{array} \right ].   		
		\label{CHSTRy}
\end{align} 
As we can see from Eq. \eqref{CHSTRy}, the controlled-Hadamard gate can be utilized to generate Bell states, for instance, \\
\noindent
$U_{\text{C}H}   [Q_0;Q_1] |10 \rangle_{Q_0Q_1} = \frac{ |01 \rangle_{Q_0Q_1}  +  |10 \rangle_{Q_0Q_1} }{\sqrt{2}} = | \Psi^+ \rangle$ and 
$U_{\text{C}H}   [Q_0;Q_1]  |11 \rangle_{Q_0Q_1} = \frac{ |00 \rangle_{Q_0Q_1}  -  |11 \rangle_{Q_0Q_1} }{\sqrt{2}} = | \Phi^- \rangle$.

Finally, let us present one more two-qubit gate called the SWAP gate. 
This is the operation such that  $| n_{Q_0} n_{Q_1}  \rangle_{Q_0Q_1}  \to | n_{Q_1} n_{Q_0}  \rangle_{Q_0Q_1} $. The matrix representation is given by
\begin{align} 
		U_{\text{SWAP}}    =
	\left [
		\begin{array}{cccc} 
		 1 & 0 & 0 & 0 \\
		 0 & 0 & 1 & 0 \\
		  0 & 1 & 0 & 0 \\
		   0 & 0 & 0 & 1 
		\end{array}
	\right ] . \label{SWAP} 
\end{align} 
Note that the SWAP gate can be created with three CNOT gates as  $ U_{\text{SWAP}}    = U_{\text{C}X}[Q_0;Q_1] \cdot U_{\text{C}X}[Q_1;Q_0] \cdot U_{\text{C}X}[Q_0;Q_1] $.

\section{ Controlled-Unitary Operators in Eq. \eqref{twoQrelaxationrotations1} } \label{Qgatesfourqubits}
In this section, let us present the mathematical representations of controlled-unitary operators given in Eq. \eqref{twoQrelaxationrotations1}
 in terms of single- and two-qubit gates.
In addition, we display the quantum circuits of them.  
First, the controlled-unitary operators in Eq. \eqref{twoQrelaxationrotations1} are represented in terms of single- and two-qubit gates as
\begin{align} 
 U_{\text{C}_2R_y \left[  \vartheta_{  \check{\boldsymbol{2}} \check{\boldsymbol{1}} }  \right] } \left[ Q_1Q_2;Q_0  \right]  &  = 
 U_{\text{C}R_y( \frac{ \vartheta_{  \check{\boldsymbol{2}} \check{\boldsymbol{1}} }}{2}) } \left[ Q_1;Q_0 \right] \cdot  U_{\text{C}X} \left[ Q_2;Q_1  \right]   \cdot 
  U_{\text{C}R^\dagger_y( \frac{ \vartheta_{  \check{\boldsymbol{2}} \check{\boldsymbol{1}} }}{2}) } \left[ Q_1;Q_0 \right]
  \cdot U_{\text{C}X} \left[ Q_2;Q_1  \right] \notag\\
  &  \cdot U_{\text{C}R_y( \frac{ \vartheta_{  \check{\boldsymbol{2}} \check{\boldsymbol{1}} }}{2}) } \left[ Q_2;Q_0 \right] ,  \notag  \\
  U_{\text{C}_2R_y \left[  \vartheta_{  \check{\boldsymbol{3}} \check{\boldsymbol{2}} }  \right]} \left[ Q_0Q_2;Q_1  \right]   & = 
 U_{\text{C}R_y( \frac{ \vartheta_{  \check{\boldsymbol{3}} \check{\boldsymbol{2}} }}{2}) } \left[ Q_0;Q_1 \right] \cdot  U_{\text{C}X} \left[ Q_2;Q_0  \right]   \cdot 
  U_{\text{C}R^\dagger_y( \frac{ \vartheta_{  \check{\boldsymbol{3}} \check{\boldsymbol{2}} }}{2}) } \left[ Q_0;Q_1 \right]
  \cdot U_{\text{C}X} \left[ Q_2;Q_0  \right] \notag\\
  &  \cdot U_{\text{C}R_y( \frac{ \vartheta_{  \check{\boldsymbol{3}} \check{\boldsymbol{2}} }}{2}) } \left[ Q_3;Q_2 \right] ,  \notag \\
U_{\text{C}_3R_y \left[ \vartheta_{  \check{\boldsymbol{3}} \check{\boldsymbol{1}} } \right] } \left[ Q_1Q_2Q_3;Q_0  \right]   & = 
U_{\text{C}R_y( \frac{ \vartheta_{  \check{\boldsymbol{3}} \check{\boldsymbol{1}} }}{2}) } \left[ Q_1;Q_0 \right] \cdot  U_{\text{C}_2X} \left[ Q_2Q_3;Q_1  \right]   \cdot 
  U_{\text{C}R^\dagger_y( \frac{ \vartheta_{  \check{\boldsymbol{3}} \check{\boldsymbol{1}} }}{2}) } \left[ Q_1;Q_0 \right]
  \cdot U_{\text{C}_2X} \left[ Q_2Q_3;Q_1  \right]  \notag\\
& \cdot U_{\text{C}_2R_y( \frac{ \vartheta_{  \check{\boldsymbol{3}} \check{\boldsymbol{1}} }}{2}) } \left[ Q_2Q_3;Q_0 \right] ,   \notag  \\
   U_{\text{C}_2R_y \left[ \frac { \vartheta_{  \check{\boldsymbol{3}} \check{\boldsymbol{1}} } }{2} \right] } \left[ Q_2Q_3;Q_0  \right]   & = 
 U_{\text{C}R_y( \frac{ \vartheta_{  \check{\boldsymbol{3}} \check{\boldsymbol{1}} }}{4}) } \left[ Q_2;Q_0 \right] \cdot  U_{CX} \left[ Q_3;Q_2  \right]   \cdot 
  U_{\text{C}R^\dagger_y( \frac{ \vartheta_{  \check{\boldsymbol{3}} \check{\boldsymbol{1}} }}{4}) } \left[ Q_0;Q_1 \right]
  \cdot U_{\text{C}X} \left[ Q_3;Q_2  \right]  \notag\\ 
  & \cdot U_{\text{C}R_y( \frac{ \vartheta_{  \check{\boldsymbol{3}} \check{\boldsymbol{1}} }}{4}) } \left[ Q_3;Q_0 \right] ,  \notag \\
  \tau_{15,16}  & = U_{\text{C}X^{ \frac{1}{2}} } \left[ Q_2;Q_3 \right] \cdot  U_{\text{C}_2X} \left[ Q_0Q_1;Q_2  \right]  
   \cdot U_{\text{C} \big{(}X^{ \frac{1}{2}} \big{)}^\dagger }  \left[ Q_2;Q_3 \right]
  \cdot U_{\text{C}_2X} \left[ Q_0Q_1;Q_2  \right]  \notag\\
  & \cdot U_{\text{C}_2 X^{ \frac{1}{2}}} \left[ Q_0Q_1;Q_3  \right]  ,  \notag \\
  \tau_{14,16}  & = U_{\text{C}X^{ \frac{1}{2}}} \left[ Q_3;Q_2 \right] \cdot  U_{\text{C}_2X} \left[ Q_0Q_1;Q_3  \right]  
   \cdot U_{\text{C} \big{(}X^{ \frac{1}{2}} \big{)}^\dagger } \left[ Q_3;Q_2 \right]
  \cdot U_{\text{C}_2X} \left[ Q_0Q_1;Q_3  \right]  \notag\\
&   \cdot U_{\text{C}_2X^{ \frac{1}{2}}} \left[ Q_0Q_1;Q_2  \right]  ,  \notag \\
  \tau_{12,16}  & =  U_{\text{C}X^{ \frac{1}{2}}} \left[ Q_0;Q_1 \right] \cdot  U_{\text{C}_2X} \left[ Q_2Q_3;Q_0  \right]  
   \cdot U_{\text{C} \big{(}X^{ \frac{1}{2}} \big{)}^\dagger }  \left[ Q_0;Q_1 \right]
  \cdot U_{\text{C}_2X} \left[ Q_2Q_3;Q_0  \right]  \notag\\ 
  & \cdot U_{\text{C}_2X^{ \frac{1}{2}}} \left[ Q_2Q_3;Q_1  \right]  ,  \notag \\
 U_{\text{C}_2X^{ \frac{1}{2}}} \left[ Q_0Q_1;Q_3  \right] & = 
  U_{\text{C}X^{ \frac{1}{4}}} \left[ Q_1;Q_3 \right] \cdot  U_{\text{C}X} \left[ Q_0;Q_1  \right]   \cdot U_{\text{C} \big{(}X^{ \frac{1}{4}} \big{)}^\dagger }  \left[ Q_1;Q_3 \right]
  \cdot U_{\text{C}X} \left[ Q_0;Q_1 \right]   \cdot U_{\text{C}X^{ \frac{1}{4}}} \left[ Q_0;Q_3  \right] ,  \notag\\
  U_{\text{C}_2X^{ \frac{1}{2}}} \left[ Q_0Q_1;Q_2  \right] & = U_{\text{C}X^{ \frac{1}{4}}} \left[ Q_1;Q_2 \right] \cdot  
  U_{\text{C}X} \left[ Q_0;Q_1  \right]   \cdot U_{\text{C} \big{(}X^{ \frac{1}{4}} \big{)}^\dagger }   \left[ Q_1;Q_2 \right]
  \cdot U_{\text{C}X} \left[ Q_0;Q_1 \right]   \cdot U_{\text{C} X^{ \frac{1}{4}}} \left[ Q_0;Q_2  \right] , \notag\\
   U_{\text{C}_2X^{ \frac{1}{2}}} \left[ Q_2Q_3;Q_1  \right] & = U_{\text{C}X^{ \frac{1}{4}}} \left[ Q_2;Q_1 \right] \cdot 
    U_{\text{C} X} \left[ Q_3;Q_2  \right]   \cdot U_{\text{C} \big{(}X^{ \frac{1}{4}} \big{)}^\dagger } \left[ Q_2;Q_1 \right]
  \cdot U_{\text{C} X} \left[ Q_3;Q_2 \right]   \cdot U_{\text{C} X^{ \frac{1}{4}}} \left[ Q_3;Q_1  \right] , \label{controlledunitaryoperatorsBellrelax}
\end{align} 
where $X^{ \frac{1}{2}} = H \cdot P_{\frac{\pi}{2}} \cdot H$ and  $X^{ \frac{1}{4}} = H \cdot P_{\frac{\pi}{4}} \cdot H$. 
Note that $\big{(} X^{ \frac{1}{2}} \big{)}^2 = X$ and $ \big{(}X^{ \frac{1}{4}} \big{)}^2=  X^{ \frac{1}{2}}.$ 
Further, $ U_{\text{C} X^{ \frac{1}{2}}} \left[ Q_i;Q_j  \right] = U_{\text{C}H}  \left[ Q_i;Q_j  \right] \cdot U_{\text{C} P_{\frac{\pi}{2}} } \left[ Q_i;Q_j  \right] \cdot U_{\text{C}H}  \left[ Q_i;Q_j  \right]$ and 
$ U_{\text{C} X^{ \frac{1}{4}}} \left[ Q_i;Q_j  \right] = U_{\text{C}H}  \left[ Q_i;Q_j  \right] \cdot U_{ \text{C} P_{\frac{\pi}{4}} } \left[ Q_i;Q_j  \right] \cdot U_{\text{C}H}  \left[ Q_i;Q_j  \right]$
$(i,j = 0,1,2,3)$.
As we explained in Sec. \ref{twoQKraus}, for instance, the operator $U_{\text{C} _2R_y(  \vartheta_{  \check{\boldsymbol{2}} \check{\boldsymbol{1}} }) } \left[ Q_1Q_2;Q_0 \right] $  
is the controlled-unitary operator such that the two qubits $Q_1$ and $Q_2$ are the control bits while $Q_0$ is the target bit. The unitary transformation to become performed is  
the rotational operation $R_y(  \vartheta_{  \check{\boldsymbol{2}} \check{\boldsymbol{1}} })$. 
Similarly, the operator $U_{\text{C}_3R_y \left[  \vartheta_{  \check{\boldsymbol{3}} \check{\boldsymbol{1}} }  \right] } \left[ Q_1Q_2Q_3;Q_0  \right] $ is the controlled-unitary operator constructed with the three controlled bits $Q_1,Q_2$ and $Q_3$ and the target bit $Q_0$.
 The unitary transformation to be performed is $R_y \left[  \vartheta_{  \check{\boldsymbol{3}} \check{\boldsymbol{1}} } \right] $. 
The operator  $ U_{\text{C} X} \left[ Q_3;Q_2  \right] $ is the controlled-$X$ operator (CNOT) composed of the control bit $Q_3$ and the target bit $Q_2$. 
Recall that $ \tau_{i,16}$ ($i = 12,14,15$) is the operator which interexchanges the $i$th row (column) and the 16th row (column) of matrix under consideration:   
When $ \tau_{i,16}$ is multiplied on the left-hand side of considering matrix $A$, the $i$th row and the 16th row of $A$ are exchanged whereas when it is multiplied on the right-hand side, 
 the $i$th column and the 16th column of $A$ get exchanged. The rest of all controlled-unitary operators appearing in the above equation are defined in the same way.     
  The quantum circuits of the controlled-unitary operators shown in Eq. \eqref{twoQrelaxationrotations1} can be constructed using the above equations. 
  We present them in Figs. \ref{UC2Ry21q1q2q0}-\ref{UC2X12q2q3q1}.
  Note that the system S is comprised of the two qubits $Q_0$ and $Q_1$ whereas the two qubits $Q_2$ and $Q_3$ constitute the environment E. 
Finally, we show the derivation of interexchange operators $\tau_{10,12}, \tau_{12,15}$ and $ \tau_{11,16}$
expressed by the interexchange operators $  \tau_{15,16} , \tau_{14,16}  $, $  \tau_{12,16}  $ and the Toffoli gates.  
By using $U_{\text{C}_2X} \left[ Q_0 Q_3;Q_2 \right] \equiv \tau_{10,12} \cdot \tau_{14,16}  $ and 
$U_{\text{C}_2X} \left[ Q_0 Q_2;Q_1 \right] \equiv \tau_{11,15} \cdot \tau_{12,16}  $,  we obtain
\begin{align} 
\tau_{10,12} & = U_{\text{C}_2X} \left[ Q_0 Q_3;Q_2 \right] \cdot   \tau_{14,16} , \notag\\
\tau_{12,15} & = \tau_{12,16} \cdot   \tau_{15,16} \cdot   \tau_{12,16} = \tau_{15,16} \cdot   \tau_{12,16} \cdot   \tau_{15,16} , \notag\\
\tau_{11,16} & =   \tau_{15,16} \cdot  \left[
  U_{\text{C}_2X} \left[ Q_0 Q_2;Q_1 \right] \cdot  \tau_{12,16} 
\right] \cdot   \tau_{15,16}.
 \label{exchangeoperatorsderivation}
\end{align} 

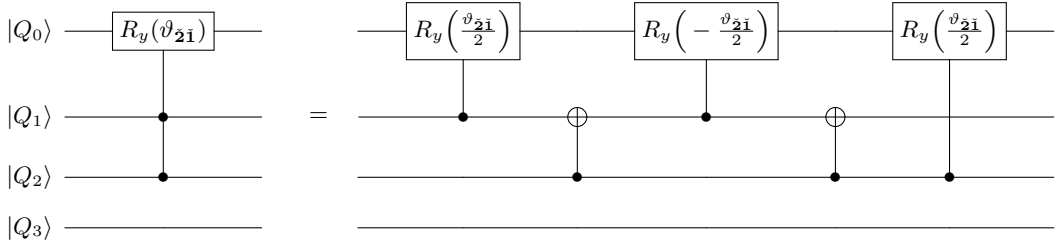
\begin{figure}[H] 
\centering
\mbox{
\Qcircuit @C=2em @R=2em {
\lstick{| Q_0 \rangle} & \gate{R_y ( \vartheta_{\check{\boldsymbol{2}} \check{\boldsymbol{1}}} ) } & \qw & \push{\rule{2em}{0em}}  & \gate{ R_y \Big{(} \frac{\vartheta_{\check{\boldsymbol{2}} \check{\boldsymbol{1}}}}{2} \Big{)}}  & \qw  
& \gate{ R_y \Big{(} -\frac{\vartheta_{\check{\boldsymbol{2}} \check{\boldsymbol{1}}}}{2} \Big{)}}  & \qw &  \gate{ R_y \Big{(} \frac{\vartheta_{\check{\boldsymbol{2}} \check{\boldsymbol{1}}}}{2} \Big{)}} & \qw \\
\lstick{| Q_1 \rangle} & \ctrl{-1} & \qw & \push{=\rule{1em}{0em}} & \ctrl{-1} & \targ     & \ctrl{-1} & \targ     & \qw  & \qw \\
\lstick{| Q_2 \rangle} & \ctrl{-1} & \qw & \push{\rule{2em}{0em}}  & \qw       & \ctrl{-1} & \qw       & \ctrl{-1} & \ctrl{-2} & \qw  \\ 
\lstick{| Q_3 \rangle} & \qw       & \qw & \push{\rule{2em}{0em}}  & \qw       & \qw       & \qw       & \qw       & \qw  & \qw \\ 
}
}
\caption{ Quantum circuit for $U_{ \text{C}_2 R_y [\vartheta_{\check{\boldsymbol{2}} \check{\boldsymbol{1}}} ]} [Q_1 Q_2; Q_0]$.}
\label{UC2Ry21q1q2q0} 
\end{figure}

\begin{figure}[H] 
\centering
\mbox{
\Qcircuit @C=2em @R=2em {
\lstick{| Q_0 \rangle} & \ctrl{1} & \qw & \push{\rule{2em}{0em}}  & \ctrl{1} & \targ & \ctrl{1} & \targ & \qw & \qw \\
\lstick{| Q_1 \rangle} & \gate{R_y ( \vartheta_{\check{\boldsymbol{3}} \check{\boldsymbol{2}}} ) } & \qw & \push{=\rule{1em}{0em}} &  \gate{ R_y \big{(} \frac{\vartheta_{\check{\boldsymbol{3}} \check{\boldsymbol{2}}}}{2} \big{)}} & \qw     
&  \gate{ R_y \Big{(} -\frac{\vartheta_{\check{\boldsymbol{3}} \check{\boldsymbol{2}}}}{2} \Big{)}}  & \qw     
&  \gate{ R_y \Big{(} \frac{\vartheta_{\check{\boldsymbol{3}} \check{\boldsymbol{2}}}}{2} \Big{)}}   & \qw \\
\lstick{| Q_2 \rangle} & \ctrl{-1} & \qw & \push{\rule{2em}{0em}}  & \qw       & \ctrl{-2} & \qw       & \ctrl{-2} & \ctrl{-1} & \qw  \\ 
\lstick{| Q_3 \rangle} & \qw       & \qw & \push{\rule{2em}{0em}}  & \qw       & \qw       & \qw       & \qw       & \qw  & \qw \\ 
}
}
\caption{ Quantum circuit for
$U_{ \text{C}_2 R_y [\vartheta_{\check{\boldsymbol{3}} \check{\boldsymbol{2}}} ]} [Q_0 Q_2; Q_1]$.}
\label{UC2Ry32q0q2q1} 
\end{figure}

\begin{figure}[H] 
\centering
\mbox{
\Qcircuit @C=2em @R=2em { 
\lstick{| Q_0 \rangle} & \gate{R_y  ( \vartheta_{\check{\boldsymbol{3}} \check{\boldsymbol{1}}}  )  } 
 & \qw & \push{\rule{2em}{0em}}  & \gate{ R_y \Big{(} \frac{\vartheta_{\check{\boldsymbol{3}} \check{\boldsymbol{1}}}}{2} \Big{)}}  & \qw  
 & \gate{ R_y \Big{(} -\frac{\vartheta_{\check{\boldsymbol{3}} \check{\boldsymbol{1}}}}{2} \Big{)}}& \qw & \gate{ R_y \Big{(} \frac{\vartheta_{\check{\boldsymbol{3}} \check{\boldsymbol{1}}}}{2} \big{)}} & \qw \\
\lstick{| Q_1 \rangle} & \ctrl{-1} & \qw & \push{=\rule{1em}{0em}} & \ctrl{-1} & \targ     & \ctrl{-1} & \targ     & \qw  & \qw \\
\lstick{| Q_2 \rangle} & \ctrl{-1} & \qw & \push{\rule{2em}{0em}}  & \qw       & \ctrl{-1} & \qw       & \ctrl{-1} & \ctrl{-2} & \qw  \\ 
\lstick{| Q_3 \rangle} & \ctrl{-1} & \qw & \push{\rule{2em}{0em}}  & \qw       & \ctrl{-1} & \qw       & \ctrl{-1} & \ctrl{-1} & \qw \\ 
}
}
\caption{ Quantum circuit for
$U_{ \text{C}_3 R_y [\vartheta_{\check{\boldsymbol{3}} \check{\boldsymbol{1}}} ]} [Q_1 Q_2 Q_3; Q_0]$.} 
\label{UC3Ry31q1q2q3q0} 
\end{figure}
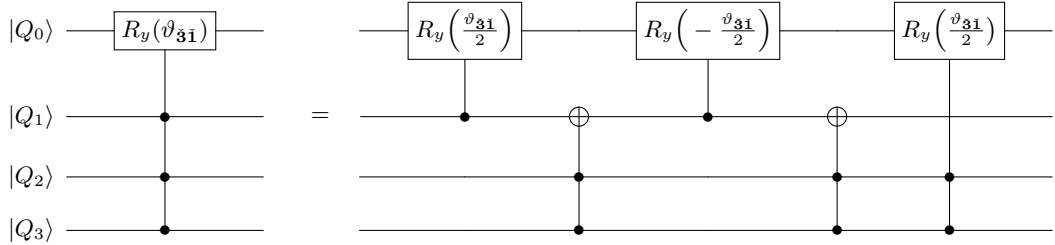

\begin{figure}[H] 
\centering
\mbox{
\Qcircuit @C=2em @R=2em {
\lstick{| Q_0 \rangle} & \gate{R_y  \Big{(} \frac{\vartheta_{\check{\boldsymbol{3}} \check{\boldsymbol{1}}}}{2} \Big{)} }   & \qw  & \push{\rule{2em}{0em}} 
 & \gate{ R_y \big{(} \frac{\vartheta_{\check{\boldsymbol{3}} \check{\boldsymbol{1}}}}{4} \big{)}}   & \qw &  \gate{ R_y \Big{(} -\frac{\vartheta_{\check{\boldsymbol{3}} \check{\boldsymbol{1}}}}{4} \Big{)}}  
 & \qw & \gate{ R_y \Big{(} \frac{\vartheta_{\check{\boldsymbol{3}} \check{\boldsymbol{1}}}}{4} \Big{)} }  & \qw \\
\lstick{| Q_1 \rangle} & \qw & \qw & \push{=\rule{1em}{0em}}  
& \qw   & \qw & \qw
 & \qw & \qw & \qw  \\
\lstick{| Q_2 \rangle} & \ctrl{-2} & \qw & \push{\rule{2em}{0em}}  
& \ctrl{-2}  & \targ  &  \ctrl{-2} 
 & \targ & \qw & \qw \\
\lstick{| Q_3 \rangle} & \ctrl{-1} & \qw & \push{\rule{2em}{0em}}  
& \qw & \ctrl{-1} & \qw 
& \ctrl{-1}  & \ctrl{-3} & \qw \\
}
}
\caption{Quantum circuit for $U_{\text{C}_2R_y \left[ \frac { \vartheta_{  \check{\boldsymbol{3}} \check{\boldsymbol{1}} } }{2} \right] } \left[ Q_2Q_3;Q_0  \right] $.
} 
\label{UC3Ry312q2q3q0} 
\end{figure}
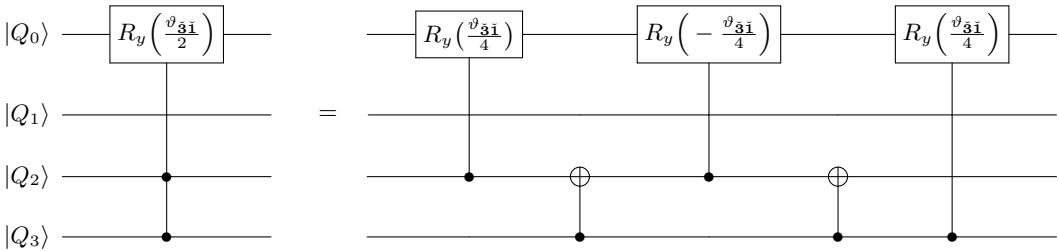

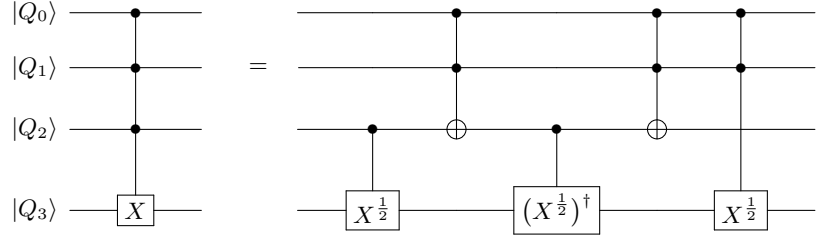
\begin{figure}[H] 
\centering
\mbox{
\Qcircuit @C=2em @R=2em {
\lstick{| Q_0 \rangle} & \ctrl{1} & \qw & \push{\rule{2em}{0em}}  & \qw             & \ctrl{1}  & \qw                     & \ctrl{1} & \ctrl{1} & \qw \\
\lstick{| Q_1 \rangle} & \ctrl{1} & \qw & \push{=\rule{1em}{0em}} & \qw             & \ctrl{1}  & \qw                     & \ctrl{1} & \ctrl{2} & \qw \\
\lstick{| Q_2 \rangle} & \ctrl{1} & \qw & \push{\rule{2em}{0em}}  & \ctrl{1}        & \targ     & \ctrl{1}                & \targ    & \qw & \qw  \\ 
\lstick{| Q_3 \rangle} & \gate{X} & \qw & \push{\rule{2em}{0em}}  & \gate{X^{\frac{1}{2} }} & \qw       & \gate{ \big{(} X^{ \frac{1}{2} } \big{)}^ \dagger } & \qw     
 & \gate{ X^{ \frac{1}{2}} } & \qw \\ 
}
}
\caption{Quantum circuit for $\tau_{15, 16}$.
} 
\label{tau1516} 
\end{figure}

\begin{figure}[H] 
\centering
\mbox{
\Qcircuit @C=2em @R=2em {
\lstick{| Q_0 \rangle} & \ctrl{1}  & \qw & \push{\rule{2em}{0em}}  & \qw             & \ctrl{1}  & \qw                     & \ctrl{1} & \ctrl{1} & \qw \\
\lstick{| Q_1 \rangle} & \ctrl{1}  & \qw & \push{=\rule{1em}{0em}} & \qw             & \ctrl{2}  & \qw                     & \ctrl{2} & \ctrl{1} & \qw \\
\lstick{| Q_2 \rangle} & \gate{X}  & \qw & \push{\rule{2em}{0em}}  &  \gate{X^{\frac{1}{2} }} & \qw       
&  \gate{ \big{(} X^{\frac{1}{2} } \big{)}^\dagger }   & \qw    &  \gate{X^{\frac{1}{2} }}  & \qw  \\ 
\lstick{| Q_3 \rangle} & \ctrl{-1} & \qw & \push{\rule{2em}{0em}}  & \ctrl{-1}       & \targ     & \ctrl{-1}               & \targ      & \qw & \qw \\ 
}
}
\caption{ Quantum circuit for $\tau_{14, 16}$.} 
\label{tau1416} 
\end{figure}
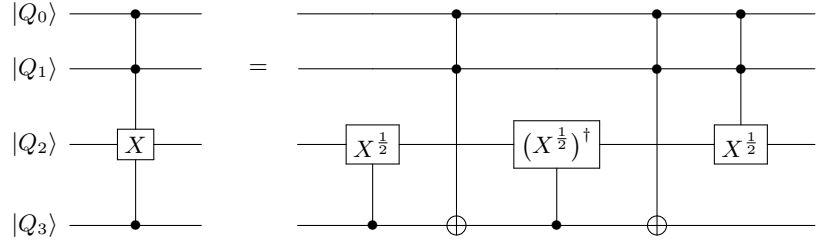

\begin{figure}[H] 
\centering
\mbox{
\Qcircuit @C=2em @R=2em {
\lstick{| Q_0 \rangle} & \ctrl{1}  & \qw & \push{\rule{2em}{0em}}  & \ctrl{1}       & \targ     & \ctrl{1}               & \targ     & \qw & \qw \\
\lstick{| Q_1 \rangle} & \gate{X}  & \qw & \push{=\rule{1em}{0em}} & \gate{X^{\frac{1}{2} }} & \qw       & \gate{ \big{(} X^{ \frac{1}{2} } \big{)}^ \dagger}  & \qw       & \gate{X^{\frac{1}{2} }} & \qw \\
\lstick{| Q_2 \rangle} & \ctrl{-1} & \qw & \push{\rule{2em}{0em}}  & \qw            & \ctrl{-2} & \qw                    & \ctrl{-2} & \ctrl{-1}  & \qw  \\ 
\lstick{| Q_3 \rangle} & \ctrl{-1} & \qw & \push{\rule{2em}{0em}}  & \qw            & \ctrl{-1} & \qw                    & \ctrl{-1} & \ctrl{-1} & \qw \\ 
}
}
\caption{ Quantum circuit for $\tau_{12, 16}$.
} 
\label{tau1216} 
\end{figure}
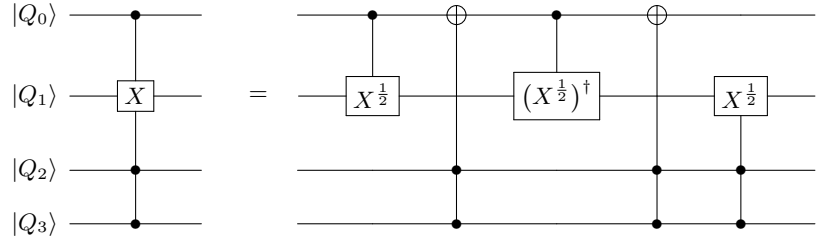

\begin{figure}[H] 
\centering
\mbox{
\Qcircuit @C=2em @R=2em {
\lstick{| Q_0 \rangle} & \ctrl{1}       & \qw & \push{\rule{2em}{0em}}  & \qw            & \ctrl{1} & \qw                    & \ctrl{1} & \ctrl{3} & \qw \\
\lstick{| Q_1 \rangle} & \ctrl{2}       & \qw & \push{=\rule{1em}{0em}} & \ctrl{2}       & \targ    & \ctrl{2}               & \targ    & \qw & \qw \\
\lstick{| Q_2 \rangle} & \qw            & \qw & \push{\rule{2em}{0em}}  & \qw            & \qw      & \qw                    & \qw      & \qw & \qw  \\ 
\lstick{| Q_3 \rangle} & \gate{X^{\frac{1}{2} }} & \qw & \push{\rule{2em}{0em}}  & \gate{X^{\frac{1}{4} }} & \qw      &  \gate{ \big{(} X^{\frac{1}{4} } \big{)}^\dagger }  & \qw      & \gate{X^{\frac{1}{4} }} & \qw \\ 
}
}
\caption{ Quantum circuit for  $U_{ \text{C}_2 X^{ \frac{1}{2}  }} [Q_0 Q_1; Q_3]$. } 
\label{UC2X12q0q1q3} 
\end{figure}
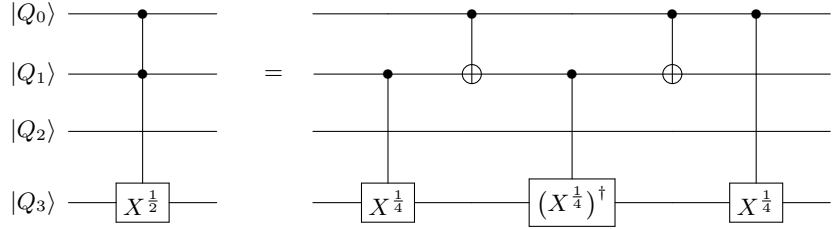

\begin{figure}[H] 
\centering
\mbox{
\Qcircuit @C=2em @R=2em {
\lstick{| Q_0 \rangle} & \ctrl{1}       & \qw & \push{\rule{2em}{0em}}  & \qw            & \ctrl{1} & \qw                    & \ctrl{1} & \ctrl{2} & \qw \\
\lstick{| Q_1 \rangle} & \ctrl{1}       & \qw & \push{=\rule{1em}{0em}} & \ctrl{1}       & \targ    & \ctrl{1}               & \targ    & \qw & \qw \\
\lstick{| Q_2 \rangle} & \gate{X^{\frac{1}{2} }} & \qw & \push{\rule{2em}{0em}}  &\gate{X^{\frac{1}{4} }} & \qw     
 & \gate{ \big{(} X^{\frac{1}{4} } \big{)}^\dagger }  & \qw      
 & \gate{X^{\frac{1}{4} }} & \qw \\ 
\lstick{| Q_3 \rangle} & \qw            & \qw & \push{\rule{2em}{0em}}  & \qw            & \qw      & \qw                    & \qw      & \qw & \qw  \\ 
}
}
\caption{ Quantum circuit for $U_{ \text{C}_2 X^{ \frac{1}{2}  }} [Q_0 Q_1; Q_2]$. } 
\label{UC2X12q0q1q2} 
\end{figure}
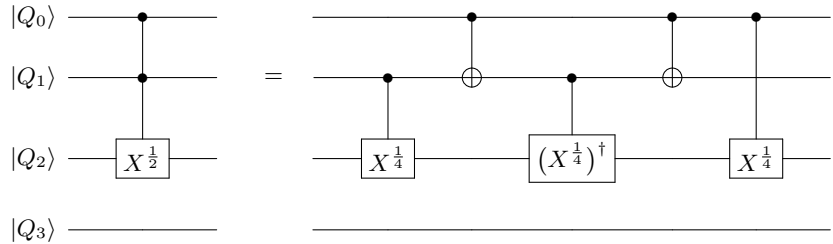

\begin{figure}[H] 
\centering
\mbox{
\Qcircuit @C=1em @R=1.5em {
    \lstick{ | Q_0 \rangle } & \qw                 & \qw &                                           
     & & \qw       & \qw       & \qw                      & \qw       & \qw  & \qw \\
    \lstick{ | Q_1 \rangle } & \gate{X^{\frac{1}{2} }} & \qw &    & 
    & \gate{X^{\frac{1}{4} }} & \qw       & \gate{(X^{\frac{1}{4} })^\dagger} & \qw       & \gate{X^{\frac{1}{4} }}  & \qw \\
    \lstick{ | Q_2 \rangle } & \ctrl{-1}        & \qw & \push{\rule{0.3em}{0em}=\rule{0.3em}{0em}}     & 
    & \ctrl{-1}      & \targ     & \ctrl{-1}                & \targ     & \qw  & \qw \\
    \lstick{ | Q_3 \rangle } & \ctrl{-1}            & \qw &                          
                      & & \qw      & \ctrl{-1} & \qw                      & \ctrl{-1} & \ctrl{-2} & \qw
}
}
\caption{ Quantum circuit for $U_{ \text{C}_2 X^{\frac{1}{2} }} [Q_2 Q_3; Q_1]$.} 
\label{UC2X12q2q3q1} 
\end{figure}
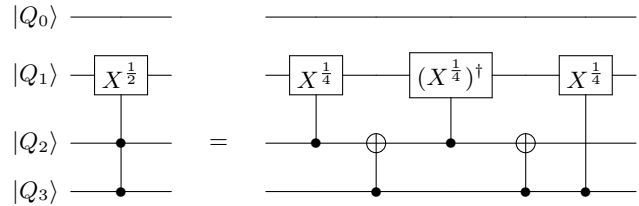

\end{widetext}

\end{document}